\documentclass[journal,comsoc,final,a4paper]{IEEEtran}

\usepackage[T1]{fontenc}
\pdfoutput=1
\IEEEoverridecommandlockouts
\usepackage[shortcuts,acronym]{glossaries}
\usepackage[linesnumbered,ruled,commentsnumbered]{algorithm2e}
\usepackage{algorithmic}
\usepackage{amsmath,amstext,amsthm,amssymb,amsfonts}
\usepackage{array}
\usepackage{balance} 
\usepackage{lastpage} 
\usepackage{bm}
\usepackage[english]{babel}
\usepackage[noadjust]{cite}
\usepackage{color}
\usepackage{esint} 
\usepackage[pdftex]{graphicx}
\usepackage{fancyhdr}
\usepackage{tablefootnote}
\usepackage{mathrsfs}
\usepackage{mathtools} 
\usepackage{multirow}
\usepackage{psfrag}
\usepackage{scalerel}
\usepackage{setspace}
\usepackage{soul}
\usepackage[usenames,dvipsnames,table]{xcolor}
\usepackage[caption=false,font=footnotesize]{subfig}
\usepackage{textcomp}
\usepackage{verbatim}
\usepackage{wasysym}

\newcommand\sfint{\scaleobj{.5}{\fint}}

\makeatletter
\newcommand{\removelatexerror}{\let\@latex@error\@gobble}
\makeatother

\newtheorem{lemma}{\bf Lemma}

\newtheorem{definition}{\bf Definition}

\newcolumntype{L}[1]{>{\raggedright\let\newline\\\arraybackslash\hspace{0pt}}m{#1}}
\newcolumntype{C}[1]{>{\centering\let\newline\\\arraybackslash\hspace{0pt}}m{#1}}
\newcolumntype{R}[1]{>{\raggedleft\let\newline\\\arraybackslash\hspace{0pt}}m{#1}}
\allowdisplaybreaks

\newcommand{\probability}{\text{Pr}}

\newcommand{\vect}{\boldsymbol}

\newcommand{\indicator}[1]{\mathbb{I}_{\{#1\}}}
\newcommand{\seta}[1]{1,\dots,#1}
\newcommand{\set}[1]{\mathcal{#1}}
\newcommand{\setSize}[1]{|#1|}

\newcommand{\indexFrame}{f}
\newcommand{\indexFrameP}{f_p}
\newcommand{\indexFrameRt}{f_r}
\newcommand{\indexFRAME}{F}
\newcommand{\indexFrameSet}{\set{\indexFRAME}}
\newcommand{\frameDeadline}{d_f}

\newcommand{\slot}{t}
\newcommand{\slotArrival}{t_a}

\newcommand{\slotTrans}{T_t}
\newcommand{\slotFrame}{T_f}
\newcommand{\RNNseqLenght}{T_P}
\newcommand{\RNNpredHoriz}{T_H}

\newcommand{\RNNModel}{M}
\newcommand{\RNNParameters}{\theta}

\newcommand{\bs}{b}
\newcommand{\BS}{B}
\newcommand{\bsSet}{\set{\BS}}

\newcommand{\ue}{u}
\newcommand{\ueOther}{\ue^{\prime}}

\newcommand{\UE}{U}
\newcommand{\ueSet}{\set{\UE}}

\newcommand{\ueTrain}{u_{tr}}

\newcommand{\ueSetTrain}{\set{\UE}_{tr}}

\newcommand{\cluster}{k}
\newcommand{\CLUSTER}{K}
\newcommand{\clusterSet}{\set{C}}
\newcommand{\clusterSetUe}{\clusterSet_\cluster}
\newcommand{\clusterSetUeFrame}{\clusterSet_\cluster^\indexFrame}

\newcommand{\video}{v}
\newcommand{\VIDEO}{V}
\newcommand{\videoSet}{\set{\VIDEO}}

\newcommand{\chunk}{c_\indexFrame}

\newcommand{\admission}{a}
\newcommand{\admissionV}{\bm{A}}
\newcommand{\admissionTot}{r}
\newcommand{\admissionTotMax}{\admissionTot_{\textrm{max}}}
\newcommand{\admissionTotUe}{\admissionTot_\ue}

\newcommand{\scheduling}{\eta}
\newcommand{\schedulingV}{\bm{\eta}}
\newcommand{\schedulingUeBsChunk}{\scheduling_{\bs\ue\chunk}}

\newcommand{\rate}{\mu}
\newcommand{\rateMax}{\rate_{\textrm{max}}}
\newcommand{\rateUeChunk}{\mu_{\ue\chunk}}
\newcommand{\rateUeSBS}{\mu_{\bs\ue}}

\newcommand{\txDelayUeFrame}{\tau_{\ue\indexFrame}}
\newcommand{\txDelayUeFrameRt}{\tau_{\ue\indexFrameRt}}

\newcommand{\MTPDelay}{\tau_{\textrm{\tiny{ MTP}}}}
\newcommand{\tMTP}{d_{\textrm{t2\tiny{MTP}}}}
\newcommand{\delayOutage}{\epsilon_{d}}

\newcommand{\queue}{q}
\newcommand{\queueUe}{\queue_\ue}

\newcommand{\auxVbleUe}{\gamma_\ue}
\newcommand{\auxVQueue}{z}
\newcommand{\auxVQueueUe}{\auxVQueue_\ue}
\newcommand{\latencyVQueue}{j}
\newcommand{\latencyVQueueUeChunk}{\latencyVQueue_{\ue\indexFrame}}

\newcommand{\lyapunov}[1]{L(#1)}
\newcommand{\lyapunovDrift}{\Delta L_t}
\newcommand{\lyapunovQueueCombined}{\chi}
\newcommand{\lyapunovConst}{\Delta_0}
\newcommand{\lyapunovWeight}{\alpha_\ue}
\newcommand{\lyapunovWeightQ}{\alpha_\queue}
\newcommand{\lyapunovWeightF}{\alpha_\latencyVQueue}
\newcommand{\lyapunovWeightClusterChunk}{\alpha_{\clusterSet_\cluster}^{\chunk}}
\newcommand{\lyapunovTradeoff}{V_\Delta}

\newcommand{\prefBS}{\succ_\bsSet}
\newcommand{\prefCluster}{\succ_\clusterSet}

\newcommand{\utilityForBSEst}{\hat{U}_{\bsSet}^{\bs,\clusterSetUe}(\slot)}
\newcommand{\utilityForBSEstNoTime}{\hat{U}_{\bsSet}^{\bs,\clusterSetUe}}
\newcommand{\utilityForClusterEst}{\hat{U}_{\clusterSet}^{\clusterSetUe,\bs}(\slot)}
\newcommand{\utilityForClusterEstNoTime}{\hat{U}_{\clusterSet}^{\clusterSetUe,\bs}}
\newcommand{\utilityForClusterEstOther}{\hat{U}_{\clusterSet}^{\clusterSet_\cluster,\Upsilon_{\clusterSet_\cluster}(\slot)}(\slot)}

\newcommand{\movAvLearningWeight}{\nu_{1}}
\newcommand{\movAvLearningSamples}{\nu_{2}}

\newcommand{\matchingClusterSide}{\Upsilon_{\clusterSet_\cluster}(\slot)}
\newcommand{\matchingBSSide}{\Upsilon_\bs(\slot)}

\newcommand{\tile}{n}
\newcommand{\TILE}{N}
\newcommand{\tileSetFOV}{\set{\TILE}}
\newcommand{\tileSetFOVPred}{\widehat{\set{\TILE}}}
\newcommand{\tileSetFOVExt}{\widetilde{\set{\TILE}}}

\newcommand{\antennaGain}{g}
\newcommand{\antennaGainTX}{\antennaGain^{\textup{Tx}}}
\newcommand{\antennaGainRX}{\antennaGain^{\textup{Rx}}}

\newcommand{\nRows}{s_r}
\newcommand{\nColumns}{s_c}

\newcommand{\pathloss}{\ell}
\newcommand{\shadowing}{S\ell}
\newcommand{\shadowingVariance}{\varsigma}
\newcommand{\blockage}{B\ell}
\newcommand{\timeBlockage}{T_{\textrm{block}}}

\newcommand{\frequency}{f_c}
\newcommand{\channel}{h}

\newcommand{\distTwoD}{d^{\text{2D}}}
\newcommand{\distThreeD}{d^{\text{3D}}}

\newcommand{\noiseAlone}{ N_0}
\newcommand{\noise}{\bandwidth_\bs \noiseAlone}
\newcommand{\bandwidth}{BW}

\newcommand{\txpower}{p}

\newcommand{\interference}{I}
\newcommand{\interferenceUe}{\interference_\ue}
\newcommand{\interferenceEst}{\hat{\interference}}
\newcommand{\interferenceEstUe}{\hat{\interference}_\ue}
\newcommand{\interferenceMov}{\tilde{\interference}}
\newcommand{\interferenceMovUe}{\tilde{\interference}_\ue}

\newcommand{\VRbaselineOne}{\scalebox{0.95}{\texttt{UREAC}}}
\newcommand{\VRbaselineTwo}{\scalebox{0.95}{\texttt{MREAC}}}
\newcommand{\VRProp}{\scalebox{0.95}{\texttt{MPROAC}}}
\newcommand{\VRPropPlus}{\scalebox{0.95}{\texttt{MPROAC+}}}

\newacronym{2d}{2D}{2-Dimensional}
\newacronym{3d}{3D}{3-Dimensional}
\newacronym{3dof}{3DoF}{3 degrees-of-freedom}
\newacronym{3gpp}{3GPP}{Third Partnetship Project}
\newacronym{4g}{4G}{Fourth Generation}
\newacronym{5g}{5G}{Fifth Generation}
\newacronym{6dof}{6DoF}{6 degrees-of-freedom}
\newacronym{ai}{AI}{artificial intelligence}
\newacronym{ar}{AR}{augmented reality}
\newacronym{bptt}{BPTT}{Backpropagation Through Time}
\newacronym{bs}{BS}{base station}
\newacronym{cdf}{CDF}{cumulative density function}
\newacronym{cnn}{CNN}{convolutional neural network}
\newacronym{csi}{CSI}{channel state information}
\newacronym{d2d}{D2D}{device-to-device}
\newacronym{da}{DA}{deferred acceptance}
\newacronym{dl}{DL}{downlink}
\newacronym{dnn}{DNN}{deep neural network}
\newacronym{dpp}{DPP}{drift-plus-penalty}
\newacronym{drnn}{DRNN}{deep recurrent neural network}
\newacronym{drl}{DRL}{deep reinforcement learning}
\newacronym{dsrc}{DSRC}{dynamic short range communication}
\newacronym{embb}{eMBB}{enhanced mobile broadband}
\newacronym{eqr}{EQR}{equirectangular}
\newacronym{fov}{FoV}{field of view}
\newacronym{gru}{GRU}{gated recurrent unit}
\newacronym{hd}{HD}{high definition}
\newacronym{hmd}{HMD}{head mounted display}
\newacronym{hrllbb}{HRLLBB}{highly reliable low latency broadband}
\newacronym{ip}{IP}{Internet Protocol}
\newacronym{lidar}{LIDAR}{Laser Imaging Detection and Ranging}
\newacronym{los}{LOS}{line-of-sight}
\newacronym{mbs}{MBS}{macro cell base station}
\newacronym{mcs}{MCS}{modulation and coding scheme}
\newacronym{ml}{ML}{machine learning}
\newacronym{mmwave}{mmWave}{millimeter wave}
\newacronym{mr}{MR}{mixed reality}
\newacronym{mtp}{MTP}{motion-to-photon}
\newacronym{nlos}{NLOS}{non-line-of-sight}
\newacronym{noma}{NOMA}{Non Orthogonal Multiple Access}
\newacronym{ofdma}{OFDMA}{Orthogonal Frequency-Division Multiple access}
\newacronym{phy}{PHY}{physical layer}
\newacronym{pf}{PF}{proportional fair}
\newacronym{poi}{PoI}{point of interest}
\newacronym{pso}{PSO}{Particle Swarm Optimization}
\newacronym{qoe}{QoE}{quality-of-experience}
\newacronym{qos}{QoS}{quality-of-service}
\newacronym{qsi}{QSI}{queue state information}
\newacronym{ran}{RAN}{radio access network}
\newacronym{relu}{ReLU}{rectified linear unit}
\newacronym{rl}{RL}{reinforcement learning}
\newacronym{rnn}{RNN}{recurrent neural network}
\newacronym{rrm}{RRM}{radio resource management}
\newacronym{rss}{RSS}{received signal strength}
\newacronym{rsu}{RSU}{road side unit}
\newacronym{sbs}{SBS}{small cell base station}
\newacronym{scn}{SCN}{small cell network}
\newacronym{sd}{SD}{standard definition}
\newacronym{sinr}{SINR}{signal-to-interference-plus-noise ratio}
\newacronym{son}{SON}{self-organizing network}
\newacronym{sue}{SUE}{small cell user equipment}
\newacronym{tdd}{TDD}{time division duplexing}
\newacronym{udn}{UDN}{ultra dense network}
\newacronym{ue}{UE}{user equipment}
\newacronym{ul}{UL}{uplink}
\newacronym{urllc}{URLLC}{ultra reliable low latency communication}
\newacronym{v2i}{V2I}{vehicle-to-infrastructure}
\newacronym{v2v}{V2V}{vehicle-to-vehicle}
\newacronym{v2x}{V2X}{vehicle-to-everything}
\newacronym{vr}{VR}{virtual reality}
\newacronym{vrx}{VRX}{vehicular receiver}
\newacronym{vtx}{VTX}{vehicular transmitter}
\newacronym{vue}{VUE}{vehicular user equipment}

\begin{document}
\bstctlcite{IEEEexample:BSTcontrol}

\title{Taming the latency in multi-user VR 360$^\circ$: A QoE-aware deep learning-aided multicast framework}
\author{Cristina~Perfecto,~\IEEEmembership{Member,~IEEE},~Mohammed~S.~Elbamby,~\IEEEmembership{Member,~IEEE},\\~Javier Del Ser,~\IEEEmembership{Senior Member,~IEEE,}~Mehdi Bennis,~\IEEEmembership{Senior Member,~IEEE}
	
	\thanks{Manuscript received April 23, 2019; revised 23 August, 2019; revised October 30th, 2019; Accepted December 26, 2019. This research  was  supported  in  part  by the Spanish Ministerio de Economia y Competitividad (MINECO) under grant TEC2016-80090-C2-2-R (5RANVIR), in part by the INFOTECH  project  NOOR, by  the  Kvantum  institute  strategic project SAFARI, by the Academy  of Finland  projects CARMA,  MISSION, SMARTER and 6Genesis Flagship  (grant  no.  318927).}
	\thanks{C. Perfecto and J. Del Ser are with the University of the Basque Country (UPV/EHU), Spain (e-mail: cristina.perfecto@ehu.eus). J. Del Ser is also with TECNALIA and with the Basque Center for Applied Mathematics (BCAM), Spain (e-mail: javier.delser@tecnalia.com).}
	\thanks{M. S. Elbamby and M. Bennis are with the Centre for Wireless Communications (CWC), University of Oulu, Finland (e-mail: \{mohammed.elbamby, mehdi.bennis\}@oulu.fi).}%
} 
\maketitle

\begin{abstract}
	Immersive virtual reality (VR) applications require ultra-high data rate and low-latency for smooth operation. Hence in this paper, aiming to improve VR experience in multi-user VR wireless video streaming, a deep-learning aided scheme for maximizing the quality of the delivered video chunks with low-latency is proposed. Therein the correlations in the predicted field of view (FoV) and locations of viewers watching 360$^\circ$ HD VR videos are capitalized on to realize a proactive FoV-centric millimeter wave (mmWave) physical-layer multicast transmission. The problem is cast as a frame quality maximization problem subject to tight latency constraints and network stability. The problem is then decoupled into an HD frame request admission and scheduling subproblems and a matching theory game is formulated to solve the scheduling subproblem by associating requests from clusters of users to mmWave small cell base stations (SBSs) for their unicast/multicast transmission. Furthermore, for realistic modeling and simulation purposes, a real VR head-tracking dataset and a deep recurrent neural network (DRNN) based on gated recurrent units (GRUs) are leveraged. Extensive simulation results show how the content-reuse for clusters of users with highly overlapping FoVs brought in by multicasting reduces the VR frame delay in 12\%. This reduction is further boosted by proactiveness that cuts by half
	the average delays of both reactive unicast and multicast baselines while preserving HD delivery rates above 98\%. 
	Finally, enforcing tight latency bounds shortens the delay-tail as evinced by 13\% lower delays in the 99th percentile.
\end{abstract}


\glsresetall
\section{Introduction}\label{sec:intro}
 	\IEEEPARstart{V}{irtual} reality \glsunset{vr}(\gls{vr}) is expected to revolutionize how humans interact and perceive media by inducing artificial sensory stimulation to the brain and immersing them into an alternative world \cite{ejder_VR_2017}.
 	Yet, given the well established fact that the quality of the visual feedback highly impacts the sense of presence \cite{VisualEffectInPresence}, for true engagement to succeed \gls{hd} content needs to be consistently streamed while the end-to-end latency or \gls{mtp} delay is kept below 15-20 milliseconds. Otherwise \gls{vr} sickness or cybersickness \textendash a phenomenon similar to motion sickness due to the exposure to low quality or delayed \gls{vr} content\textendash~ might ruin the experience.
 	For this reason, high-end \gls{vr} manufacturers have been long compelled to using wired connections between the \glspl{hmd} and \gls{vr} servers with high processing and storage capabilities. 
 	However, this constraint physically limits the movement of \gls{vr} users and hence, degrades the \gls{qoe}, thereby calling for further development of mobile/wireless solutions that are able to provide both convenience and a high-quality \gls{vr}.
 	Moreover, as access to social \gls{vr} experiences surges, driven by location-based \gls{vr} and 360$^\circ$ formats \cite{ciscoVisualIndexGlobalMobile}, the gap between the available bandwidth and the prohibitively high demands by mobile immersive \gls{vr} is likely to prevail. 
 	Clearly, disruptive content provisioning paradigms are needed to unleash the plethora of new business opportunities for leisure/entertainment industry that mobile interconnected \gls{vr} will bring.
 	In this context, mobile \gls{vr} spearheads the newly coined \gls{hrllbb} use cases sitting across \gls{embb} and \gls{urllc} service categories in \gls{5g} networks \cite{Network_VR}. 
 	The distinctive feature of \gls{hrllbb}, if compared to \gls{urllc}~\cite{jrn:URLLC_Mehdi}, is the need to reliably provide massive data delivery to multiple users with low-latency. 

\begin{figure}[!t]
		\centering
		\includegraphics[width=.85\linewidth]{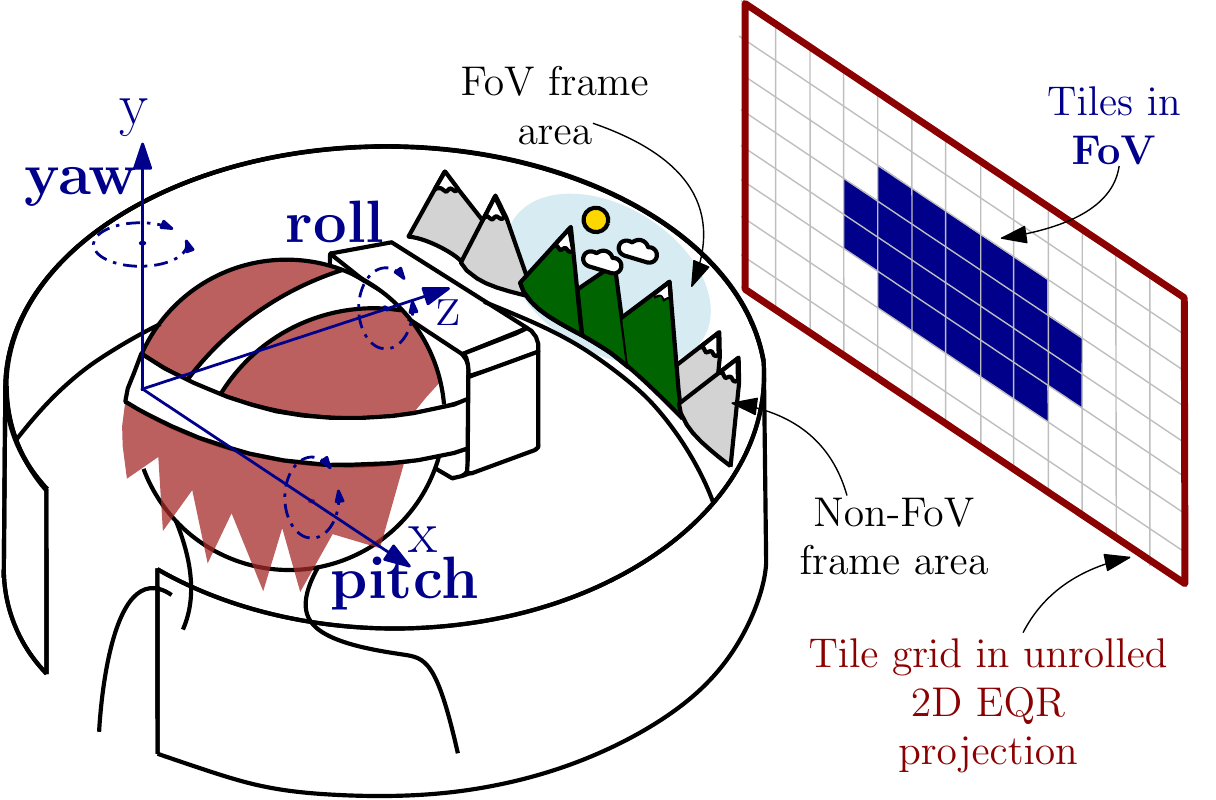}
		\caption{Tiled-FoV mapping of a user's \gls{3dof} pose in the equirectangular (\gls{eqr}) projection of a 360$^\circ$ video frame.}\label{fig:VR_user_FoV}
	\end{figure}

 	Focusing on omnidirectional 360$^\circ$ or spherical video format~\cite{PIEEE_VR} and reducing its bandwidth consumption, video coding solutions that adapt the streaming to users' attention by tracking their visual region of interest are abundant in the literature. 
 	Their common goal is to stream in \gls{hd}\footnote{\label{note1}In the context of 360$^\circ$, 4K is widely viewed as the minimum resolution in current \glspl{hmd}, and ideally 8K or higher is desired.} only users' \emph{viewports}, i.e., the portion of sphere in a user's \gls{fov} while wearing an \gls{hmd} and, optionally, in lower quality the rest.
 	To do so, foveated, tile-based, and projection/viewport-based \gls{fov}-streaming are the most commonly adopted approaches, all requiring real-time tracking of either users' eye-gaze or head angles i.e., the \gls{3dof} \emph{pose} expressed by yaw, pitch and roll angles\footnote{Pitch, yaw and roll head orientation angles represent rotations over the x, y and z-axis, respectively.} as represented in Fig.~\ref{fig:VR_user_FoV}. 
 	Foveated solutions as per \cite{doppler_EUCNC_2017,jnl:GazeTrack_2018} are conditioned to availability of advance eye-gaze tracking mechanisms in the \glspl{hmd} and real-time frame rendering in the servers, whereas for tile and viewport-based solutions, regular head-tracking is enough. 
 	The main drawback of projection-streaming \cite{JChakareski_ICC17_viewport} lies in its large server storage needs, given that for each frame multiple viewpoints are kept. Lastly, in tile-based streaming approaches as per \cite{hosseini_divideconquer_2016, qian_optimCell_2016, Xiao_OptTiling360_2017, Ghosh_tileRateAdapt_2017}, the video frame is divided in a grid of regular tiles, each encoded in both \gls{hd} and at a lower resolution. 
 	Then, only the tiles within a user's \gls{fov} region are streamed in \gls{hd}. 
 
	Real-time tile-based \gls{fov} streaming to a network of \gls{vr} users involves a number of time-consuming steps: edge controllers/servers need to first acquire pose data, process it to compute the set of tiles within the \gls{fov}, and lastly schedule their transmission. 
	Then, on-\gls{hmd} processing will be performed to compose \textendash\emph{stich}\textendash and display the corresponding portion of the video frame. The end-to-end delay of the process is non-negligible.
	Thus, as the number of users in the network increases, operating this cycle within the \gls{mtp} delay budget for each frame for every user becomes challenging; even more so if the server/edge controllers and users are not wired. 
	The latter realization calls for unconventional solutions at different network levels. Among them, the use of proactive transmission of \gls{fov} tiles \textendash contingent on the availability of prediction results on the \gls{fov} for the upcoming frames\textendash~can significantly improve the efficiency of content delivery even with non-perfect prediction accuracy, especially in fluctuating channel environments, as illustrated by the following numerical example:
	Let the tile size be $1$ Mb and the \gls{fov} of the user in each frame have a total of $32$ tiles. This means that without proactivity, a rate budget of $32$ Mb on each frame transmission interval is needed to fully deliver a user's \gls{fov} tiles. Assume that a user's rate budget in the transmission period of two subsequent frames $f_1$ and $f_2$ be $70$ Mb and $10$ Mb, respectively. This will lead to an outage in the second frame. Assume that during the transmission period of frame $f_1$, the \gls{sbs} predicts the user's future \gls{fov} of frame $f_2$ with an accuracy of $75\%$, i.e., $24$ tiles out of the $32$ are correctly predicted. Then, using proactive transmission, the \gls{sbs} can deliver during the first transmission period the $32$ tiles of frame $f_1$ and the predicted $32$ tiles of frame $f_2$. When the second transmission period starts and the user reports its actual \gls{fov} for frame $f_2$, the \gls{sbs} will have to only transmit the remaining $25\%$ of the \gls{fov}, i.e., $8$ tiles, which the user's rate budget allows handling without experiencing any outage.
	
	\begin{figure}[t!]
		\centering
		\includegraphics[width=1\linewidth]{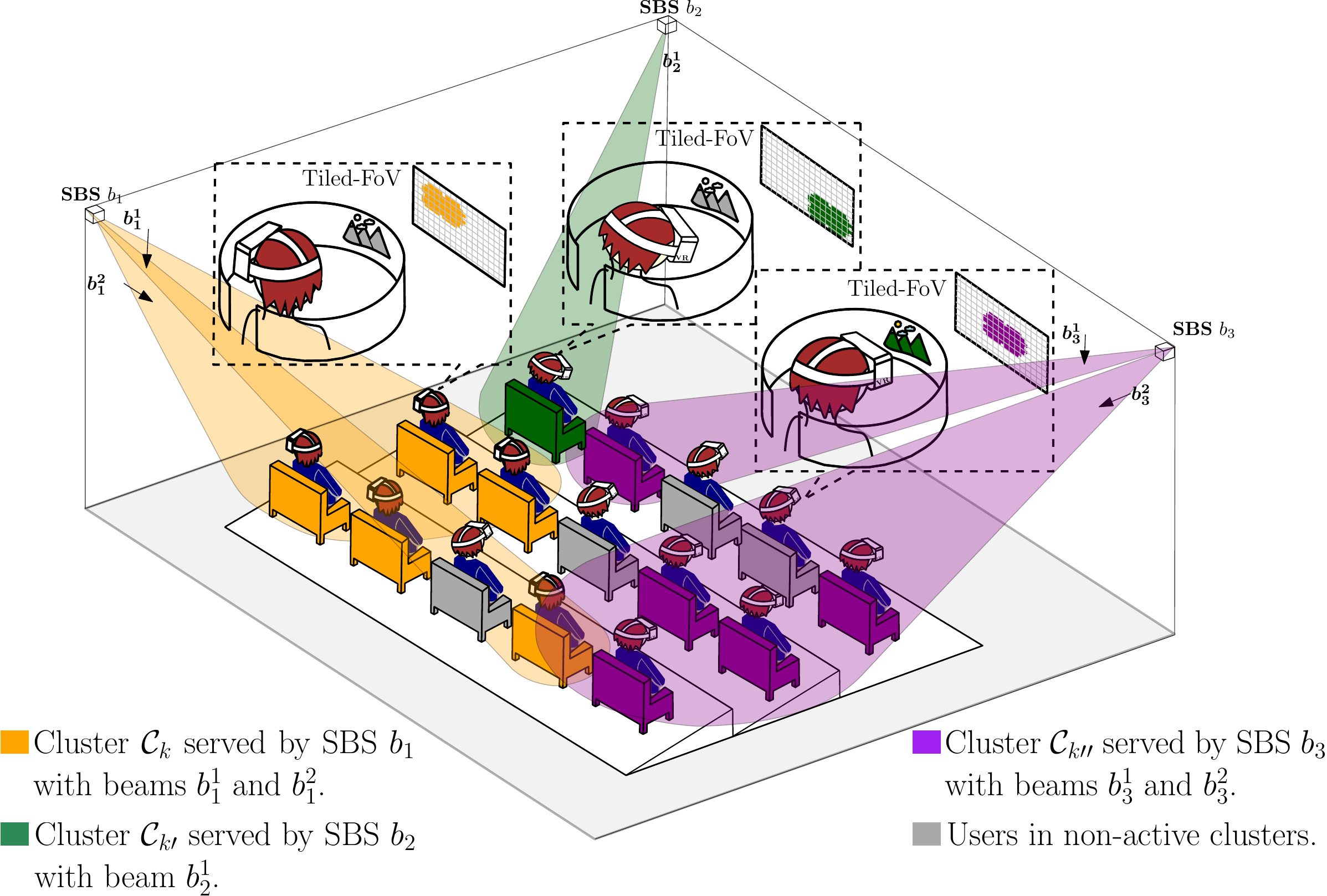}
		\caption{Schematic representation of tiled-FoV mmWave multicast scheduled transmission to VR users with overlapping FoVs. Users belonging to a given cluster are served by a single mmWave \gls{sbs} through different non-overlapping beams of variable beamwidth.}
		\label{fig:multi-beam_cluster_tx}
	\end{figure}
	
	\subsection{Related Work}\label{subsec:rel_work}
 	The need to predict users' fixation opens the door for harnessing \gls{ml} in wireless systems~\cite{walid_machine_learning}. 
 	Moreover, public availability of extensive pose and saliency datasets for 360$^\circ$ video equips an invaluable input to optimize \gls{vr} content streaming through supervised and unsupervised deep learning applied to \gls{fov} prediction.
	To that end, \textit{content-based}~\cite{Cornia2018PredictingLSTMSaliency} or \textit{trajectory-based}~\cite{qian_optimCell_2016,Bao_MotionPredic360VR_2016} strategies can be adopted: the former tries to mimic human capacity to detect the most salient regions of an upcoming video frame i.e., its \glspl{poi}, and predominantly rely on \glspl{cnn} to do so. 
	The latter tracks head (or eye-gaze) movement of \gls{vr} users through their \glspl{hmd} and largely relies on  the use of \glspl{rnn}. However, due to the distinctive nature of 360$^\circ$ videos, the use of content-based information alone might not be suitable since the \glspl{poi} in a video frame might fall out of a user's current \gls{fov} and not affect the viewing transition~\cite{saliency:HowDoPeopleExploreInVR}. 
	Similarly, both having several \glspl{poi} or none within the \gls{fov} might be problematic. 
	Nevertheless, the availability of proper video saliency, could strongly improve the accuracy of the head movement prediction by combining saliency-based head movement prediction with head orientation history, as exemplified by~\cite{nguyen2018yourAttentionIsUnique,nguyen2019_360saliencyDataset,VR_video_dataset_Appl,Li2019_LongTermFoVPrediction}.
	Recently \gls{drl} i.e., the hybridization of deep-learning with \gls{rl}, have provided the required tractability to deal with highly-dimensional state and action space problems. Hence, its application to 360$^\circ$ video streaming has been facilitated. %
	Good illustration of this trend are the works in \cite{Xu2018DRLPredictingHM} and~\cite{Zhang2019DRLfor360VR}, where a \gls{drl}-based head movement prediction and a general video-content independent tile rate-allocation framework are respectively presented.
	Despite the rich recent literature on \gls{fov} prediction, most of these and other works provide bandwidth-aware adaptive rate schemes with a clear focus on Internet \gls{vr} video streaming. 
	Hence, they do not address the particularities of the mobile/cellular environment.

	Furthermore, \glspl{hmd} have limited computing power and storage capacity, so the above predictions need to be offloaded to the network edge. 
	In this sense, the emerging role of edge computing for wireless \gls{vr} and its relation with communication has been largely covered in the recent literature  \cite{Network_VR,Dey_VRAR_2017,mangiante_VREdge_2017,JChakareski_cachingEdgeVR_2017,WCNC_mmWave_VR,MTao_ComputingCachingVR_2018}. Therein, \cite{Network_VR} outlined the challenges and the technology enablers to realize a reliable and low-latency immersive mobile \gls{vr} experience.
	Whereas  \cite{Dey_VRAR_2017,mangiante_VREdge_2017,JChakareski_cachingEdgeVR_2017,WCNC_mmWave_VR,MTao_ComputingCachingVR_2018}, explore different optimization objectives while investigating some fundamental trade-offs between edge computing, caching, and communication for specific \gls{vr} scenarios. 
	Yet, most of the few works considering \gls{vr} multi-user scenarios focus either on direct \cite{conf:NokiaWCNC_Prasad_2018} or \gls{d2d}-aided \cite{conf:Nokia5G-WF_Prasad_2018} content broadcasting, disregarding the potentials for bandwidth savings and caching of existing correlations between \gls{vr} users or contents~\cite{carlsson2019hadUlookedWhereIdid}.
	Also the few works that leverage content correlation through \gls{ml}, such as \cite{Walid_Asilomar_datacorrelation}, none capitalizes prediction related information to perform a proactive mmWave multicast transmission. 
	Furthermore, to the best of the authors' knowledge, imposing high-reliability and low-latency constraints on such wireless \gls{vr} service problem has not been studied so far.

	\begin{table*}[!ht]
		\centering
		\scriptsize
		\renewcommand{\arraystretch}{.95}
		\caption{Summary of Main Notations}\label{tab:Notation}
		\resizebox{0.9\textwidth}{!}{
			\begin{tabular}{|C{2cm}L{6cm}|C{2.2cm}L{5.9cm}|}
				\hline
				\multicolumn{1}{|c|}{\textbf{Symbol}} & \multicolumn{1}{c|}{\textbf{Description}} & \multicolumn{1}{c|}{\textbf{Symbol}} & \multicolumn{1}{c|}{\textbf{Description}} \\ \hline
				\multicolumn{2}{|l|}{\cellcolor{lightgray}\textbf{Time and Indexing}} & \multicolumn{2}{l|}{\cellcolor{lightgray}\textbf{Channel Model}} \\ 
				$\slot$, $\slotTrans$ & Time index and slot duration & $\channel_{\bs\ue}$ & \multicolumn{1}{l|}{Channel gain for user $\ue$ from SBS $\bs$} \\
				$t_a$ & Frame request arrival time & $P\ell_{bu}$, $S\ell_{bu}$, $B\ell_{bu}$ & \multicolumn{1}{l|}{Pathloss, shadowing and blockage from $\bs$ to user $u$} \\
				$\indexFrame$ & Video frame index & $P\ell_{\text{LOS}}$, $P\ell_{\text{NLOS}}$ & \multicolumn{1}{l|}{LOS and NLOS pathloss} \\
				$\slotFrame$ & Time between frames & $\shadowingVariance_{\text{LOS}}^{SF}$, $\shadowingVariance_{\text{NLOS}}^{SF}$ & \multicolumn{1}{l|}{LOS and NLOS shadowing variance} \\
				$\indexFrameRt$ & Real-time frame index & $\frequency$ & \multicolumn{1}{l|}{Normalized central frequency} \\
				$\indexFrameP$ & Prediction frame index & $d_{\bs\ue}^\textup{2D}$, $d_{\bs\ue}^\textup{3D}$ & \multicolumn{1}{l|}{Azimuth and elevation plane distance} \\ 
				
				\multicolumn{2}{|l|}{\cellcolor{lightgray}\textbf{Sets}} & $\timeBlockage$ & \multicolumn{1}{l|}{Time between blockage events} \\ 
				$\ueSet$, $\ueSetTrain$ & VR users (test set) and training users & 
				
				\multicolumn{2}{l|}{\cellcolor{lightgray}\textbf{Communication Model}} \\ 
				$\bsSet$ & SBSs & $\antennaGainTX_{\bs\ue}$, $\antennaGainRX_{\bs\ue}$ & \multicolumn{1}{l|}{Transmit/receive antenna gains of SBS $b$ to user $u$} \\
				$\videoSet$ & Videos & $\varphi^{t_x}$, $\varphi^{r_x}$ & Transmit and receive beamwidths\\
				$\indexFrameSet$ & Frames indexes in a video & $\vartheta_{bu}^{t_x}$, $\vartheta_{bu}^{r_x}$ & \multicolumn{1}{l|}{Transmit and receive beams angular deviation.} \\
				$\clusterSet$ & VR clusters & $\txpower_{\bs}$, $\txpower_{\bs'}$ & \multicolumn{1}{l|}{Transmit powers of SBS $b$ and interfering SBSs $b'$} \\
				$\clusterSet_\cluster^\indexFrame$ & VR users in $\cluster$-th cluster for video frame index $\indexFrame$ & $\bandwidth_{\bs}$ & \multicolumn{1}{l|}{Bandwidth for SBS $b$ in mmWave band} \\
				$\tileSetFOV_{\ue}^{f}$, $\tileSetFOV_{\clusterSet_\cluster}^{f}$ & FoV tiles of user $\ue$ and cluster $\clusterSet_\cluster$ for frame index $\indexFrame$ & $\textrm{SINR}_{bu}$ & SINR for user $u$ being served by SBS $b$\\
				$\tileSetFOVPred_{\ue}^{f}$, $\tileSetFOVPred_{\clusterSet_\cluster}^{f}$ & Predicted tiles in the FoV of user $\ue$ and of cluster $\clusterSet_\cluster$ at frame index $\indexFrame$ &\smash{$\textrm{SINR}_{\bs,\clusterSetUeFrame}^{\chunk}$}&Multicast SINR of chunk $\chunk$ from SBS $b$ to cluster $\clusterSetUeFrame$\\

				\multicolumn{2}{|l|}{\cellcolor{lightgray}\textbf{Problem Formulation}}&$\interferenceUe$ & Instantaneous interference for user $\ue$ \\ 
				
				$\admissionTotUe(\slot)$ & Total traffic admission for user $u$ &$\noiseAlone$ & \multicolumn{1}{l|}{Noise power spectral density} \\ 
				$L_{c_{f}}$ & Data size of chunk $c_{f}$&\multicolumn{2}{l|}{\cellcolor{lightgray}\textbf{Lyapunov Optimization Framework}}\\ 
				
				$\admission_{\ue_\indexFrame}$, $\scheduling_{\bs\ue\chunk}$ & Chunk admission and scheduling variables&$\queueUe$ & \multicolumn{1}{l|}{Traffic queue of user $u$}\\
				
				$\tau_{uf}$ & Transmission delay of frame $f$ to user $u$&$\auxVQueueUe$, $\latencyVQueueUeChunk$ & \multicolumn{1}{l|}{Virtual queues of time averaged constraints} \\ 
				$\delayOutage$ & \multicolumn{1}{l|}{Delay reliability metric}& $\gamma_{u}$ & \multicolumn{1}{l|}{Auxiliary variables for the EOP} \\
				$\rateUeChunk$ & Rate of delivering chunk $c_{f}$ to user $u$ & $\lyapunov{\cdot}$, $\lyapunovDrift$ & \multicolumn{1}{l|}{Lyapunov and Lyapunov-drift functions} \\
				$\tMTP$& Time left to schedule before MTP is exceeded& $\lyapunovTradeoff$ &\multicolumn{1}{l|}{Lyapunov drift-plus-penalty trade-off variable} \\ 
				$\MTPDelay$ & Motion-to-photon latency&\multicolumn{2}{l|}{\cellcolor{lightgray}\textbf{FoV Prediction and DRNN}} \\ 
				
				\multicolumn{2}{|l|}{\cellcolor{lightgray}\textbf{Matching Theory}}&$\RNNModel_{\bm{\RNNParameters}}^{\video,\RNNpredHoriz}$&Supervised learning model for video $\video$ and $\RNNpredHoriz$\\
				$\bm{\Upsilon}$, $\succ_\bs$, $\succ_{\clusterSetUe}$ & Matching function and preference relations&$\RNNModel$, $\theta$&  Learning algorithm and parameter set\\
				$U_{\bsSet}^{\bs,\clusterSetUe}$, $U_{\clusterSet}^{\clusterSetUe,\bs}$& Clusters and SBSs utilities&$\bm{X_{tr}}^{\video}$, $\bm{Y_{tr}}^{\video,\RNNpredHoriz}$ & Training dataset and binary-encoded matrix of target tiles\\ 
				$\utilityForBSEstNoTime$, $\utilityForClusterEstNoTime$&Modified utilities over the estimated parameters&$\RNNpredHoriz$, $\RNNseqLenght$ & \multicolumn{1}{l|}{Prediction horizon and input sequence length} \\ 
				$\movAvLearningWeight$, $\movAvLearningSamples$&Weight and sample number of moving average procedure&$\boldsymbol{p}_{3\ueTrain}^{\indexFrame}$, $\boldsymbol{p}_{3\ue}^{\indexFrame}$& 3DoF pose vectors of a training user $\ueTrain$ and of a user $\ue$\\
				$\interferenceEstUe$, $\interferenceMovUe^{\nu_2}$ & Estimated/moving-average interference for user $\ue$ &$r$, $\Gamma$& Reset and update gates of the GRU\\
				\multicolumn{2}{|l|}{\cellcolor{lightgray}\textbf{User Clustering}}&$\bm{h}_{f-1}$, $\bm{\tilde{h}}_f$, $\bm{h}_{f}$ &Previous, candidate and new hidden states in GRU\\
				$\tilde{d}_{\ue,\ueOther}^{\indexFrameP}$,$d_{\ue,\ueOther}^{\indexFrameP}$&FOV and FOV+ user location based clustering distances&$\alpha$, $\beta_1$, $\beta_2$ &Learning rate and parameters for Adam algorithm\\
				\hline
			\end{tabular}
		}
	\end{table*}
 %
   Therefore, this manuscript proposes to incorporate \gls{ml} and multicasting into the optimization problem of maximizing the streaming quality of FoV-based 360$^\circ$ videos with \gls{hrllbb} guarantees.
   The use of \gls{ml} to predict users' \gls{fov} in advance and leverage inter-user correlations is pivotal to the system. 
   Then, building upon the aforementioned correlations, multicast transmissions aimed for clusters of users with partially or fully overlapping FoVs will be proactively scheduled such that strict latency bounds are kept. 
   Moreover, the adoption of \gls{mmwave} frequency band communications \textendash where at each time slot a given \gls{sbs} will steer multiple spatially orthogonal beams towards a cluster of users\textendash~ to transmit contents is key to benefiting from high transmission rates that contribute to a reduced on-the-air delay. 
 
 \noindent For clarity, we summarize the main contributions of this paper as follows:
	\begin{itemize}
			\item To provide high capacity while ensuring bounded latencies in wireless 360$^\circ$ \gls{vr} streaming, we propose a proactive physical-layer multicast transmission scheme that leverages future content and user location related correlations.
			\item We model the problem as a quality maximization problem by optimizing the users' \gls{hd} frame request admission and scheduling to \glspl{sbs} and, by borrowing tools from dynamic stochastic optimization, we recast the problem with traffic load stability and latency bound considerations. Subsequently, to solve it, a low complexity and efficient algorithm based on matching theory is proposed. 
			\item To validate our proposed scheme, for \gls{fov} prediction we developed a \gls{drnn} based on \glspl{gru} that is trained with a dataset of real 360$^\circ$ \gls{vr} poses. The predicted \glspl{fov} for a given time horizon are thus available to dynamically cluster users based on their \gls{fov} overlap and proximity within a VR theater, as well as to provide insights on how much the scheme is affected by the accuracy of the prediction and the clustering decisions. 
			\item Extensive simulations are conducted to investigate the effects of the prediction horizon, of the \gls{vr} frame size, the clustering, and of the network size, which conclude that the proposed approach outperforms considered reference baselines by delivering more \gls{hd} quality frames, while ensuring tight transmission delay bounds.
	\end{itemize}

	The remaining of this manuscript is organized as follows: In Section \ref{sec:sys_mod} the system model and the underlying assumptions are described. 
	The optimization problem of wireless \gls{vr} video content delivery is formulated in Section~\ref{sec:prob_form}.
	Section~\ref{sec:Matching Theory} presents our proposed matching theory algorithm to schedule wireless multicast/unicast transmission resources for \gls{vr} video chunks under latency constraints. 
	A detailed description of the \gls{fov} prediction and adopted user clustering schemes is provided in Section~\ref{sec:DRNN}. 
	Simulation results and performance evaluation are presented in Section~\ref{sec:sim_results}. Finally, Section~\ref{sec:concl} concludes the paper. 

	\textit{Notations}: Throughout the paper, lowercase letters, boldface lowercase letters, (boldface) uppercase letters and italic boldface uppercase letters represent scalars, vectors, matrices, and sets, respectively. $\mathbb{E}[\cdot]$ denotes the expectation operator and $\probability(\cdot)$ the probability operator. Function of $z$ and utility of $z$ are correspondingly expressed as ${\displaystyle\sfint}(z)$ and by $U(z)$. $\indicator{z}$ is the indicator function for logic $z$ such that $\indicator{z}=1$ when $z$ is true, and $0$, otherwise. The cardinality of a set $\mathcal{S}$ is given by $S\!=\!|\mathcal{S}|$. Moreover, ${[z]}^{+}\!\triangleq\!\max(z,0)$ and $\overline{z}$ stands for the time average expectation of quantity $z$, given by $\overline{z}=\lim_{T\rightarrow\infty}\frac{1}{T}\sum_{t=1}^{T}\mathbb{E}[z(t)]$. Lastly, $\circledast$ represents the Hadamard product, \texttt{tanh($z$)}$=\frac{e^{z}-e^{-z}}{e^{z}+e^{-z}}$ is the hyperbolic tangent and 
 	$\sigma(z)=\frac{1}{1+e^{-z}}$ the sigmoid activation functions for the neural network.

\section{System Model}\label{sec:sys_mod} 
	In this section we introduce the system model which encompasses the considered deployment scenario, as well as the adopted wireless channel and communication models. 
	For ease of reading, a non-comprehensive list of the notations used throughout the rest of the manuscript is provided in Table \ref{tab:Notation}.

	\subsection{Deployment Scenario}\label{subsec:SM_Scenario}
 	We consider a \gls{vr} theater with seats arranged in $\nRows$ rows and $\nColumns$ columns, and where a network of \gls{vr} users $\ueSet$, all wearing \gls{mmwave} \glspl{hmd}, are located.
 	In this scenario, each user chooses to watch an \gls{hd} $360^\circ$ \gls{vr} video $\video\in\videoSet$, with $\videoSet$ denoting the set of available \gls{vr} videos in the catalog. 
 	Due to their large size and to limited storage capacity in the \glspl{hmd}, the \gls{hd} frames of videos are cached in the edge network and are delivered to users through $\BS=\setSize{\bsSet}$ \glspl{sbs} distributed around the theater. To ensure timely and smooth streaming experience, a lower-quality \gls{sd} version of the video frames are assumed to be pre-cached in the user's \gls{hmd} to be streamed if the HD frames are not successfully delivered on time.
 	The \glspl{sbs} operate in the \gls{mmwave} band and are endowed with multi-beam beamforming capabilities to boost physical layer multicast transmission \cite{MulticastPHY2006} of shared video content to users grouped into clusters. This  setting is graphically represented in Fig.~\ref{fig:multi-beam_cluster_tx}.

 	In the network edge, without loss of generality, we assume that all videos in the catalog are encoded at the same frame rate $1/\slotFrame$ \textendash with $\slotFrame$ the time between frames\textendash~and have the same length i.e., consist of a set of frames $\indexFrameSet\!=\!\{f\}_{\indexFrame=1}^\indexFRAME\!\subset\!\mathbb{N}$. 
 	Moreover, the frames from the spherical videos are unwrapped into a 2D \glsreset{eqr}\gls{eqr}\label{page_eqr} or lat-long projection with pixel dimensions $P_{H}\!\times\!P_{V}$ and divided into $\tileSetFOV\!=\!\{\seta{\TILE}\}$ partitions or \emph{tiles}  arranged in an $N_{H}\!\times\!N_{V}$ regular grid, so that $\TILE\!=\!\TILE_{H}\cdot\TILE_{V}$ and each tile is sized $P_{H}/\TILE_{H}\!\times\! P_{V}/\TILE_{V}$ pixels.
 	Therefore, when watching any given video  $\video\in\videoSet$, the \gls{fov} of user $\ue\in\ueSet$ during frame $\indexFrame\in\indexFrameSet$ can be expressed as a tile subset $\tileSetFOV_{\ue}^{\indexFrame}\subseteq\tileSetFOV$.

	In a nutshell, two main indices are used in our manuscript: one refers to the decision slots indexed by $t\!=\!\{1,2,\cdots\}$ and slot duration $\slotTrans$ which is the prevailing time scale enforced for transmission/scheduling purposes, whereas the second index $\indexFrame$ refers to the video frame index. Hence, the slotted time index $t$ and frame index $\indexFrame$ satisfy $\indexFrame\!\triangleq\!\lceil\frac{t*\slotTrans}{\slotFrame}\rceil$ so that \smash{$\indexFrame\!=\!\{1,2,\cdots\}\!\in\!\mathbb{N}$}. With this division in mind, \emph{chunk} hereafter denotes the part of the video that corresponds to one frame in time and one tile in \gls{eqr} space.

	\subsection{FoV and Spatial Inter-user Correlation}\label{subsec:Fov+SpatialCorrelation}
\begin{figure*}[!ht]
	\centering
	\subfloat[][]{
		\centering
		\includegraphics[width=.75\linewidth]{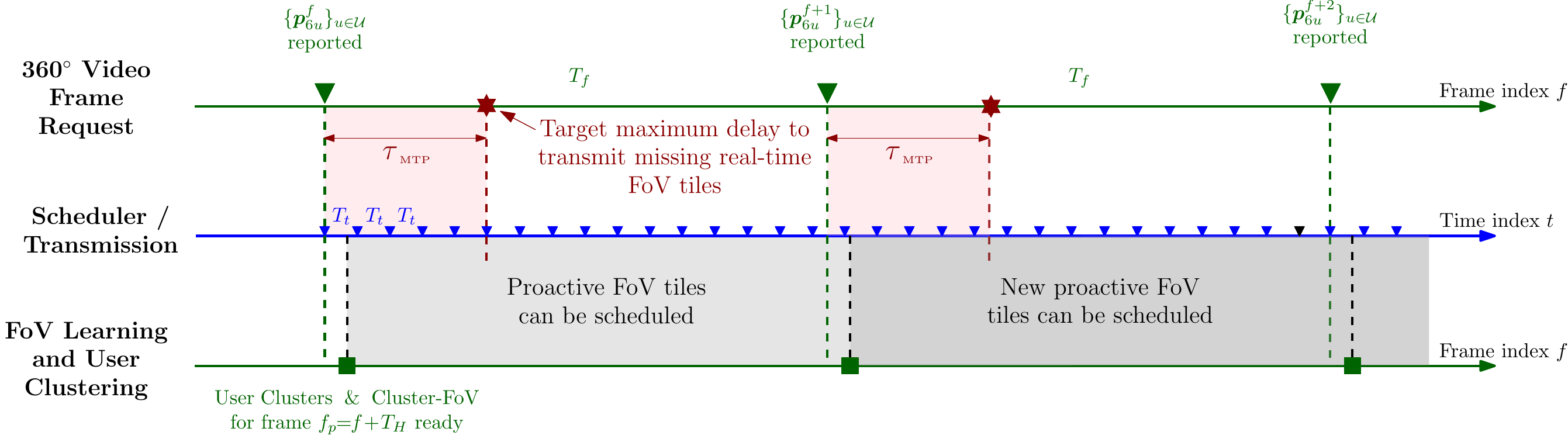}
		\label{fig:timing}}
	\\
	\subfloat[][]{
		\centering
		\includegraphics[width=.8\linewidth]{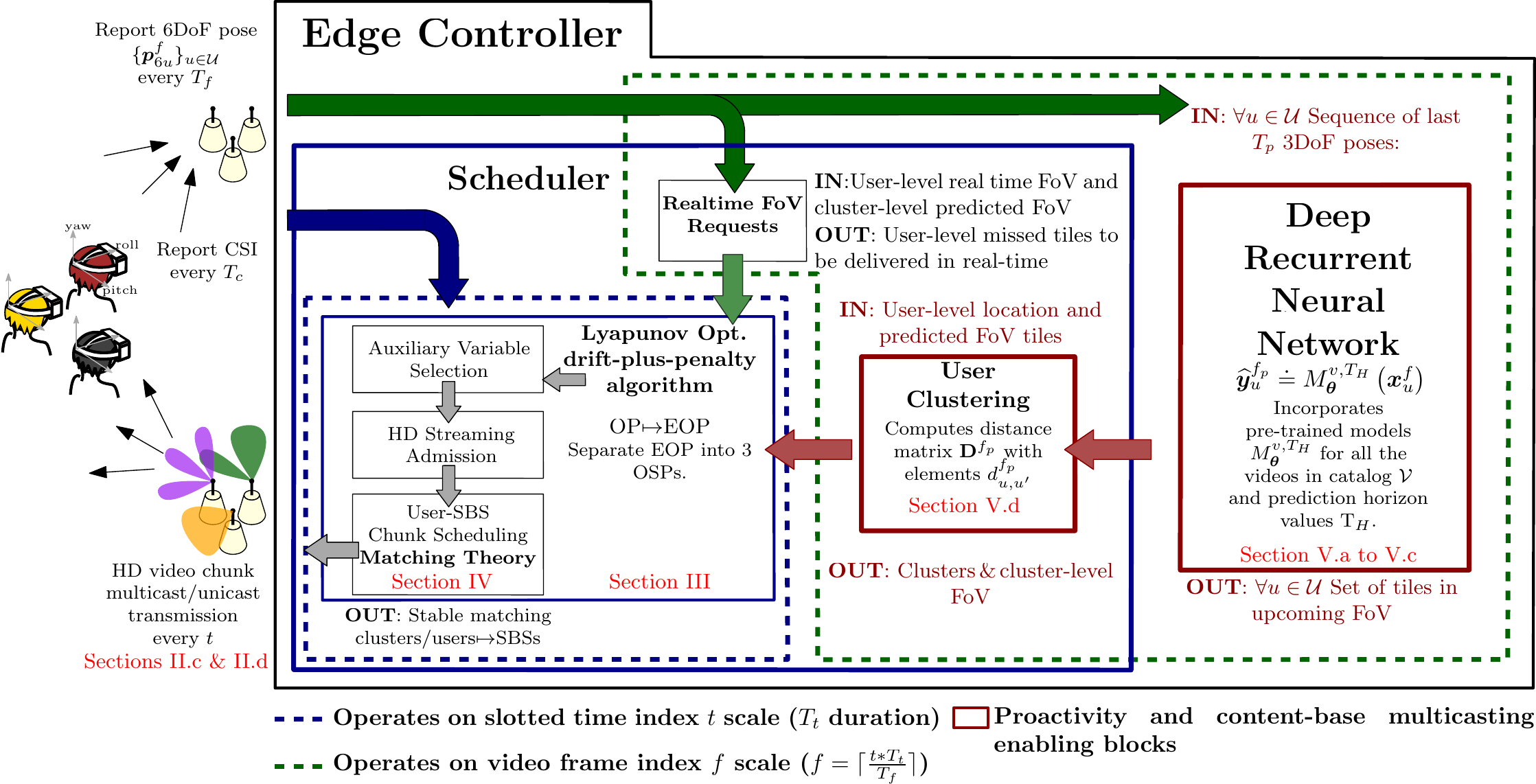}
		\label{fig:block_flow}}
	\caption{\protect\subref{fig:timing} Timing sequence to showcase operation at transmission/scheduling level and at videoframe level. \protect\subref{fig:block_flow} Flowchart of the proposed DRNN aided wireless scheduling of HD 360$^\circ$ video FoV chunks. Once user pose has been reported in the UL, FoV prediction and user clustering follow. Subsequently, a scheduler balances HD chunk admission vs. queue stability subject to latency constraints. Finally a matching theory algorithm associates scheduled chunks to \gls{sbs}-user clusters pairs leveraging multi-beam mmWave multicast transmission in the DL.}
	\label{fig:block_flow_complete}
	\end{figure*}

	To leverage \gls{fov} and spatial correlations between users in the \gls{vr} theater deployment from Section \ref{subsec:SM_Scenario}~, and as outlined in Fig. \ref{fig:block_flow}~, we assume that users report in the \gls{ul} their video index $\video$ and their \gls{6dof} pose every $\slotFrame$ ms. This \gls{6dof} pose includes head orientation angles and x, y and z-axis coordinates\footnote{\label{note2}We notice here that 360$^\circ$ videos are displayed from a centered perspective whereby users are only allowed to look around. Therefore, as users' physical location do not affect their view, for \gls{fov} prediction purposes only the \gls{3dof} head orientation needs to be considered, even if the whole \gls{6dof} pose is reported.}. 
	The information is then forwarded from the \glspl{sbs} to the edge controller where \gls{fov} prediction, user-clustering and scheduling decisions take place.
	
	New real-time/proactive chunk scheduling decisions will be taken every $\slotTrans$ such that for all users chunks are scheduled for \gls{dl} transmission by the \glspl{sbs}, and delivered before the frame deadline $\frameDeadline$ expires. In this regard, the chunks that correspond to  real-time i.e., $\indexFrame\!\!=\!\!\indexFrameRt$, and to predicted future \glspl{fov} will be given by $\tileSetFOV_{\ue}^{\indexFrameRt}$ and by $\{\tileSetFOVPred_{\ue}^{\indexFrame}\}_{\indexFrame\!=\!\indexFrameRt\!+\!1}^{\indexFrameP}$, respectively. 
			
	To provide the estimated \glspl{fov}, let a supervised \gls{ml} model $\RNNModel_{\bm{\RNNParameters}}^{\video,\RNNpredHoriz}$ be defined in the edge controller \textendash with $\RNNModel$ denoting the model's learning algorithm and $\bm{\RNNParameters}$ its parameter set\textendash~associated to each video $\video\!\!\in\!\!\videoSet$ and time horizon\footnote{\label{note3}Without loss of generality, that the time horizon for the prediction $\RNNpredHoriz$ is measured as an integer multiple of the frames.} $\RNNpredHoriz$ for which a model is constructed as 
	\begin{equation}\label{eq:learningModel}
	\widehat{\bm{y}}_{\ue}^{\indexFrameP}\triangleq \RNNModel_{\bm{\RNNParameters}}^{\video,\RNNpredHoriz}(\bm{x}_{\ue}^{\indexFrame}).
	\end{equation}
	
	Once the model's offline training, as detailed in Section \ref{subsec:DRNN_training}, has been completed and its parameters are known, given a sequence $\bm{x}_\ue^{\indexFrame}$ of $\RNNseqLenght$ length of past \gls{3dof} poses\footnote{\label{note4}This sequence consists of the user's last $\RNNseqLenght$ recorded head angles, as per $\boldsymbol{p}_{3\ue}^{\indexFrame}$ with video frame indexes \smash{$\indexFrame\!\!\in\!\{\indexFrameRt\!-\!\RNNseqLenght\!+\!1,\ldots,\indexFrameRt\}$}.} collected in $\bm{x}_\ue^{\indexFrame}$, the model will produce the vector of labels  $\widehat{\bm{y}}_{\ue}^{\indexFrameP}\triangleq\{\widehat{y}_{\ue,\tile}^{\indexFrameP}\}_{\tile=1}^{\TILE}$ for frame index $\indexFrameP=\indexFrame+\RNNpredHoriz$ as per \eqref{eq:learningModel}. The corresponding set of tiles in the predicted \gls{fov} is a mapping such that $\{\tileSetFOVPred_{\ue}^{\indexFrameP}=\forall \tile\in [\seta{\TILE}]\hspace{-1mm}:\widehat {y}_{\ue,\tile}^{\indexFrameP}=1\},  \forall\ue\in\ueSet$.
	
	Subsequently, the predicted \glspl{fov} and reported poses will be fed into a user-clustering module whereby users watching the same \gls{vr} video $\video$ will be grouped together based on their \gls{fov} and spatial correlation. The inputs for the scheduler will therefore be: $\forall\ue\!\in\!\ueSet$ the real-time \gls{fov} tile-sets $\tileSetFOV_{\ue}^{\indexFrameRt}$ for the current index frame $\indexFrame\!=\!\indexFrameRt$ as well as the predicted $\CLUSTER$ user-clusters $\{\mathcal{C}_{k}^{f_p}\}_{k=1}^{\textup{K}}\mid \bigcup_{k=1}^\textup{K}\mathcal{C}_{k}^{\indexFrameP}=\mathcal{U}$ with their corresponding cluster-level predicted FoVs \smash{$\{\tileSetFOVPred_{\mathcal{C}_k}^{\indexFrameP}=\bigcup_{u\in\mathcal{C}_k^{f_p}}\tileSetFOVPred_{\ue}^{\indexFrameP}\}_{k=1}^K$}.
	 
	Since spherical videos are not locally cached, a huge imbalance emerges in the amount of information being sent in the \gls{ul} vs. \gls{dl}. Therefore, for the purpose of this manuscript we will only focus on the effect of the \gls{dl} \gls{hd} $360^\circ$ video transmission from edge \glspl{sbs} to \gls{vr} users, and assume that enough \gls{ul} resources are available to users for timely pose update and \gls{csi} report.
	
 	Also, in the remaining of the manuscript and for a given time horizon $\RNNpredHoriz$, following the \gls{ul} report of users' pose, the availability of the real-time and predicted FoVs as well as of user-clustering partitioning results is assumed. The detailed description of the proposed \gls{fov} prediction and user-clustering schemes with their algorithmic implementation details to produce such inputs are provided in Section \ref{sec:DRNN}. Next, the wireless channel model and the communication model between the \glspl{sbs} and the VR users are presented.

\subsection{mmWave Channel and Communication Model}\label{subsec:channel_model}

	At \gls{mmwave} frequencies, due to the quasi-optical nature of electromagnetic wave propagation, signals are highly directional. For that reason channels are composed of a single-path propagation component for the dominant path and a set of multi-path components. For tractability and without loss of generality, in this paper we will neglect the multi-path components and consider only the dominant path for the purpose of \gls{vr} wireless streaming. 

	In this single-path, we adopt the \glsdisp{3gpp}{3GPP} contributed channel model \cite{ChannelModel2018} which is valid for frequencies ranging from 0.5 to 100 GHz and bandwidths of up to 10\% of the center frequency not exceeding 2 GHz. Among the different scenarios therein, typical indoor deployment cases including office environments and shopping malls, are showcased. Selecting the indoor open office scenario, with user devices and \glspl{sbs} located at 1 m and 3 m height respectively, a distance dependent \gls{los} probability is defined as 
	\begin{equation}\label{eq:LOS_prob}
		\probability(\text{\scalebox{0.75}{\gls{los}}})=
		\begin{cases}
			1, & \distTwoD_{\bs\ue}\leq 5\textup{ m},\\
			\text{exp}\bigl(-\frac{\distTwoD_{\bs\ue}-5}{70.8}\bigr), & 5\textup{ m}<d_{\bs\ue}^{\text{2D}}\leq 49\textup{ m},\\
			0.54\text{exp}\bigl(-\frac{\distTwoD_{\bs\ue}-49}{211.7}\bigr), & 49\textup{ m}< d_{\bs\ue}^{\text{2D}},\\
		\end{cases}
	\end{equation}
  	where $\distTwoD_{\bs\ue}$ stands for the distance in meters between the \gls{sbs} and the user in the azimuth plane. Subsequently, results from $\probability(\text{\scalebox{0.75}{\gls{los}}})$ are exploited to calculate the large-scale fading effects in the channel. Specifically, pathloss $\pathloss$ is given (in dB) as follows,
	\begin{align}
		\pathloss_{\text{LOS}} &= 32.4+17.3\cdot \text{log}_{\text{10}}{}\distThreeD_{\bs\ue}+20\cdot \text{log}_{\text{10}} \frequency,\label{eq:PathlossLOS}\\
		\pathloss_{\text{NLOS}}^{prev} &= 38.3\cdot \text{log}_{\text{10}} \distThreeD_{\bs\ue}+17.3+24.9\cdot \text{log}_{\text{10}} \frequency ,\label{eq:PathlossNLOS}\\  
		\pathloss_{\text{NLOS}}&= \max(\pathloss_{\text{LOS}},\pathloss_{\text{NLOS}}^{prev}),
	\end{align}
 	\noindent with $\distThreeD_{\bs\ue}$, $\frequency$ in \eqref{eq:PathlossLOS} and \eqref{eq:PathlossNLOS} representing the distance in meters between the \gls{sbs} and the user in the elevation plane and the channel's central frequency normalized with 1 GHz, respectively. A log-normally distributed shadowing fading loss $\shadowing$, with standard deviation for the \gls{los} and \gls{nlos} cases of $\shadowingVariance_{\text{\gls{los}}}^{\shadowing}=3$ dB and $\shadowingVariance_{\text{\gls{nlos}}}^{\shadowing}=8.03$ dB respectively, supplements the large-scale fading calculations. 
 
 	In addition to the pathloss and shadowing fading, the channel intermittency due to sporadic human blockage is also accounted for. This blockage $\blockage(\slot_B)$ could hinder the communication by bringing 20-30 dB penalty upon the channel gain. To that end, based on the spatial location of users within the \gls{vr} theater, the prospective human blockers that might obstruct the direct ray in the azimuth plane between a given user and each of the available \glspl{sbs} are counted. Thereupon, the count-weighted probabilistic arrival of blockage-events is evaluated every $\timeBlockage$, with the value of $\timeBlockage$ chosen such that $\timeBlockage/T_t\gg 1$ and the index $\slot_B$ satisfy $\slot_B\!\triangleq\!\lceil\frac{t*\slotTrans}{\timeBlockage}\rceil$. The reason for operating on larger time-scale lies on correlation between blockage events along several transmission intervals due to the relative slowness of human head and body limb movement with respect to other channel fading effects. Indeed, human blockage durations of few hundreds of ms or more are reported in the literature~\cite{jnl:StatBlockageModelling_Raghavan2019}. By combining the channel as per \cite{ChannelModel2018} with the statistic blockage model, both dynamic channel fluctuations and the arrival of sporadic and longer-lasting human blockage events are suitably captured. Accordingly, the channel gain $h_{bu}(\slot)$ in dB from \gls{sbs} $b\!\in\!\mathcal{B}$ to user $u\!\in\!\mathcal{U}$ is given as
 	\begin{equation}\label{eq:channel_loss}
 		\channel_{\bs\ue}(\slot)=\pathloss_{\bs\ue}(\slot)+\shadowing_{\bs\ue}(\slot)+\blockage_{\bs\ue}(\slot_B).  
 	\end{equation}

 	To benefit from multi-beam transmission, we assume that \glspl{sbs} are equipped with a limited number of radio frequency (RF) chains, whereas users' \glspl{hmd} will have a single RF chain, limiting their beamforming and combining capability. 
 	These assumptions are grounded on current high costs and power consumption of analog-to-digital converters for \gls{mmwave} frequencies.
 	For tractability, the actual antenna radiation pattern is approximated by a 2D sectored antenna model  \cite{wildman_2DsecAntenna_2014} in the \glspl{sbs} and in the \glspl{hmd}. In this model antenna gains are considered constant within the mainlobe, and equal to a smaller constant in the sidelobes.
 	 Let $g_{bu}^{\textup{Tx}}(\varphi^{\textup{Tx}},\vartheta_{bu}^{\textup{Tx}}(t))$ and $g_{bu}^{\textup{Rx}}(\varphi^{\textup{Rx}},\vartheta_{bu}^{\textup{Rx}}(t))$ denote the transmission and reception antenna gains from SBS $b$ to the \gls{hmd} of \gls{vr} user $u$ while using beams of beamwidth $\varphi$, given by
 		\begin{equation}
			g_{bu}^{\varangle}(\varphi^{\varangle},\vartheta_{bu}^{\varangle}(\slot))=\left\lbrace 
			\begin{array}{ll}
				\frac{2\pi-\left(2\pi-\varphi^{\varangle}\right)g_{sl}}{\varphi^{\varangle}}\text{,} & |\vartheta_{bu}^{\varangle}(t)|\leq\frac{\varphi^{\varangle}}{2},\\
				\vspace*{-0.1cm}g_{sl}, & \text{otherwise,}
			\end{array}\right.\label{eq:antennaTx}
		\end{equation}
 	with $\varangle\in\{\textup{Tx},\textup{Rx}\}$, where $\vartheta_{bu}^{\varangle}(t)$ stands for the angular deviation from the boresight directions of \gls{sbs} $b$ and of \gls{vr} user $u$, and $g_{sl}$ is the constant sidelobe gain with $g_{sl} \in [0,1)$.
 	High directionality of \gls{mmwave} communication often implies a search process to find the best steering directions.
 	In our system model, full knowledge of the seating area layout and of the fixed locations of the $B$ \glspl{sbs} is assumed.
 	Moreover, as stated before, users in $\mathcal{U}$ will report their \gls{6dof} pose information in the \gls{ul} with $T_f$ periodicity. 
 	With these assumptions, even if both the \glspl{sbs} and users are aware of each other's location and know a priori what their respective optimal beams' boresight directions are, the above antenna model effectively captures subtle misalignment errors arriving from the limited availability of unique beam patterns common in codebook-based beam alignment approaches under analog beamforming.
 In other words, the angular deviation $\vartheta_{bu}^{\varangle}(t)$ in \eqref{eq:antennaTx} reflects the inability of the analog beamformer to direct the mainbeam at any arbitrary location.  
 The \gls{sinr} for user $u$ served by \gls{sbs} $b$ is thus given by
 \begin{equation}
	\textrm{SINR}_{bu}(\slot)=\frac{ \txpower_{\bs} \channel_{\bs\ue}(\slot) \antennaGainRX_{\bs\ue}(\slot)\antennaGainTX_{\bs\ue}(\slot)}{\interference_{\ue}(\slot)+\noise},\label{eq:sinr}
 \end{equation}
 where the numerator represents the power of the received signal at user $u$ from \gls{sbs} $b$ under transmit power $p_b$, and the denominator is the sum of the interference power and Gaussian noise power. In our system $\interference_{\ue}(t) = \sum_{\bs'\in \bsSet\setminus\{\bs\} }  \txpower_{\bs'} \channel_{\bs'\ue}(t) \antennaGainRX_{\bs'\ue}(t)\antennaGainTX_{\bs'\ue}(t)$ is the interference that arrives from the transmission of other \glspl{sbs} reaching user $\ue$ through channel, transmit and receive antenna gains, and power level $\channel_{\bs'\ue}(t)$, $\antennaGainRX_{\bs'\ue}(t)$, $\antennaGainTX_{\bs'\ue}(t)$ and $\txpower_{\bs'}$, respectively. The noise power is given by the noise power spectral density $\noiseAlone$ in watts per hertz multiplied by the system bandwidth $\bandwidth_\bs$. 

 Note that since multicast transmission is considered, 
 the achievable rate of user \smash{$\ue\!\in\!\clusterSet_\cluster^\indexFrame$} depends on the composition of $\clusterSetUe$ for each frame index $\indexFrame$, with the assumptions behind this composition being the \gls{fov} and spatial correlation as detailed in Section \ref{subsec:Clustering}.

 \section{Problem Formulation}\label{sec:prob_form}
 In this section, building upon the multi-user \gls{vr} scenario described in Section \ref{sec:sys_mod}, we formulate the network-wide optimization problem of scheduling the \gls{fov} contents of an \gls{hd} 360$^\circ$ video frame by a deadline $\frameDeadline$ such that \gls{vr} sickness can be avoided. The problem formulation explicitly incorporates the proactive/real-time nature of content requests, as well as multicast/unicast transmission capabilities in the \glspl{sbs}. 
 We pose the problem as a frame quality maximization, while the latency constraints and transmission queues stability are maintained.

	For admission purposes, in our scenario a user $\ue$ associated during video frame index $\indexFrame\!\!\!=\!\!\!\indexFrameP$ to a cluster $\clusterSet_{\cluster}^{\indexFrameP}\subset\clusterSet$ is allowed to request chunks that belong either to the cluster-level predicted \gls{fov} \smash{$\tileSetFOVPred_{\clusterSetUe}^{\indexFrameP}$} or to its real-time \gls{fov} $\tileSetFOV_\ue^{\indexFrameRt}$. In this sense, the admission of the predicted \gls{fov} proactive requests allows to leverage cluster-level multicast transmissions of shared \gls{fov} chunks, i.e. content reuse. On their behalf, real-time \gls{fov} chunk requests are the result of missed tiles, i.e. \smash{$\tileSetFOV_\ue^{\indexFrameRt} \setminus \tileSetFOVPred_{\clusterSetUe}^{\indexFrameRt} \neq \emptyset$}, due to imperfect prediction accuracy in the \glsunset{drnn}\gls{drnn} module. In the latter case, these requests need to be expedited to meet \glsunset{mtp}\gls{mtp} latency related constraints and provide a smooth \gls{vr} experience.
	
	Let $\admissionTotUe(\slot)$ be the total traffic admission for user $\ue\in\clusterSet_{\cluster}^{\indexFrame}$ with  $\bigcup_{k=1}^\textup{K}\mathcal{C}_{k}^{\indexFrame}=\mathcal{U}, \forall \indexFrame\in\indexFrameSet$
	\begin{align}\label{eq:traffic_admision}
	\admissionTotUe(\slot) & =\!\!\!\sum_{\indexFrame\in\indexFrameSet}\Bigl(\indicator{\indexFrame=\indexFrameRt}\admission_{\ue\indexFrame}(\slot)\!\!\!\!\sum_{\chunk\in\tileSetFOV_\ue^\indexFrame}\!\!\!L_{\chunk}+\indicator{\indexFrame=\indexFrameP}\admission_{\ue\indexFrame}(\slot)\!\!\!\!\sum_{\chunk\in\tileSetFOVPred_{\clusterSetUe}^{\indexFrame}}\!\!\!L_{\chunk}\Bigr),
	\end{align}
	where $L_{\chunk}$ is the data size of chunk $\chunk$, and $\admission_{\ue\indexFrame}(\slot)$ is a binary variable that indicates if the video frame $\indexFrame$ is admitted for offloading to user $\ue$. The aforementioned two-fold nature of the chunk admission requests is evinced through the two separate terms in \eqref{eq:traffic_admision}. Moreover, we notice here that the value of $\admissionTotUe(\slot)$ in \eqref{eq:traffic_admision} is upper bounded by the maximum value $\admissionTotMax$ such that $\admissionTotUe(\slot)\!\!=\!\admissionTotMax\!\Rightarrow\admission_{\ue\indexFrame}\!=\!\!1,\;\forall\indexFrame\!\!\in\!\indexFrameSet$ that, in practice, represents a situation where the system's traffic load is so low that all frames can be admitted in \gls{hd} without risking queue stability.

 Extending the notation of the admission to consider unicast and multicast transmission of chunks in real-time and proactive respectively, the rate of delivering chunk $\chunk$ to user $\ue$ is 
 	\begin{equation}
		\rateUeChunk(\slot)\!=\!
		\begin{cases}
			\sum\limits_{\bs\in\bsSet}\!\schedulingUeBsChunk(\slot)\rateUeSBS(\slot), & \indexFrame\!=\!\indexFrameRt,\\
			\sum\limits_{\bs\in\bsSet}\!\schedulingUeBsChunk(\slot)\!\!\!\min\limits_{\forall\ueOther\in\clusterSetUeFrame | \chunk\in\tileSetFOVPred_{\ueOther}^\indexFrame}\! \rate_{\bs\ueOther}(\slot),& \text{otherwise},
		\end{cases}
	\label{eq:chunk_rate}
	\end{equation}
 with $\rateUeSBS(\slot)=\bandwidth_\bs\log_{2}\bigl(1\!+\!\textrm{SINR}_{\bs\ue}(t)\bigr)$ the unicast real-time rate and $\schedulingUeBsChunk(\slot)$ the binary scheduling variable for chunk $\chunk$ to user $\ue$ from base station $\bs$. 
 For the proactive multicast case with $\indexFrameRt\!\!<\indexFrame\leq\!\!\indexFrameP$, the \gls{sbs} will adapt its rate to match that of the worst user in the cluster \smash{$\clusterSetUeFrame$} to whom user $\ue$ has been assigned for the frame index at hand\footnote{\label{note5}In practice, this is accomplished by adapting the \glsunset{mcs}\gls{mcs} that reflects users' perceived channel quality.}. This way it guarantees that the chunk will be correctly decoded by all the interested cluster-users.
 Moreover, we remark here that the value of $\rateUeChunk(\slot)$ in \eqref{eq:chunk_rate} is bounded above by a maximum achievable service rate $\rateMax$ that in practice is determined by the highest available \glsreset{mcs}\gls{mcs} index.

 For notational compactness, to express that a requested chunk $\chunk$ corresponds either to the user's real-time \gls{fov} or to the user's cluster-level predicted \gls{fov}, we will hereafter denote the targeted \gls{fov} chunk set as  \smash{$\tileSetFOVExt_\ue^{\indexFrame}=\indicator{\indexFrame=\indexFrameRt}\tileSetFOV_\ue^{\indexFrame}+(1-\indicator{\indexFrame=\indexFrameRt})\tileSetFOVPred_{\clusterSetUeFrame}^\indexFrame $}.  Then $\queueUe(\slot)$ \textendash namely, the traffic queue of a user $\ue$, \smash{$\forall\ue\!\in\!\clusterSetUeFrame$}\textendash~evolves as
 \begin{equation}
 	\queueUe(\slot+1)=\Bigr[\queueUe(\slot)-\!\!\!\!\!\!\sum_{\indexFrame=[\indexFrameRt,\indexFrameP]}\!\sum_{\chunk\in\tileSetFOVExt_\ue^{\indexFrame}}\rateUeChunk(\slot)\Bigl]^{+}+\admissionTotUe(\slot).\label{eq:queue_update} 
 \end{equation}

 We remark here that although only chunks for frame indexes $\indexFrame\!\!=\!\!\{\indexFrameRt,\indexFrameP\}$ are admitted, the range of frame indexes corresponding to chunks co-existing in a user's queue at a given time may span to values $\indexFrame\!=\![\indexFrameRt,\indexFrameRt+1,\cdots,\indexFrameP]$. A scheduling policy, that is aware of the \gls{vr} specific latency-reliability constraints, will timely determine which chunks need be expedited from these priority-based queues. Without loss of generality, we assume that the video streaming and \gls{fov} chunk scheduling start simultaneously, and denote the transmission delay of the video frame $\indexFrame=\indexFrameRt$ to user $\ue$ as $\txDelayUeFrame(\slot)$. Let $\tMTP(\slot)$ represent the available time to schedule the frame before the considered \gls{mtp} delay deadline is exceeded and given by 
 \begin{equation}
 	\tMTP(\slot)=[\slotArrival+\MTPDelay-t]^+,\label{eq:MTPdeadline}
 \end{equation}
 where $\slotArrival$ corresponds to the timestamp when the chunk was requested, and $\MTPDelay$ is the constant \gls{mtp} latency-aware\footnote{\label{note6}To realize a mobile \gls{vr} operating within \gls{mtp} bounds, a holistic approach that considers transmission, computing, and control latency performance should be considered. Yet, 360$^\circ$ videos require substantially higher data rates to provide acceptable quality to users and pose a greater challenge on the content delivery process. Hence, in this work we argue that the question of how this high data rate transmission can be realized with such latency constraint is on itself important to study and focus on the transmission delay saving a $d_f\!-\!\MTPDelay$ budget for other edge-server computing or on-\gls{hmd} processing delays.} maximum transmission latency. In this regard, the following \gls{hrllbb} constraint is imposed to ensure that the transmission delay of the current playing frame does not exceed the \gls{mtp} delay with high probability: 
 \begin{equation}
 	\lim_{T\rightarrow\infty}\frac{1}{T}\sum_{t=1}^{T}\probability(\tau_{\ue\indexFrameRt}(\slot)\geq \tMTP(t))\leq\delayOutage,\label{eq:rel_const_1a}
 \end{equation}
 \noindent where $\delayOutage\ll1$ is a predefined delay reliability metric. We then recast the probability in  \eqref{eq:rel_const_1a} as the expectation over an indicator function, i.e., the constraint is rewritten as:
 \begin{equation}
 	\lim_{T\rightarrow\infty}\frac{1}{T}\sum_{\slot=1}^{T}\mathbb{E}[\indicator{\txDelayUeFrameRt(\slot)\geq \tMTP(t)}]\leq\delayOutage.\label{eq:rel_const_1}
 \end{equation}

 Collecting the \gls{hd} frame admission and the binary scheduling variables as 
	\smash{$\admissionV(\slot)\hspace{-.5mm}=\hspace{-.5mm}\{\admission_{\ue\indexFrame}\!\!:\!\forall \ue\in\ueSet,\forall\indexFrame\in\{\indexFrameP,\indexFrameRt\}\}$} and	\smash{$\schedulingV(\slot)=\{\scheduling_{\bs\ue\chunk}(\slot):\forall \bs\in\bsSet, \forall \ue\in\ueSet,\forall \chunk\in\tileSetFOVExt_\ue^\indexFrame\}$}, our optimization problem is to maximize the user's quality by optimizing the HD frame admission and scheduling policies subject to the latency constraints:
 \begin{subequations}\label{eq:OP} 
	\begin{align}
	\textrm{\textbf{OP:}}\!\!\!\max_{\schedulingV(\slot),\admissionV(\slot)}\!& U\bigl(\{\overline{\admissionTotUe}\}\bigr)=\sum_{\ue\in\ueSet}\bigl(\sfint (\overline{\admissionTotUe})\bigr)\label{eq:OP_utility} \\
	\text{s.t.}\quad\quad& \overline{\queueUe}\leq\infty,\;\forall\ue\!\in\!\ueSet,\label{eq:const_queue}\\
	&\admission_{\ue\indexFrame}(\slot)\in\{0,1\},\;\forall\ue\!\in\!\ueSet,\forall\indexFrame\in\{\indexFrameP,\indexFrameRt\},\\
	&\admissionTotUe(\slot)\!\leq\!\admissionTotMax,\;\forall \ue\!\in\!\ueSet,\\
	&\rateUeChunk(\slot)\!\leq\!\rateMax,\;\forall\ue\!\in\!\ueSet,\forall\chunk\!\in\!\tileSetFOVExt_\ue^\indexFrame,\forall\!\indexFrame\!\!\in\!\indexFrameSet,\\
	&\scheduling_{\bs\ue\chunk}\negmedspace(\slot)\!\in\!\{0,1\},\;\forall\bs\!\in\!\bsSet,\forall\ue\!\in\!\ueSet,\forall\chunk\!\in\!\tileSetFOVExt_\ue^\indexFrame,\forall \indexFrame\!\in\!\indexFrameSet,\\
	& \lim_{T\rightarrow\infty}\!\frac{1}{T}\!\sum_{\slot=1}^{T}\!\mathbb{E}[\indicator{\txDelayUeFrameRt(\slot)\geq \tMTP(t)}]\!\leq\!\delayOutage,\;\! \forall\ue\!\in\!\ueSet.\label{eq:consts_rel}
	\end{align}
 \end{subequations}
 where $\fint(\cdot)$ is a non-decreasing and concave function that can be adjusted to reflect different optimization objectives.
 
 To find a tractable solution for the above stochastic optimization problem, we first define a set of auxiliary variables $\{\gamma_{u}(t)\},\forall \ue\in\ueSet$. Accordingly, the stochastic optimization problem in (\ref{eq:OP}) can be transformed from a utility function of time-averaged variables into an equivalent optimization problem of time-averaged utility function of instantaneous variables:
 \begin{subequations}\label{eq:EOP} 
	\begin{align}
	\textrm{\textbf{EOP:}}\quad\max_{\admissionV(\slot),\schedulingV(\slot),\{\auxVbleUe(\slot)\}} & \overline{U\bigl(\{\auxVbleUe(\slot)\}\bigr)} =\overline{\sum_{\ue\in\ueSet}\Bigl(\sfint(\auxVbleUe)\Bigr)}\nonumber \\
	\text{s.t.}\qquad & \overline{\auxVbleUe}\leq\overline{\admissionTotUe}\label{time_av_const1},\;\forall \ue\in\ueSet,\\
	& \auxVbleUe(\slot)\leq \admissionTotMax,\;\forall \ue\in\ueSet,\label{time_av_const2}\\
	& (\ref{eq:const_queue})-(\ref{eq:consts_rel})
	\end{align}
 \end{subequations}
 
 Next, by invoking the framework of Lyapunov optimization \cite{neely_lyapunov}, \emph{virtual queues} are constructed to help satisfy the time-average inequality constraints. By ensuring that these queues are stable, the time average constraints, namely \eqref{eq:consts_rel} and \eqref{time_av_const1}, are guaranteed to be met. Therefore, we define $\auxVQueueUe(\slot)$ and $\latencyVQueueUeChunk(\slot)$ virtual queues that correspond to the constraints over the auxiliary variables and over the transmission delay, respectively. Accordingly, the virtual queues are updated as follows: 
 \begin{equation}
 	\auxVQueueUe(\slot+1)={\big[\auxVQueueUe(\slot)-\admissionTotUe(\slot)+\gamma_{u}(t)\big]}^{+},\label{eq:aux_Q}
 \end{equation}
 \begin{equation}
	\latencyVQueueUeChunk(\slot\negmedspace+\! 1)={\big[\latencyVQueueUeChunk(\slot)+(\indicator{\txDelayUeFrameRt(\slot)\geq \tMTP(t)}-\epsilon_{d})\queueUe(\slot\!+\! 1)\big]}^{+}.\label{eq:delay_Q}
 \end{equation}
 Notice that the virtual queue in \eqref{eq:delay_Q} is built after having scaled-up the constraint in \eqref{eq:rel_const_1} by multiplying both sides of it with the actual queue size. Hereinafter, for readability reasons $\indicator{\tau_{uc_{f}}(\slot)\geq\tMTP(t)}$ will be shortened to  $\indicator{\tMTP(t)}$ to denote $\tau_{uc_{f}}$ exceeding the \gls{mtp} delay.

 Let \smash{$\bm{\chi}(t) = \{\queueUe(\slot), \auxVQueueUe(\slot), \latencyVQueueUeChunk(\slot): \ue\in\ueSet, \indexFrame\in\indexFrameSet\}$} be the vector of combined traffic and virtual
 queues with $\vect{\chi}(\slot) = [\bm{\chi}_{\ue}(\slot)]_{\ue\in\ueSet}$. Then, to represent a scalar metric of the congestion, let the quadratic Lyapunov function be given by
 \begin{equation}
 	L(\bm{\chi}(t))\triangleq\frac{1}{2}\!\sum_{\ue\in\ueSet}{\queueUe(\slot)}^{2}+\frac{1}{2}\!\sum_{\ue\in\ueSet}{\auxVQueueUe(\slot)}^{2}+\frac{1}{2}\!\sum_{\ue\in\ueSet}\sum_{\indexFrame\in\indexFrameSet}{\latencyVQueueUeChunk(\slot)}^{2},
 \end{equation}
 and the one-timeslot Lyapunov drift function be \smash{$\lyapunovDrift\negmedspace=\negmedspace \lyapunov{\bm{\lyapunovQueueCombined}(\slot \negmedspace+\negmedspace1)}\negmedspace-\negmedspace\lyapunov{\bm{\lyapunovQueueCombined}(\slot)}$}. Hence, we leverage the drift-plus-penalty algorithm to find the control actions that greedily minimize a bound of the drift function minus a scaled utility function, i.e., $\lyapunovDrift-\lyapunovTradeoff \mathbb{E}\{U(\{\auxVbleUe(\slot)\})\}$, where $\lyapunovTradeoff$ is the parameter that controls the trade-off between minimizing the queue backlog and approaching the optimal solution. 
 \begin{lemma}
	At each time instant $t$, the following bound satisfies the drift-plus-penalty function $\lyapunovDrift-\lyapunovTradeoff \mathbb{E}\{U(\{\auxVbleUe(\slot)\})\}$ under any queue state and control strategy:
	\begin{align}
	\lyapunovDrift\!-\!\lyapunovTradeoff \mathbb{E}\bigl\{U(\{\auxVbleUe(\slot)\})\bigr\}\!& \leq \lyapunovConst(\slot)\nonumber\\
	& 
	-\sum_{\ue\in\ueSet}\Bigl[\auxVQueueUe(\slot)\auxVbleUe(t)-\lyapunovTradeoff U(\{\auxVbleUe(\slot)\})\Bigr]_{\#1}\nonumber \\
	&  -\sum_{\ue\in\ueSet}\Bigl[\Big(\lyapunovWeight(\slot)-\auxVQueueUe(\slot)\Big)\admissionTotUe(\slot)\Bigr]_{\#2}\nonumber \\
	&  -\sum_{\ue\in\ueSet}\Bigl[\lyapunovWeight(t)\sum_{\indexFrame\in\indexFrameSet}\sum_{\chunk\in\tileSetFOVExt_\ue^\indexFrame}\rateUeChunk\Bigr]_{\#3},\label{eq:drift_plus_penalty_bound}
	\end{align}
	where $\lyapunovConst(\slot)$ is an upperbounded constant parameter at each time slot $t$ in \eqref{eq:drift_plus_penalty_bound}, 
	and $\lyapunovWeight(\slot)$ collects the terms related to the traffic queue and to the transmission delay virtual queue as
	\begin{equation}\label{eq:lyapunovWeight}
	\lyapunovWeight(\slot)=\lyapunovWeightQ(\slot)+\indicator{ \tMTP(t)}\lyapunovWeightF(\slot),
	\end{equation}
	which are given as \smash{$\lyapunovWeightQ(\slot)=\queueUe(\slot)(1+\delayOutage^2)-\delayOutage\sum_{\indexFrame\in\indexFrameSet}\latencyVQueueUeChunk(\slot)$} and $\lyapunovWeightF(\slot)=\sum_{\indexFrame\in\indexFrameSet}\latencyVQueueUeChunk(\slot)+(1-2\delayOutage)\queueUe(\slot)$.
 \end{lemma}
 \noindent \begin{IEEEproof}
	See Appendix A
 \end{IEEEproof}
 The solution to (\ref{eq:EOP}) can be found by greedily minimizing the right-hand side of (\ref{eq:drift_plus_penalty_bound}) for each time slot. Instead, since the optimization variables are decoupled in (\ref{eq:drift_plus_penalty_bound}), we split the optimization problem into three disjoint subproblems that are solved  concurrently based on the observation of the traffic and the virtual queues.

 It can be also seen from \eqref{eq:drift_plus_penalty_bound} that the optimization problem at a given time instant $\slot$ is only function of the current states of the traffic and the virtual queues. This means that a solution to the problem does not require the knowledge of the traffic statistics or queue evolution probabilities. The equation also shows that the \emph{separability criteria}~\cite{book:Lyapunov_Georgiadis2006} is met i.e., that the admission and scheduling are not dependent on the variables of each other.

 \subsection{Auxiliary Variable Selection}\label{subsec:OSP1-VbleSel}
 The first subproblem is the minimization of the term~$\#1$ in (\ref{eq:drift_plus_penalty_bound}) i.e., the selection of the auxiliary variables. The problem can be decoupled on a per user basis as follows:
 \begin{subequations}\label{eq:OSP1} 
	\begin{align}
	\textrm{\textbf{OSP1:}}\hspace{.15em}\max_{\{\auxVbleUe\}}\; & V_\Delta U(\auxVbleUe(\slot))-\auxVQueueUe(\slot)\auxVbleUe(\slot)\\
	\text{s.t.}~ & \auxVbleUe(\slot)\leq\admissionTotMax,\;\forall \ue\in\ueSet.
	\end{align}
 \end{subequations}

 By selecting a linear utility function, i.e., $U(\gamma_{u}(t))=\gamma_{u}(t)$, the optimal value of the auxiliary variable is found to be:
 \begin{equation}
 	\auxVbleUe(\slot)=\begin{cases}
 		\admissionTotMax, & \auxVQueueUe(\slot)\leq V_{\Delta},\\
		0, & \text{otherwise}.\label{eq:OSP1-sol}
	\end{cases}
 \end{equation}
 It is worth noting here that different utility functions can be selected depending on the network optimization objective. For example, a logarithmic function of the admission rates could be selected to provide proportional fairness between users.

 \subsection{HD Streaming Admission}\label{subsec:OSP2-HDAdmission}
 Next, the \gls{hd} chunk admission problem is optimized by solving the subproblem given by the term~$\#2$ of (\ref{eq:drift_plus_penalty_bound}). The optimization subproblem is formulated as:
 \begin{subequations}\label{eq:OSP2} 
	\begin{align}
	\textrm{\textbf{OSP2:}}\hspace{.15em}\max_{\admissionV(\slot)} & \sum_{\ue\in\ueSet}\bigl(\auxVQueueUe(\slot)-\lyapunovWeight(\slot)\bigr)\admissionTotUe(\slot)\\
	\text{s.t.}~&\admission_{\ue\indexFrame}(\slot)\in\{0,1\},\;\forall\ue\!\in\!\ueSet,\forall\indexFrame\in\{\indexFrameP,\indexFrameRt\},
	\end{align}
 \end{subequations}
 The above admission rate maximization problem is convex and its optimal solution is:
 \begin{equation}
 	\admission_{\ue\indexFrame}(\slot)=\begin{cases}
		1 & \auxVQueueUe(\slot)\geq\lyapunovWeight(\slot),\\
		0 & \text{otherwise}.
 	\end{cases}
 \end{equation}

 In other words, the optimal \gls{hd} chunk admission control is to either admit or discard the whole frame, depending on the physical and virtual queue state.

 \subsection{User-SBS Chunk Scheduling}\label{subsec:OSP3-ChunkSched}

 The third subproblem aims at scheduling user requests of \gls{hd} video chunks to base stations. The optimization subproblem is formulated by maximizing the term $\#3$ in (\ref{eq:drift_plus_penalty_bound}) as follows:
 \begin{subequations}\label{eq:OSP3}
	\begin{align}
	\textrm{\textbf{OSP3:}}\hspace{.15em}\max_{\schedulingV(\slot)} & \sum_{\ue\in\ueSet}\lyapunovWeight(\slot)\!\!\!\!\sum_{\indexFrame=\{\indexFrameRt,\indexFrameP\}} \!\sum_{\chunk\in\tileSetFOVExt_\ue^\indexFrame}\rateUeChunk(\slot)	\label{eq:OSP3-decision}\\ 
	\text{s.t.}~& \rateUeChunk(\slot)\!\leq\!\rateMax,\;\!\!\forall \chunk\!\in\!\tileSetFOVExt_{\ue}^{\indexFrame},\forall\! \indexFrame\!\in\!\indexFrameSet,\;\!\forall \ue\!\in\!\!\ueSet,	\label{eq:OSP3-const1}\\
	& \scheduling_{\bs\ue\chunk}(\slot)\!\in\!\{0,1\},\!\!\forall \bs\!\in\!\bsSet,\forall \ue\!\in\!\ueSet,\forall \chunk\!\in\!\tileSetFOVExt_\ue^\indexFrame,\forall \indexFrame\!\!\in\!\indexFrameSet. \label{eq:OSP3-const2}
	\end{align}
 \end{subequations}

 We emphasize that OSP3 is a combinatorial problem where video chunks for users need to be scheduled by \glspl{sbs} using a \gls{mmwave} multicast transmission. Subsequently, a matching algorithm is designed to associate chunk scheduling requests arising either from clusters of users or from individual users to the set of \glspl{sbs} operating in \gls{mmwave} band in the theater. 

\section{A Matching Theory approach to HD chunk scheduling}\label{sec:Matching Theory}
 The use of Matching Theory \cite{gale_shapley_1992} \textendash a mathematical framework from labor economics that attempts to describe the formation of mutually beneficial relationships over time\textendash, has recently garnered a considerable interest in the context of resource allocation for wireless networks \cite{walid_matching_2015}. However, for the sake of completeness we will first provide several definitions to properly address the fundamentals of this framework adapted to the problem at hand. Then, we will formulate the utility functions that lie at its core for both sets of agents.
  
 \subsection{Matching Theory Preliminaries}
 \begin{definition}A matching game is defined by two sets of players ($\clusterSet$,$\bsSet$) and two preference profiles denoted by $\prefBS$ and $\prefCluster$, allowing each player $\bs\in\bsSet$, $\clusterSetUe\in\clusterSet$ to accordingly rank the players in the opposite set. 
 \end{definition}

 \begin{definition}
 The output of a matching game is a matching function $\bm{\Upsilon}(t)=\{\Upsilon_{b,\clusterSetUe}(t)\}$ that bilaterally assigns players $\bm{\Upsilon}_\bs(\slot)\triangleq\{\bs\in \bsSet:\Upsilon_{\bs,\clusterSetUe}(\slot)=1\}$ and $\bm{\Upsilon}_{\clusterSetUe}(\slot)\triangleq\{\clusterSetUe\in\clusterSet:\Upsilon_{\bs,\clusterSetUe}(\slot)=1\}$ such that $|\bm\Upsilon_{\clusterSetUe}(\slot)|\leq q_\clusterSet$ and $|\bm\Upsilon_\bs(t)|\leq q_\bsSet$ are fulfilled, with $q_\bsSet$, $q_\clusterSet$ the \textit{quota} of the players which, for a one-to-one matching game satisfy $q_\bsSet=q_\clusterSet=1$.
 \end{definition}
 
 \begin{definition}
 A preference $\succ$ is a complete, reflexive and transitive binary relation between the players in $\bsSet$ and $\clusterSet$. Therefore, for any \gls{sbs} $\bs\in\bsSet$ a preference relation $\succ_\bs$ is defined over the set of clusters $\clusterSet$ such that for any two clusters $(\clusterSetUe,\clusterSetUe^\prime)\in \clusterSet\times\clusterSet$ with $\clusterSetUe\neq\clusterSetUe^\prime$, and two matchings $\bm{\Upsilon}(t)$, $\bm{\Upsilon}^\prime(t)$ so that ${\Upsilon}_\bs(t)=\clusterSetUe$ and $\Upsilon_{\bs}^\prime(t)=\clusterSetUe^\prime$:
 	\begin{equation}
	\left(\clusterSetUe,\bm{\Upsilon}(\slot)\right)\succ_\bs\left(\clusterSetUe^\prime,\bm{\Upsilon}^\prime(\slot)\right)\Leftrightarrow U_{\bsSet}^{b,\clusterSetUe}(\slot)>U_{\bsSet}^{b,\clusterSetUe^\prime}(\slot).\label{eq:SBS_preference}	
	\end{equation}
 Similarly, for any cluster of users $\clusterSetUe\in\clusterSet$ a preference relation $\succ_{\clusterSetUe}$ is defined over the set of \gls{sbs} $\bsSet$ such that for any two \glspl{sbs} $(b,b^\prime)\in\bsSet\times\bsSet$ with $b\neq b^\prime$, and two matchings $\bm{\Upsilon}(\slot)$, $\bm{\Upsilon}^\prime(t)$ we have that ${\Upsilon}_{\clusterSetUe}(\slot)=\bs$ and ${\Upsilon}_{\clusterSetUe}^\prime(t)=\bs^\prime$:
	\begin{equation}
	\left(b,\bm{\Upsilon}(\slot)\right)\succ_{\clusterSetUe}\left(b^\prime,\bm{\Upsilon}^\prime(\slot)\right)\Leftrightarrow U_{\clusterSet}^{\clusterSetUe,\bs}(\slot)>U_{\clusterSet}^{\clusterSetUe,\bs^\prime}(\slot), \label{eq:cluster_preference}	
	\end{equation}
	where $U_{\bsSet}^{\bs,\clusterSetUe}(\slot)$ and $U_{\clusterSet}^{\clusterSetUe,\bs}(\slot)$ denote the utility of cluster $\clusterSetUe$ for \gls{sbs} $\bs$ and the utility of \gls{sbs} $\bs$ for cluster $\clusterSetUe$, correspondingly.
\end{definition}

\subsection{Matching Utility Formulation}\label{subsec:matchingUtilityFormulation}

 The \gls{hd} chunk scheduling subproblem in \eqref{eq:OSP3} is formulated as a matching game between the \glspl{sbs} and the clusters of users. As such, both sides seek to greedily maximize the overall \gls{vr} experience by efficiently allocating the \gls{mmwave} transmission resources while \gls{vr} \gls{qoe} related constraints are met. Hence, each timeslot with updated information on channel and queue state, new scheduling requests for video chunk transmission will be prioritized in each cluster and in each \gls{sbs}, and new sets of matching pairs will be found using the proposed approach. With the above principles in mind, we formulate the utilities for both sets.
 
 The utility of serving a given cluster of users with at least one pending chunk request from the \glspl{sbs} point of view will essentially reflect two aspects: the \textit{priority} and \textit{relevance} of the chunk $c_f$ at hand. The priority of the whole frame to which the requested chunk belongs to is controlled by the dynamics of $\queueUe(\slot)$ and $\latencyVQueueUeChunk(\slot)$ as per \eqref{eq:queue_update} and \eqref{eq:delay_Q} through $\tMTP(t)$ as given by \eqref{eq:MTPdeadline}. The relevance of the chunk within the cluster \gls{fov} is related to its popularity i.e., how many of the cluster members have requested this chunk. Intuitively, the cluster-level multicast approach decreases the wireless network load by transmitting each chunk falling into the cluster-level \gls{fov} only once. Moreover, transmitting first the most relevant chunks also contributes to increasing the overall system welfare. Therefore, \glspl{sbs} will build their preference profile using the following utility function:
  	\begin{align}\label{eq:util_cluster}
		U_{\bsSet}^{\bs,\clusterSetUe}(\slot)&=\sum_{\ue\in\clusterSet_\cluster}\indicator{c_f\in\tileSetFOVExt_\ue^\indexFrame}\lyapunovWeight(\slot)\\
		&=\sum_{\ue\in\clusterSet_\cluster}\indicator{c_f\in\tileSetFOVExt_\ue^\indexFrame}\bigl\{\lyapunovWeightQ(\slot)+\indicator{ \tMTP(t)}\lyapunovWeightF(\slot)\bigr\}\nonumber.
	\end{align}

 Notice that in \eqref{eq:util_cluster} by definition, $\indicator{\tMTP(t)}$ can only be non-zero for the currently playing frame index. 
 Similarly, the utility of a \gls{sbs} from the clusters' perspective will depend on the goodness of the transmission opportunity through the offered rate in \eqref{eq:chunk_rate}. In other words, we define the utility as 
    \begin{align}\label{eq:util_bs} 
		 U_{\clusterSet}^{\clusterSetUe,\bs}(\slot)=&\indicator{\indexFrame=\indexFrameRt}\!\!\min\limits_{\forall\ue\in\clusterSetUeFrame | \chunk\in\tileSetFOV_{\ue}^\indexFrame} \rate_{\bs\ue}(\slot)\nonumber\\
		 &+(1-\indicator{\indexFrame=\indexFrameRt})\!\!\min\limits_{\forall\ue\in\clusterSetUeFrame | \chunk\in\tileSetFOVPred_{\ue}^\indexFrame} \rate_{\bs\ue}(\slot), 
	\end{align}

\subsection{Stability of the Matching}\label{subsec:MatchingStability}
Next, the notion of stability is introduced and an interference estimation method is proposed to guarantee that the \gls{hd} chunk scheduling game converges to a stable matching.

\begin{definition}\label{def:blocking_pair}
	Given a matching $\Upsilon$ with $\Upsilon_\bs=\clusterSetUe$ and $\Upsilon_{\clusterSetUe}=\bs$, and a pair $(\bs^\prime,\clusterSetUe^\prime)$ with $\Upsilon_\bs(\slot)\neq \cluster^\prime$ and $\Upsilon_{\clusterSetUe}\neq \bs'$, $(\bs',\cluster')$ is said to be blocking the matching $\Upsilon$ and form a blocking pair if: 1) $\bs'\succ_{\cluster}\bs$, 2) $\cluster'\succ_{\bs}\cluster$. A matching $\Upsilon*$ is stable if there is no blocking pair.
\end{definition}
Gale-Shapley's \gls{da} algorithm \cite{gale_shapley_1962} provides a polynomial time solution  guaranteed to be two-sided stable for one-to-one canonical matchings i.e., those matching games where the preference profiles of the players are not affected by any other player's decisions. The influence of a given player's matching over another's is referred to as \textit{externality}. As the game evolves, the existence of externalities triggers dynamic updates in the values of the perceived utilities and, consequently, ensuring the stability of the matching is challenging. 

The above matching game cannot be directly solved using \gls{da}; the utilities in \eqref{eq:util_cluster}-\eqref{eq:util_bs} are function of the instantaneous service rate which, in turn depends on the interference \textendash a well-known source of externalities\textendash{} through the \gls{sinr}. Moreover, in the context of directional communications, the arrival direction of the interference caused by other \glspl{sbs}\footnote{\label{note7}We remark here that by matching each \gls{sbs} to a single cluster with orthogonal non-overlapping beams for the multicast transmission, only the impairment due to inter-\gls{sbs} interference needs to be considered.} greatly impacts the service rate. Hence, evaluating the instantaneous interference and casting preferences accordingly implies full knowledge of the system-wide current matching state by all the players, which is impractical in terms of signaling overhead. 
Different approaches are used in the literature to handle the matching externalities either through finding preference profiles that do not induce externality~\cite{ZhouMatchingD2D16}, or through centralized approval methods that evaluate how each matching change affects other players before allowing it~\cite{DiMatchingNoma16}. Alternatively, we consider a distributed and computationally efficient algorithm that suits the low latency nature of the considered application by replacing the instantaneous values of the service rate in the utilities with estimated ones.

Let the measured inter-\gls{sbs} interference at user $u$ in the previous time instant $t\negmedspace-\negmedspace1$
be denoted as ${\interference}_{\ue}(t\negmedspace-\negmedspace1)$, and $\interferenceEst_{\ue}(t)$ the estimated inter-\gls{sbs} interference at time instant $t$. Adopting an interference estimation procedure with learning parameter $\nu_{1}$ and moving average inference $\interferenceMov_{\ue}^{\nu_{2}}(\slot-1)$ with a window of $\nu_{2}$ samples, the estimated interference is given by
\begin{alignat}{1}\label{eq:inteference_est}
\interferenceEst_{\ue}(\slot) & =\movAvLearningWeight\interference_{\ue}(\slot-1)+(1-\movAvLearningWeight)\interferenceMov_{\ue}^{\movAvLearningSamples}(\slot-1).
\end{alignat}

Let \smash{$\utilityForBSEst$}, \smash{$\utilityForClusterEst$} be the new expressions for the utilities which exploit the estimated service rate $\hat{\rate}_{\bs\ue}(\slot)$=$\bandwidth_\bs\log_{2}\Bigl(1\!+\!\!\frac{ \txpower_{\bs} \channel_{\bs\ue}(\slot) \antennaGainRX_{\bs\ue}(\slot)\antennaGainTX_{\bs\ue}(\slot)}{\interferenceEst_{\ue}(\slot)+\noise}(\slot)\Bigr)$ through $\hat{\rate}_{\ue\chunk}(\slot)$, such that 

	\begin{align}
	\utilityForClusterEst&=\indicator{\indexFrame=\indexFrameRt}\!\!\min\limits_{\forall\ue\in\clusterSetUeFrame | \chunk\in\tileSetFOV_{\ue}^\indexFrame} \hat{\rate}_{\bs\ue}(\slot)\nonumber\\
	&+(1-\indicator{\indexFrame=\indexFrameRt})\!\!\min\limits_{\forall\ue\in\clusterSetUeFrame | \chunk\in\tileSetFOVPred_{\ue}^\indexFrame} \hat{\rate}_{\bs\ue}(\slot),\label{eq:util_bs_est}\\
	\utilityForBSEst&=\sum_{\ue\in\clusterSet_\cluster}\indicator{c_f\in\tileSetFOVExt_\ue^\indexFrame}{\hat\alpha_\ue(\slot)}(\slot)\label{eq:util_cluster_est}\nonumber\\
	&=\sum_{\ue\in\clusterSet_\cluster}\indicator{c_f\in\tileSetFOVExt_\ue^\indexFrame}\bigl\{\hat\alpha_Q(\slot)+\indicator{ \tMTP(t)}\hat\alpha_F(\slot)\bigr\}.
	\end{align}
	\begin{center}
	\removelatexerror
	\LinesNumberedHidden{
		\begin{algorithm}[t]
			\footnotesize
			\caption{HD chunk scheduling between \glspl{sbs} and User-clusters}
			\label{alg:matching1} 
			\textbf{Phase I - Interference learning and candidate chunk selection}
			\begin{itemize}{\scriptsize 
					\item Each $\ue\in\ueSet$, updates $\interferenceEst_\ue(\slot)$ as per \eqref{eq:inteference_est} and reports channel in the UL.
					\item In the edge controller, queues in $\{\bm\lyapunovQueueCombined(t)\}_{\ue\in\ueSet}$ are updated by solving \eqref{eq:OSP1},   \eqref{eq:OSP2}.
					\item For each $\clusterSet_{\cluster}\!\!\in\!\clusterSet$ a cluster-level chunk request pool is created and each request therein is assigned an urgency tag \smash{$\lyapunovWeightClusterChunk$}$=\sum_{\ue\in\clusterSet_\cluster | \chunk\in\tileSetFOV_\ue^\indexFrame}\lyapunovWeight(\slot)$ with $\lyapunovWeight(\slot)$ as per \eqref{eq:lyapunovWeight}.Then, the request pool is sorted in descending order of \smash{$\lyapunovWeightClusterChunk$}.}   
			\end{itemize}
			
			\textbf{Phase II - Matching game construction}
			\begin{itemize}{\scriptsize 
					\item Each cluster $\clusterSetUe\!\!\in\!\clusterSet$, updates $\utilityForClusterEstNoTime\!\!$ over the SBSs in $\bsSet$ as per \eqref{eq:util_bs_est}.
					\item Each SBS $\bs\!\!\in\!\bsSet$, updates $\utilityForBSEstNoTime\!\!$ over  $\{\clusterSet_{\cluster}\}_{\cluster=1}^{\CLUSTER}$ as per (\ref{eq:util_cluster_est}) evaluating the cluster utility by its most urgent chunk-request, i.e. by $\max\{\lyapunovWeightClusterChunk\}$.}
			\end{itemize}
			\textbf{Phase III - Deferred Acceptance for SBS-Cluster allocation}
			\begin{itemize}{\scriptsize 
					\item For each SBS $\bs$, initialize the subset of eligible clusters, $\mathcal{E}_{\clusterSet}^\bs\subseteq\clusterSet$ so that  initially $|\mathcal{E}_{\clusterSet}^\bs|=|\clusterSet|$. 
					\item For each SBS $\bs$, each cluster $\clusterSet_\cluster$, initialize the subset of unmatched clusters $\mathcal{S}_{\clusterSet}\subseteq\clusterSet$ and SBS $\mathcal{S}_{\bsSet}\subseteq\bsSet$, so that initially $|\mathcal{S}_{\clusterSet}|=|\clusterSet|$, $|\mathcal{S}_{\bsSet}|=|\bsSet|$.}
			\end{itemize}
			\While{$|\mathcal{S_\bsSet}|\not=\emptyset \textup{ and} \sum_{\bs \in \bsSet}|\mathcal{E}_\clusterSet^\bs|\not=\emptyset $}{
				\vspace{2mm}
				Pick a random SBS $\bs\in\bsSet$;\\
				\If{$|\mathcal{E}_{\clusterSet}^\bs|\not=\emptyset$}{\scriptsize
					SBS $\bs$ sends a chunk scheduling proposal to its best ranked \textit{eligible} cluster $\clusterSet_\cluster$, $\clusterSet_\cluster\in\mathcal{E}_{\clusterSet}^\bs$;\\
					\eIf{$\clusterSet_\cluster\in\mathcal{S}_\clusterSet$}{ 
						Match $\bs$ and $\clusterSet_\cluster$ setting $\Upsilon_\bs(t)=\clusterSet_\cluster$ and $\Upsilon_{\clusterSet_\cluster}(t)=\bs$;\\
						Remove $\bs$ and $\clusterSet_\cluster$ from $\mathcal{S}_{\bsSet}$ and $\mathcal{S}_{\clusterSet}$ respectively;\\
					}{ 
						\eIf{$\utilityForClusterEst>\utilityForClusterEstOther$}{ 
							Reject proposal from $\matchingClusterSide$; add back $\matchingClusterSide$ to $\mathcal{S}_\bsSet$ and remove $\clusterSet_\cluster$ from $\mathcal{E}_\clusterSet^{\matchingClusterSide}$;\\
							Match $\bs$ and $\clusterSet_\cluster$ setting $\matchingBSSide=\clusterSet_\cluster$ and $\matchingClusterSide=\bs$;\\
							Remove $\bs$ from $\mathcal{S}_\bsSet$;
						}{ 
							Refuse proposal from $\bs$;\\
							Remove $\clusterSet_\cluster$ from $\mathcal{E}_{\clusterSet}^\bs$; 
						}
					}
				}
			}
			\textbf{Phase IV - Stable matching}
		\end{algorithm}
	}
\end{center}

Under this new utility formulation there are no longer externalities in the system. Therefore, the \gls{hd} chunk scheduling matching game can be solved using \gls{da}, which is guaranteed to converge to a stable matching $\bm\Upsilon(\slot)$ once the inter-cell interference learning process converges. 
The process described above as well as the details of the matching rounds are described in Algorithm \ref{alg:matching1}.

\section{DRNN FoV Prediction and FoV+Location Aware User Clustering}\label{sec:DRNN}
In this section, the DRNN that predicts VR users' \glspl{fov} for upcoming video frames and the clustering scheme that leverages \gls{fov} and spatial inter-user correlations are described. 
We first motivate the selection of the adopted sequential learning model and briefly summarize its operation dynamics. Following that, the DRNN architecture implementation details are provided and the training process is explained. Finally, the distance metric driving the user-clustering partitioning and its algorithmic implementation are specified.

\subsection{Sequential Deep Learning Model Operation}\label{subsec:GRUdynamics}
 Predicting a VR user's tiled-\gls{fov} is an instance of movement prediction where an input sequence of a user's past and current pose vectors is mapped to a multi-label output. In the output, each label represents one tile in the video frame and its value provides an estimate over the likelihood of the tile belonging to the user's future \gls{fov}.
  
 To build our sequential learning model we adopt the \gls{gru} \cite{GRUseminal} architecture, a variant of \glspl{rnn} that uses two simple gating mechanisms whereby long-term dependencies are effectively tackled and the memory/state from previous activations is preserved. Hence, compared to other models such as long short-term memory (LSTM) units \cite{LSTMseminal}, \glspl{gru} are faster to train and have proven to perform better for small datasets \cite{GRUseminal2}, the case considered in Section \ref{subsec:Dataset_FOVAccuracy}. Specifically, for every operation time step \textendash which is measured in terms of video frames and therefore indexed with $\indexFrame\in\indexFrameSet$\textendash~the \gls{gru} units 
 update the value of their hidden state $\bm{h}_{f}$ as a non-linear function of an input sequence $\bm{x}_\ue^\indexFrame$ and of the previous hidden state $\bm{h}_{f-1}$. The non-linear function is parameterized by $\bm{\theta}$ following a recurrence relation $\bm{h}_{f}={\scaleobj{.75}\fint}\bigl(\bm{h}_{f-1},\bm{x}_\ue^f;\bm{\theta}\bigr)$ that is visually sketched in Fig.\ref{fig:detailed-GRU} and formally described by the following model equations:
 
 \begin{align}
	 \Gamma_\indexFrame=&\sigma\bigl(\bm{W}_{\Gamma}\bm{x}_{\ue}^{\indexFrame}+\bm{Z}_{\Gamma}\bm{h}_{\indexFrame-1} +\bm{b}_\Gamma\bigr)\label{GRU_update_unit_level}\\
	 r_\indexFrame=&\sigma\bigl(\bm{W}_r \bm{x}_{\ue}^{\indexFrame}+\bm{Z}_r \bm{h}_{\indexFrame-1}+\bm{b}_r\bigr)\label{GRU_reset_unit_level}\\
	 \bm{h}_\indexFrame=&(1-\bm{\Gamma}_\indexFrame)\circledast \bm{h}_{\indexFrame-1}\nonumber\\&+\bm{\Gamma}_\indexFrame \circledast \texttt{tanh}\bigl(\bm{W}\bm{x}_\ue^{\indexFrame}+\bm Z\bigl(\bm{r}_\indexFrame \circledast \bm{h}_{\indexFrame-1}\bigr)+\bm{b}_h \bigr),\label{GRU_outputState2}
 \end{align}
 
 where weight matrices $\bm{W}_{\Gamma}$, $\bm{Z}_{\Gamma}$, $\bm{W}_r$, $\bm{Z}_r$, $\bm{W}$, $\bm{Z}$ and bias terms $\bm{b}_\Gamma$, $\bm{b}_r$, $\bm{b}_h$ represent the model parameters comprised in $\bm{\theta}$ that, with those of the fully connected neural layer in Fig. \ref{fig:DRNN-BlockDiagram}, are learned during the \gls{drnn} offline training process.

 The value of the update gate vector $\bm{\Gamma}_f$, as per \eqref{GRU_update_unit_level}, governs through the linear interpolation in \eqref{GRU_outputState2} the amount of the previous hidden state $\bm{h}_{f-1}$ and of the new hidden state candidate $\bm{\tilde{h}}_f=\texttt{tanh}\bigl(\bm{W}\bm{x}_\ue^{\indexFrame}+\bm Z\bigl(\bm{r}_\indexFrame \circledast \bm{h}_{\indexFrame-1}\bigr)+b_h \bigr)$ contributing to the next hidden state activation $\bm{h}_f$. 
 Likewise, the reset gate vector $\bm{r}_f$, as per \eqref{GRU_reset_unit_level}, controls the degree of the contribution of the previous hidden state $\bm{h}_{f-1}$ preserved for the new hidden state candidate $\bm{\tilde{h}}_f$. When the contribution from the previous state is deemed irrelevant, the next hidden state  $\bm{\tilde{h}}_f$ is reset and will depend only on the input sequence. 
\vspace*{-0.45cm}
\subsection{DRNN architecture}\label{subsec:DRNN_architecture}
\begin{figure}[!t]
	\centering
	\begin{subfloat}
		\centering
		\includegraphics[width=.85\columnwidth]{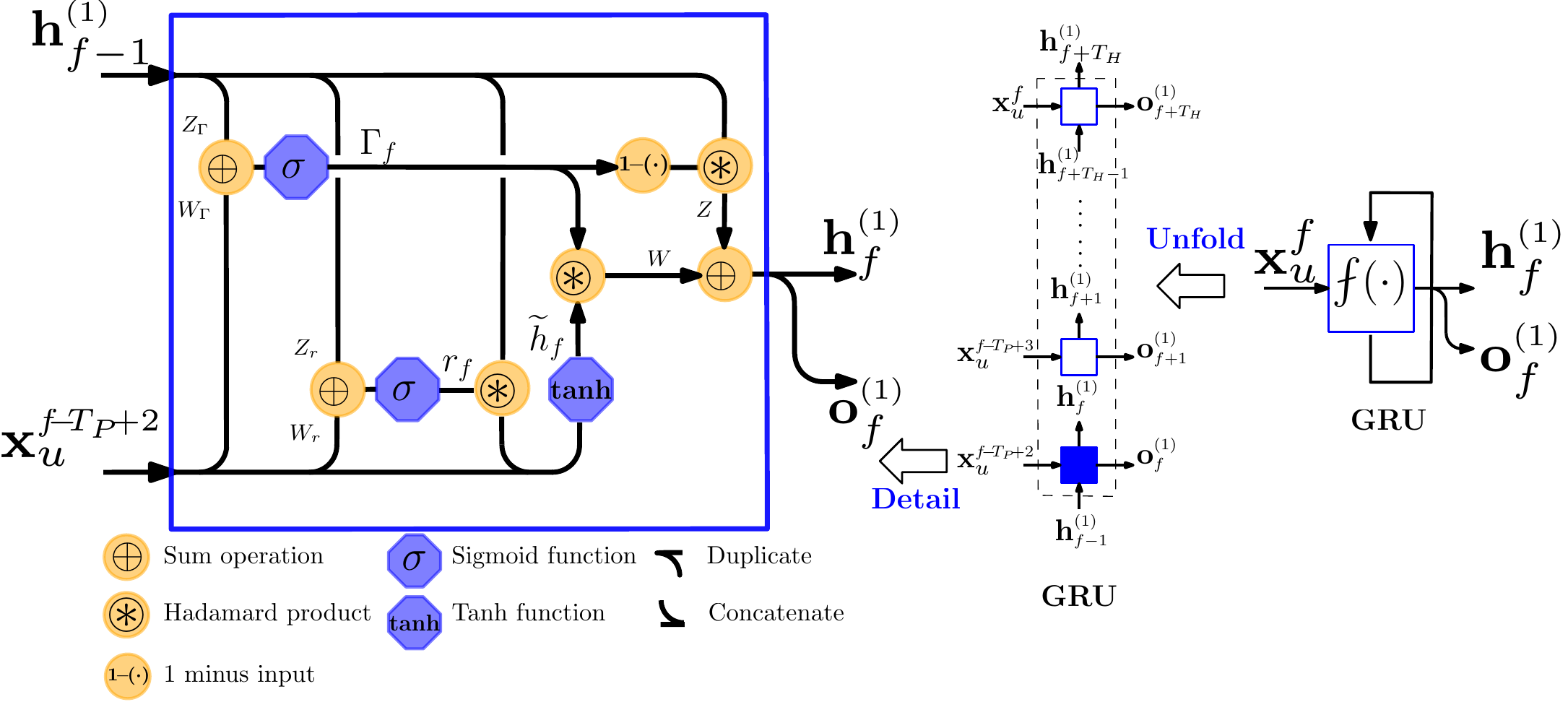}
		\caption{Detailed graphical representation of GRU unfolding and of the \smash{$\bm{h}_{f}^{(1)}$} computation in the unfolded GRU cell. The notation $(\cdot)^{(1)}$ indicates that the GRU at hand belongs to the first layer, which is highlighted in blue in the DRNN architecture from Fig. \ref{fig:DRNN-BlockDiagram}.}
		\label{fig:detailed-GRU}
	\end{subfloat}	
	\vspace{.2cm}
	\begin{subfloat}
		\centering
		\includegraphics[width=.94\columnwidth]{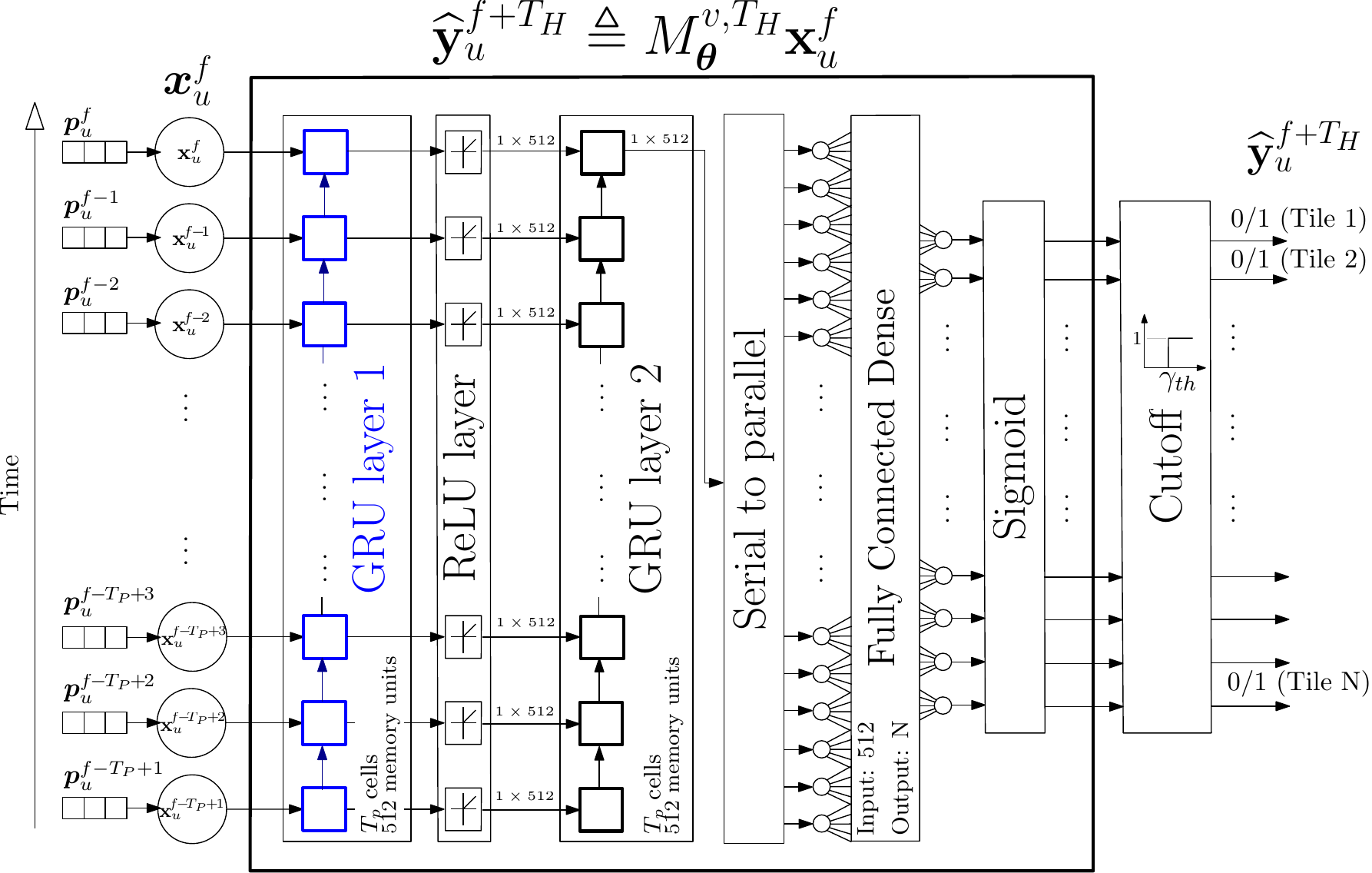}
		\caption{Block diagram of the deep learning model for the edge controller. The DRNN  predicts the tiles in the FoV of user $u$ at frame index $\indexFrameP=f_r+\RNNpredHoriz$, i.e. $\RNNpredHoriz$ frames ahead.}
		\label{fig:DRNN-BlockDiagram}
	\end{subfloat}
\end{figure}

The building blocks of our proposed deep recurrent learning model $M_{\bm{\theta}}^{v,T_{H}}$ based on \gls{gru} layers and implemented using Keras \cite{chollet2015keras}, a high-level neural networks API running on top of a Tensorflow backend,  are represented in Fig. \ref{fig:DRNN-BlockDiagram}.

\textbf{Input representation:} Every $\slotFrame$, an input sequence of size $\RNNseqLenght$ corresponding to \gls{3dof} pose vectors $\bm{x}_\ue^\indexFrame\!\triangleq\!\{\boldsymbol{p}_{3\ue}^{\indexFrame}\}_{\indexFrame=\indexFrameRt-\RNNseqLenght+1}^{\indexFrameRt}$ is fed to the first \gls{gru} layer. 

\textbf{Sequence processing:} The input is then processed following model equations \eqref{GRU_update_unit_level}-\eqref{GRU_outputState2} in Section \ref{subsec:GRUdynamics} by a $\RNNseqLenght$ time-step \gls{gru} cell with a hidden state size equal to 512 examples. Following a \gls{relu} activation that performs a $[z]^+$ operation, the output of the first \gls{gru} layer goes through a second \gls{gru} layer with the same characteristics. The output state from this second layer  $\bm{h}_\indexFrame^{(2)}\triangleq\bm{o}_\indexFrame^{(2)}$ is then fed to a serial to parallel (S/P) layer before going across a dense neural layer that connects with the $N$ output neurons. 

\textbf{Output representation:} Given the multi-label nature of our learning model, a sigmoid activation layer is used to map the $N$ sized dense output into $N$ probability values or logits $\{\probability(n)=\sigma(\bm{W}_{d}\bm{h}_\indexFrame^{(2)}+\bm{b}_d)_n\}_{n=1}^N$ that are Bernoulli distributed, i.e., the probability of each label is treated as independent from other labels' probabilities. The output of the sigmoid is then binarized with a cutoff layer such that 
\begin{equation}
	\widehat {y}_{u,n}^{\indexFrameP}= 
	\begin{cases}
		1, & \sigma(\bm{W}_{d}\bm{h}_\indexFrame^{(2)}+\bm{b}_d)_n\geq \gamma_{th},\\
		0, & \text{otherwise},
	\end{cases}
\end{equation}
where $\bm{W}_{d}$, $\bm{b}_d$ are the weights and biases of the dense fully-connected layer and $\gamma_{th}$ is the threshold value for the cutoff layer, which is chosen to balance accuracy and recall. After the binarization, the predicted \gls{fov} for a user $u$ and frame index \smash{$\indexFrameP=f+\RNNpredHoriz$} is retrieved as $\tileSetFOVPred_{\ue}^{\indexFrameP}=\{ n\in [1,...,N]\hspace{-1mm}:\widehat {y}_{u,n}^{\indexFrameP}=1\}$.
\vspace*{-0.2cm}
\subsection{DRNN Training}\label{subsec:DRNN_training}
The aim of the training in the supervised deep recurrent learning model $M_{\bm{\theta}}^{v,T_{H}}$ is to iteratively find the $\bm\theta$ parameters that minimize a binary cross-entropy loss function $\mathcal{L}(\bm\theta)$ for all training instances. This loss function, for model parameters $\bm\theta$, labels \smash{$y_{u_{tr}^{v},n}^{\indexFrameP}$} and logits $\{\probability(n)\}_{n=1}^N$ captured from the output of the sigmoid layer in Fig. \ref{fig:DRNN-BlockDiagram}, is expressed as
\begin{equation}\label{eq:DRNN_lossfunction}
\mathcal{L}(\bm\theta)\negmedspace=\negmedspace-\frac{1}{N}\negmedspace\sum_{n=1}^N \left[y_{u_{tr}^{v},n}^{\indexFrameP}\!\log(\probability(n))\negmedspace+\negmedspace (1\!-\negmedspace y_{u_{tr}^{v},n}^{\indexFrameP})\log(1\!-\negmedspace\probability(n))\right].
\end{equation}

During the model offline training, \gls{bptt} algorithm \cite{werbosBPTT} and Adam algorithm \cite{AdamAlgo} are used to optimize the gradients. Adam is set with learning rate $\alpha\!=\!0.01$, parameters $\beta_1\!=\!0.9$, $\beta_2\!=\!0.999$ and no decay. The gradient backpropagation is performed over data batches of size 512 and during 20 training epochs. 

Next, the information related to users' physical location and to their predicted \glspl{fov} is leveraged to develop a user clustering scheme.

\subsection{Proposed \gls{fov} and Location Aware User Clustering}\label{subsec:Clustering}
 Once the predictions of the entire set of users $\ueSet$ for a frame index $\indexFrameP$ are ready, users viewing the same video $\video$ are grouped into clusters based on their spatial and content correlations.  
 
 Mathematically, let $\clusterSet_\cluster^{\indexFrameP}$ denote the $\cluster$-th cluster in which the set of users $\ueSet$ is partitioned for frame index \smash{$\indexFrameP=\indexFrame\!+\!\RNNpredHoriz$} such that $\bigcup_{\cluster=1}^\CLUSTER \clusterSet_\cluster^{\indexFrameP}\!\! = \ueSet$. Here, the cluster partitioning can be obtained by i) computing the $|\ueSet|\times|\ueSet|$ distance matrix $\bm{D}^{\indexFrameP}$, whose  $d_{\ue,\ueOther}^{\indexFrameP}=\tilde{d}_{\ue,\ueOther}^{\indexFrameP} \left(d_{\ue\ueOther}^{\text{2D}}/d_{min}^{\text{2D}}\right)$ element results from quantifying the \gls{fov}-related distance or dis-similarity between any pair of users $\{\ue,\ueOther\}\!\in\!\ueSet$ which is given by $\tilde{d}_{\ue,\ueOther}^{\indexFrameP}\!=\!1\!-\!\sum_{\tile=1}^\TILE\mathbb{I}_{\{\tile\in\tileSetFOVPred_{\ue}^{\indexFrameP}\}} \mathbb{I}_{\{\tile\in\tileSetFOVPred_{\ueOther}^{\indexFrameP}\}}/(\TILE\!-\!\sum_{\tile=1}^\TILE\mathbb{I}_{\{\tile\notin\tileSetFOVPred_{\ue}^{\indexFrameP}\}} \mathbb{I}_{\{\tile\notin\tileSetFOVPred_{\ueOther}^{\indexFrameP}\}})$; and ii) scaling it by their relative physical distance $d_{\ue\ueOther}^{\text{2D}}$ divided by $d_{min}^{\text{2D}}$, that denotes the minimum value for such relative distance as per the theater dimensions and seat arrangements.
 
 To implement the clustering scheme that builds on the above distance metric, a hierarchical agglomerative clustering with average linkage has been considered. This clustering scheme allows operating over the constructed dendrogram to increase/decrease the number of resulting clusters and thereby investigating the trade-offs in terms of communication resource utilization versus achieved performance when many/few clusters, as per $\CLUSTER$, are used.  

 Once the clusters $\{\mathcal{C}_k^{\indexFrameP}\}_{k=1}^{K}$ have been estimated using the specific clustering strategy, 
 the cluster-level \gls{fov} is built and ready to be leveraged in the proposed multicast/unicast scheduling strategy as $\tileSetFOVPred_{\clusterSet_{\cluster}}^{\indexFrameP}\!\!=\!\!\bigcup_{u\in\mathcal{C}_k^{\indexFrameP}}\tileSetFOVPred_{\ue}^{\indexFrameP}$.

\section{Simulation and Performance Evaluation}\label{sec:sim_results}
In this section, we numerically validate the effectiveness of the proposed solution. For that purpose, we start by describing the dataset with real head-tracking information for 360$^\circ$ videos and the DRNN \gls{fov} prediction accuracy results which will impact the performance evaluation of the mmWave multicast transmission. Following that, the deployment scenario and the considered baseline schemes are described. Finally, the performance evaluation of the proposed approach is evaluated and some insightful results are discussed\footnote{\label{note8}For the interested reader, a demo showcasing the qualitative results achieved comparing our approach to the baselines described in Section \ref{subsec:scenarios_baseline} is available at https://youtu.be/djt9efjCCEw.\label{demoVideo}}. 

\subsection{360$^\circ$ Video Head-tracking Dataset and DRNN \gls{fov} Prediction Accuracy Results}\label{subsec:Dataset_FOVAccuracy}
To validate our proposed approach, the information fed into the \gls{drnn} for training and simulation corresponds to \gls{3dof} traces from the dataset in \cite{VR_video_dataset} whereby the pose of 50 different users while watching a catalog of $V\!\!=\!\!10$ \gls{hd} 360$^\circ$ videos from YouTube were tracked. The selected videos are 60~s long, have 4K resolution and are encoded at 30~fps. A  $100^\circ\!\times\!100^\circ$ \gls{fov} is considered and, to build the tiled-\gls{fov}, the \gls{eqr} projection of each of the video frames has been divided into $N=200$ square tiles of $192\times192$ pixels arranged in a regular grid of $N_{V}\!=\!10$ and $N_{H}\!=\!20$ tiles. The dataset provides the ground-truth labels after mapping the \gls{3dof} poses to their corresponding tiled FoVs.
In view of the size and characteristics of the dataset, the original 50 users have been split into disjoint $\mathcal{U}_{tr}$ and $\mathcal{U}$ sets for training and for test purposes with cardinalities $|\mathcal{U}_{tr}|\!=\!35$ and $|\mathcal{U}|\!=\!15$, respectively.

	\begin{table}[t!]
		\centering
		{\scriptsize
			\caption{FOV Test Accuracy: Effect of Prediction Horizon}
			\label{table:DRNN_accuracy}
			\resizebox{\columnwidth}{!}{
				\begin{tabular}{|c|c|c|c|c|c|}
					\hline
					\multicolumn{1}{|c|}{\multirow{2}{*}{\textbf{Video}}}&\multicolumn{1}{c|}{\multirow{2}{*}{\textbf{Category}\footnotemark}} & \multicolumn{4}{c|}{\textbf{Jaccard similarity index} $J^{\RNNpredHoriz}_{\video}$ (mean $\pm$ std. dev.)}\\
					\cline{3-6}
					&& $\RNNpredHoriz=5$ & $\RNNpredHoriz=10$ & $\RNNpredHoriz=20$ & $\RNNpredHoriz=30$ \\
					\hline \hline
					SFRSport &NI, SP		& 0.70$\pm$0.06 & 0.69$\pm$0.04	& 0.63$\pm$0.03 & 0.50$\pm$0.05\\
					MegaCoaster &NI, FP	& 0.68$\pm$0.06 & 0.65$\pm$0.05	& 0.64$\pm$0.07 & 0.61$\pm$0.05\\
					RollerCoaster &NI, FP	& 0.74$\pm$0.05 & 0.70$\pm$0.05 & 0.64$\pm$0.04 & 0.63$\pm$0.05 \\
					SharkShipwreck	&NI, SP	& 0.53$\pm$0.03	& 0.48$\pm$0.03 & 0.44$\pm$0.03 & 0.36$\pm$0.03 \\
					Driving &NI, FP	& 0.76$\pm$0.04	& 0.71$\pm$0.04 & 0.63$\pm$0.03 & 0.58$\pm$0.02 \\
					ChariotRace &CG, FP	& 0.71$\pm$0.02 & 0.71$\pm$0.02 & 0.68$\pm$0.02 & 0.65$\pm$0.03 \\
					KangarooIsl&NI, SP	& 0.69$\pm$0.04 & 0.65$\pm$0.03 & 0.63$\pm$0.03 & 0.58$\pm$0.03 \\
					Pac-man&CG, FP		& 0.83$\pm$0.03 & 0.73$\pm$0.05 & 0.67$\pm$0.05 & 0.66$\pm$0.06 \\
					PerilsPanel&NI, SP		& 0.69$\pm$0.02 & 0.65$\pm$0.02 & 0.56$\pm$0.03 & 0.53$\pm$0.03 \\
					HogRider&CG, FP		& 0.68$\pm$0.04 & 0.66$\pm$0.04 & 0.65$\pm$0.04 & 0.57$\pm$0.05 \\
					\hline
				\end{tabular}
			}
			\medskip
			\caption{FOV Test Accuracy: Effect of the number of GRU Layers}
			\label{table:DRNN_accuracy_hidden}
			\begin{tabular}{|c|c|c|c|c|}
				\hline
				\multicolumn{1}{|c|}{\multirow{3}{*}{\textbf{Video}}}   & \multicolumn{4}{c|}{\textbf{Jaccard similarity index} $J^{\RNNpredHoriz}_{\video}$ (mean $\pm$ std. dev.)}\\
				\cline{2-5} 
				&\multirow{2}{*}{$\bm\RNNpredHoriz$} &\multicolumn{3}{c|}{\textbf{Number of GRU layers}} \\ \cline{3-5} 
				&                        & 1                  	  & 2                  & 3      \\
				\hline\hline
				\multirow{4}{*}{MegaCoaster} & 5 & 0.52$\pm$0.08      & 0.68$\pm$0.06      & 0.68 $\pm$ 0.04\\
				& 10& 0.50$\pm$0.07      & 0.65$\pm$0.05      & 0.65 $\pm$ 0.03\\
				& 20& 0.46$\pm$0.07      & 0.64$\pm$0.07      & 0.63 $\pm$ 0.05\\
				& 30& 0.32$\pm$0.04      & 0.61$\pm$0.05      & 0.61 $\pm$ 0.06\\ \hline
				\multirow{4}{*}{Pac-man}  & 5 & 0.82$\pm$0.04      & 0.83$\pm$0.03      & 0.76 $\pm$ 0.04\\
				& 10& 0.60$\pm$0.07      & 0.73$\pm$0.05      & 0.73 $\pm$ 0.04\\
				& 20& 0.53$\pm$0.05      & 0.67$\pm$0.05      & 0.67 $\pm$ 0.05\\
				& 30& 0.49$\pm$0.06      & 0.66$\pm$0.06      & 0.65 $\pm$ 0.05\\
				\cline{1-5} 
			\end{tabular}
		}
	\end{table}
	\footnotetext{With category codes: NI=Natural Image, CG=Computer Generated, SP=Slow-paced, FP=Fast-paced.}
	

Results in Table \ref{table:DRNN_accuracy} represent the accuracy of the prediction models for different values of $\RNNpredHoriz$ in terms of the Jaccard similarity index, which is defined for each user $u$ viewing a frame $\indexFrame$ of a video $\video$ as the intersection over the union between the predicted and the actual \gls{fov} tile sets  $J(\tileSetFOVPred_{\ue}^{\indexFrame},\tileSetFOV_{\ue}^{\indexFrame})=\setSize{\tileSetFOVPred_{\ue}^{\indexFrame}\cap \tileSetFOV_{\ue}^{\indexFrame}}/\setSize{\tileSetFOVPred_{\ue}^{\indexFrame}\cup \tileSetFOV_{\ue}^{\indexFrame}}$. In Table \ref{table:DRNN_accuracy}, this index has been first averaged over the frames of the video at hand, and then over  all the test users, i.e., $J^{\RNNpredHoriz}_{\video}=\frac{1}{\setSize{\ueSet}\setSize{\indexFrameSet}}\sum_{\ue\in\ueSet}\sum_{\indexFrame\in\indexFrameSet}\frac{\setSize{\tileSetFOVPred_{\ue}^{\indexFrame}\cap \tileSetFOV_{\ue}^{\indexFrame}}}{\setSize{\tileSetFOVPred_{\ue}^{\indexFrame}\cup \tileSetFOV_{\ue}^{\indexFrame}}}. $
The results in the table confirm the anticipated decrease of the accuracy as the prediction horizon moves further away from the last reported pose. Similarly, results in Table \ref{table:DRNN_accuracy_hidden} show that increasing the depth of the \gls{drnn} by adding more \gls{gru} layers is counter-productive; it overfits the training data and unnecessarily increases the complexity of the model. 

\subsection{Deployment Details and Reference Baselines}\label{subsec:scenarios_baseline}
 We consider two \gls{vr} theater settings, a small and a medium size capacity theaters with dimensions \smash{$\{s_r,s_c\}\!=\!\{5,10\}$} and \smash{$\{s_r,s_c\}\!=\!\{10,15\}$} seats, respectively. In both configurations the seats are separated from each other by 2 m, and there is a 4 m distance from the seat area to the walls of the enclosure. As detailed in Section \ref{sec:sys_mod}, \glspl{sbs} are located at ceiling level in the upper 4 corners of the theater. 
 A total of $7$ different scenarios are studied for simulation: scenarios sT-\$v correspond to the small theater with $10$ users per video with \smash{$\$=\VIDEO\!=\!\{1,3,5\}$} videos being played; scenarios bT-\$v correspond to the big theater with $15$ users per video with \smash{$\$=\VIDEO\!=\!\{1,3,5,10\}$} videos being played. The set of default parameter values for simulations is provided in Table~\ref{tab:SimParameters}. For benchmarking purposes, the following baseline and proposed schemes are considered:
 \begin{itemize}
	\item $\VRbaselineOne$: Chunk requests are scheduled in real-time for \gls{mmwave} unicast transmission. 
	\item $\VRbaselineTwo$: Chunk requests are scheduled in real-time and multi-beam \gls{mmwave} multicast transmission is used.
	\item $\VRProp$: Chunk requests are proactively scheduled and multi-beam \gls{mmwave} multicast transmission is used.
	\item $\VRPropPlus$: Corresponds to the \emph{proposed} approach which considers $\VRProp$ and the \gls{hrllbb} constraint in the scheduler. 
 \end{itemize}
  	\begin{table}[t!]
 		\renewcommand{\arraystretch}{.95}
 		\caption{Main Simulation Parameters}\label{tab:SimParameters}\vspace{-2mm}
 		\centering
 		\resizebox{0.9\columnwidth}{!}{
 			\begin{tabular}{|L{5.2cm}|C{3.6cm}|}
 				\hline \centering \textbf{Parameter}&\textbf{Value}\\
 				\hline
 				\hline Simulation time& 60000 ms\\
 				Channel coherence time ($T_c$)& 1ms\\
 				Blockage re-evaluation time ($\timeBlockage$) &100 ms\\
 				Transmission slot duration ($\slotTrans$) &0.25 ms\\
 				RF chains & 1 per \gls{hmd}; 4 per \gls{sbs} \\
 				Beam-level Rx beamwidth & 5$^\circ$\\
 				Beam-level Tx beamwidths & [5$^\circ$:5$^\circ$:$45^\circ$]\\
 				Carrier frequency ($\frequency$)& 28 GHz\\
 				Bandwidth ($\bandwidth_\bs$)& 0.85 GHz\\
 				Noise spectral density ($N_0$)&-174 dBm/Hz\\
 				Noise figure & 9 dB\\
 				\gls{sbs} transmit power ($p_{i}$) & 15 dBm\\
 				Motion-to-photon delay ($\MTPDelay$) & 10 ms\\
 				Delay reliability metric $\delayOutage$ & 0.01\\
 				Tiles per frame ($N$) & 200\\
 				Video frame duration ($\slotFrame$)& 33 ms (30 fps video)\\
 				Videos catalog size ($\VIDEO$) & [1, 3, 5, 10]\\
 				Users per video & [10, 15]\\
 				Number of clusters ($K$)& [$2\cdot\VIDEO$, $3\cdot\VIDEO$, $4\cdot\VIDEO$]\\
 				Prediction horizon ($\RNNpredHoriz$)& [5, 10, 20, 30] frames\\
 				DRNN input sequence ($\RNNseqLenght$)& 30 pose values\\
 				DRNN cutoff value ($\gamma_{th}$)& 0.5\\
 				\hline
 			\end{tabular}
 		}
 	\end{table}
 
	\begin{figure*}[!ht]
		\centering
		\includegraphics[width=.87\textwidth]{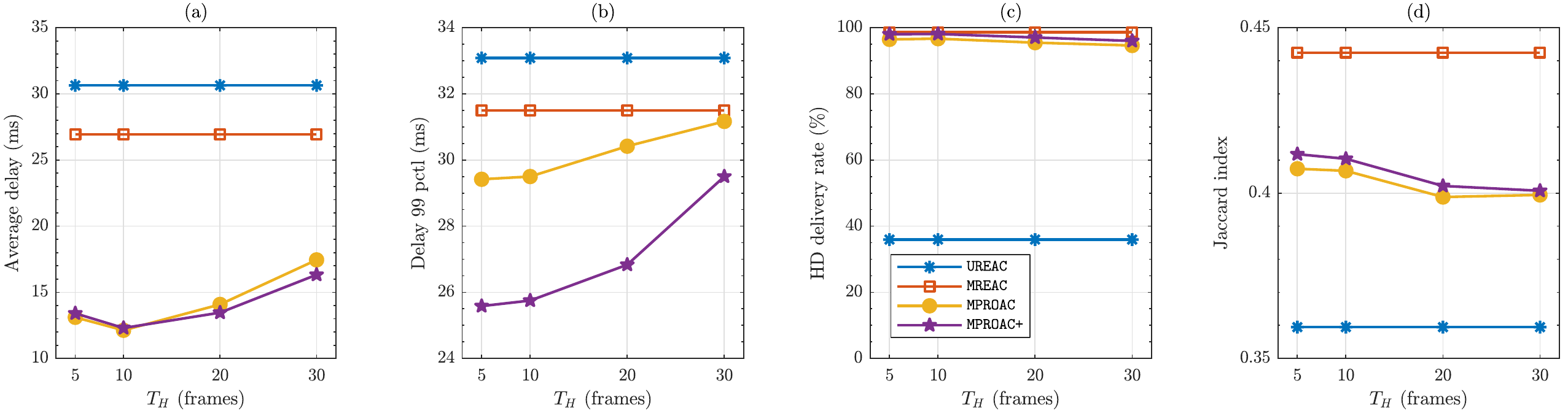}
		\caption{(a) Average delay, (b) $99$th percentile delay, and (c) HD delivery rate and (d) Jaccard index performance versus the prediction horizon (in frames) in \texttt{bT-3v} with a load of $1$ Gbps per user ($0.972$ Mb chunk size), and $\lyapunovTradeoff$ = $1\cdot10^8$.}
		\label{plot:prediction_horizon}
	\end{figure*}
	\begin{figure*}[!ht]
		\centering
		\begin{subfloat}
			\centering
			\includegraphics[width=.87\textwidth]{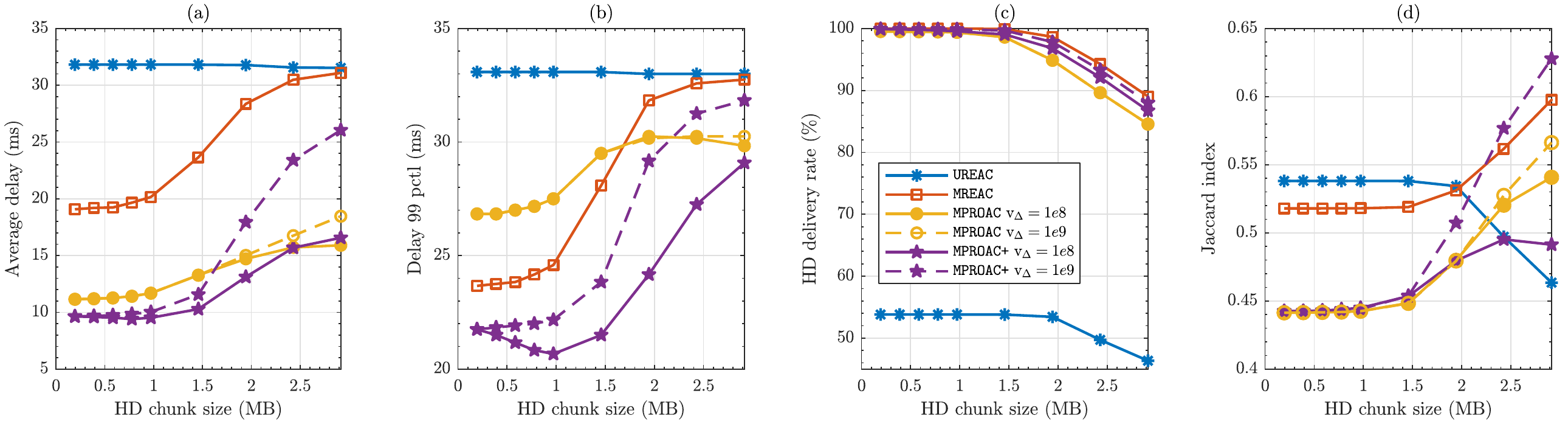}
			\label{plot:rate_case_1}
		\end{subfloat}	
		\begin{subfloat}
			\centering
			\includegraphics[width=.87\textwidth]{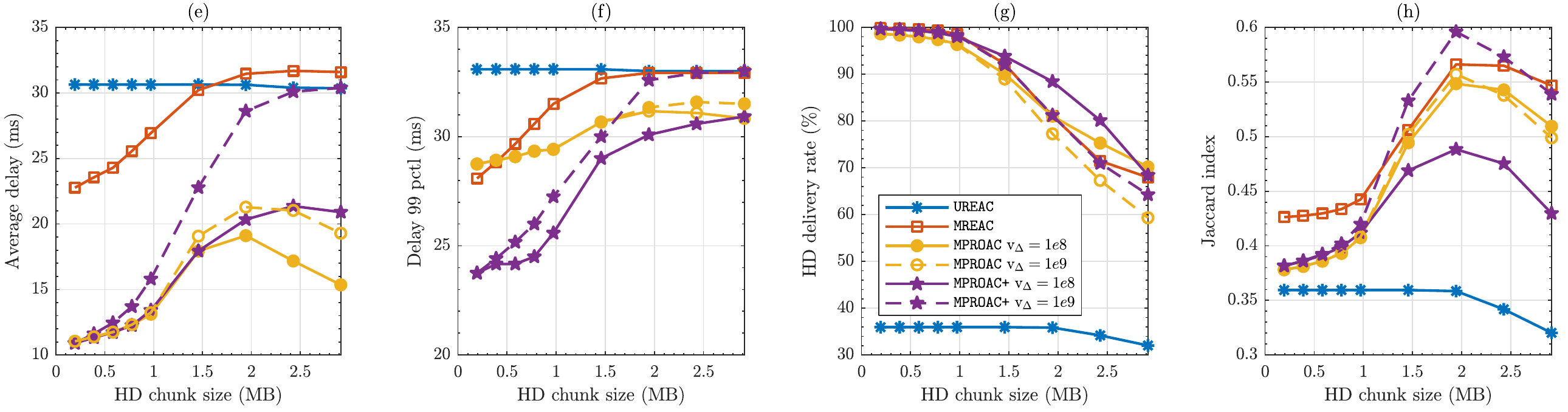}
			\label{plot:rate_case_2}
		\end{subfloat}
		\vspace{-.3cm}
		\caption{(a) and (e) Average delay, (b) and (f) $99$th percentile delay, (c) and (g) HD delivery rate and (d) and (h) Jaccard index performance in \texttt{sT-3v} and \texttt{bT-3v}, respectively, as a function of the HD chunk size, for $\VIDEO\!=\!3$ videos, $\CLUSTER\!=\!2\times\!\VIDEO$ clusters, $\RNNpredHoriz\!=\!5$ frames, and Lyapunov trade-off $\lyapunovTradeoff\!=\!1\!\cdot\!10^8$ and $\lyapunovTradeoff\!=\!1\!\cdot\!10^9$.}	
		\label{plot:rate_case}
	\end{figure*}
	\begin{figure*}[!ht]
		\centering
		\includegraphics[width=.87\textwidth]{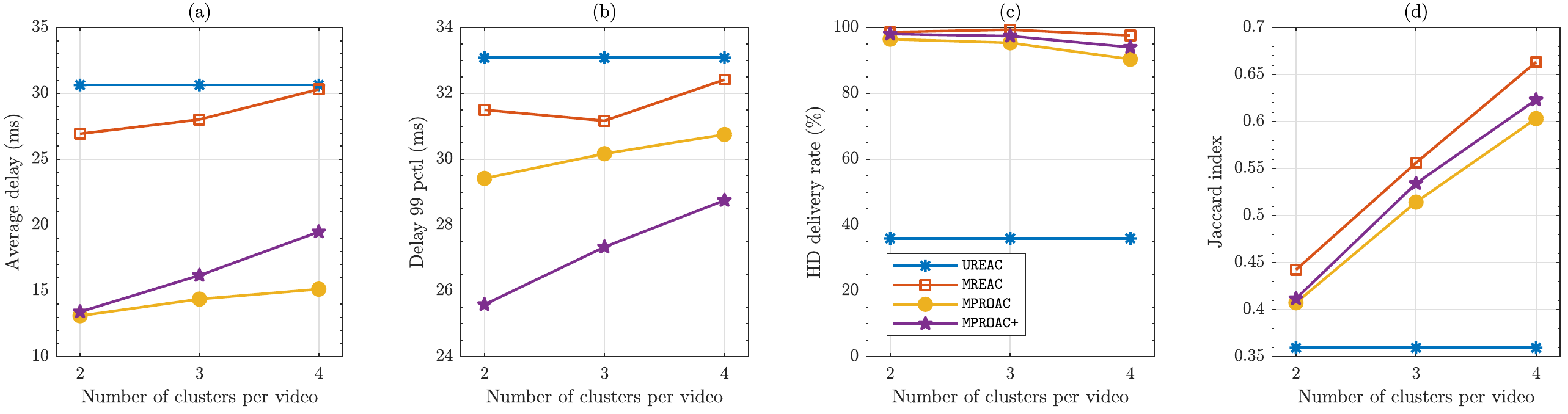}
		\caption{(a) Average delay, (b) $99$th percentile delay, and (c) HD delivery rate and (d) Jaccard index performance versus cluster per video, in \texttt{bT-3v} with a load of $1$ Gbps per user ($0.972$ Mb chunk size),, and $\lyapunovTradeoff$ = $1\cdot10^8$.}
		\label{plot:clusters}
	\end{figure*}

\subsection{Discussion}\label{sim:discussion}
	Next, the impact of the \gls{fov} prediction horizon, the requested video quality, the maximum number of clusters and, the network size are evaluated. 
	To that end, the performance of each scheme is evaluated through its average and the $99$th percentile (delay 99 pctl) transmission delay, calculated as the delay until the last tile in the \gls{fov} of the requesting user has been delivered. 
	We observe here that focusing merely on the average system performance would fail to capture the real-time aspects of the problem. 
	Consequently, to show that the adopted Lyapunov online control method is able to keep the latency bounded below the $\MTPDelay$ threshold with the desired probability $1-\epsilon$, the $99$th percentile delay is provided too. 
	Furthermore, alongside with the $99$th percentile delay plots, the HD successful delivery rate metrics highlight the trade-off between the utility (maximizing the quality) and the probabilistic latency constraint. 

	Lastly, for each user $\ue$ and frame index $\indexFrame$ the Jaccard similarity index between the successfully delivered chunks and the actual \gls{fov} is computed.
	The index is given by 
	$J(\breve{\tileSetFOV}_{\ue}^{\indexFrame},\tileSetFOV_{\ue}^{\indexFrame})\!=\!\setSize{\breve{\tileSetFOV}_{\ue}^{\indexFrame}\cap \tileSetFOV_{\ue}^{\indexFrame}}/\setSize{\breve\tileSetFOV_{\ue}^{\indexFrame}\cup \tileSetFOV_{\ue}^{\indexFrame}}$
	with $\breve{\tileSetFOV}_{\ue}^{\indexFrame}\subseteq\bigcup\{\tileSetFOVPred_{\clusterSet_{\cluster}}^{\indexFrame},\tileSetFOV_{\ue}^{\indexFrame}\}$ denoting the set of tiles correctly decoded at user $\ue\in\clusterSetUeFrame$. 
	We notice here that $\breve{\tileSetFOV}_{\ue}^{\indexFrame}\setminus\tileSetFOV_{\ue}^{\indexFrame}$ represents a measure of the multicasting overhead owing to operating with cluster-level predicted \gls{fov} chunk requests that, hence, will grow larger the less correlated the \glspl{fov} of the cluster-members are. 
	Similarly, $\tileSetFOV_{\ue}^{\indexFrame}\setminus\breve{\tileSetFOV}_{\ue}^{\indexFrame}\neq\emptyset$ hints missed tiles. 
	Therefore the evolution of this index, averaged over all the users and frame indices, globally captures the trade-off related to \gls{fov} correlation among cluster members.

\subsubsection{Impact of the FoV Prediction Horizon}\label{pe_PredHorizon}
	We first consider the impact of the \gls{drnn} prediction horizon $\RNNpredHoriz$ on the performance of the proposed approaches $\VRPropPlus$ and $\VRProp$, and compare it with the reactive baselines $\VRbaselineOne$ and $\VRbaselineTwo$, whose performance is not affected. 
	Intuitively, in our scheme longer $\RNNpredHoriz$ allow the scheduler to schedule future frames earlier, but increases the overall amount of data to be transmitted due to having lower prediction accuracy, as shown in Table \ref{table:DRNN_accuracy}. %
	In Fig.~\ref{plot:prediction_horizon}, it can be seen that the scheduler can maintain high \gls{hd} quality streaming even with long prediction horizons.
	The frame delay is shown to first decrease, due to having more time to schedule chunks in advance, then increases again, due to having to schedule a higher number of chunks in real-time that were missed by the predictor. 
	Transmitting more real time leads to lower utilization of the user's predicted \gls{fov}, which decreases the Jaccard index.  

\subsubsection{Impact of the Requested Video Quality}\label{pe_requestQuality}
	We move on now to investigate the impact of the requested \gls{hd} video quality on the performance of the proposed scheme. %
	To that end, looking into both the sT-$3$v and bT-$3$v scenarios, we evaluate the impact of the quality through the \gls{hd} \emph{chunk size} of the frame shown to the user.  
	The performance metrics of each scheme are depicted in Fig.~\ref{plot:rate_case}(a)-(d) and Fig.~\ref{plot:rate_case}(e)-(h) for the small and big theater scenarios, respectively. 
	The figures clearly show the trade-off between frame delay and \gls{hd} streaming rate. As the chunk size increases, the average and 99th percentile delays increase for the different schemes. 
	Moreover, comparing $\VRbaselineOne$ with the other schemes, it is shown that multicasting brings $40-50\%$ increase in the HD rate and $33-70\%$ latency reduction through the utilization of common FoVs of different users. 
	At high chunk sizes, the higher network load clearly increases the service delay. 
	By delivering the predicted frames in advance, both the  $\VRPropPlus$ and $\VRProp$ minimize the average delay without sacrificing the HD quality rate. 
	The proposed $\VRPropPlus$ scheme is shown to also keep the worst delay values bounded due to imposing the  
	latency constraint, as compared to the $\VRProp$. 
	Further comparing $\VRbaselineOne$ to $\VRbaselineTwo$, it is shown that multicasting significantly reduces the delay due to the utilization of common \glspl{fov} of different users. 

	As the performance of the  $\VRPropPlus$ and $\VRProp$ schemes for different values of the Lyapunov parameter $\lyapunovTradeoff$ are provided, the trade-off between the frame delay and the quality is further illustrated.
	Indeed, the results in Fig.~\ref{plot:rate_case} show that as the $\lyapunovTradeoff$ increases, the scheduling algorithm prioritizes maximizing users' \gls{hd} delivery rate, whereas at lower values of $\lyapunovTradeoff$, the scheduler prioritizes stabilizing the traffic and virtual queues i.e., keeping the delay bounded with high probability. 
	This comes at the expense of having lower \gls{hd} delivery rate. The $\VRPropPlus$ approach also achieves 17-37\% reduction in the 99th percentile latency as compared to $\VRProp$ and $\VRbaselineTwo$ schemes, respectively.  

	Furthermore, the Jaccard similarity in Fig.~\ref{plot:rate_case}(d) and Fig.~\ref{plot:rate_case}(h) illustrates  the trade-offs of utility and latency versus transmission utilization. 
	At low traffic loads, high quality rate and low latency result in lower Jaccard index, which is due to the large amount of extra data sent due to sending an estimated \gls{fov}. 
	As the traffic load increases, the proactive schemes transmits more real-time frames, which increases the Jaccard index. 
	The Jaccard index decreases again at higher traffic loads as the effect of missed frames goes up (the average delay approaches the deadline as can be seen in Fig.~\ref{plot:rate_case}(a), Fig.~\ref{plot:rate_case}(e)) 

\subsubsection{Impact of the Number of Clusters}\label{pe_clusterNum}
	Subsequently, we examine how the maximum number of clusters affects the performance of the multicast schemes, as compared to the $\VRbaselineOne$ scheme, which operates in unicast. 
	Fig.~\ref{plot:clusters} shows that a lower number of clusters allows for higher content reuse of the overlapping areas in the \glspl{fov} of users, which results in lower delay and higher \gls{hd} quality rate. 
	By allowing a higher number of clusters per video, however, higher Jaccard similarity indices are scored, as less unnecessary chunks are sent to users, as shown in Fig.~\ref{plot:clusters}(d).

	\begin{figure}[!t]
		\centering
		\includegraphics[width=.89\columnwidth]{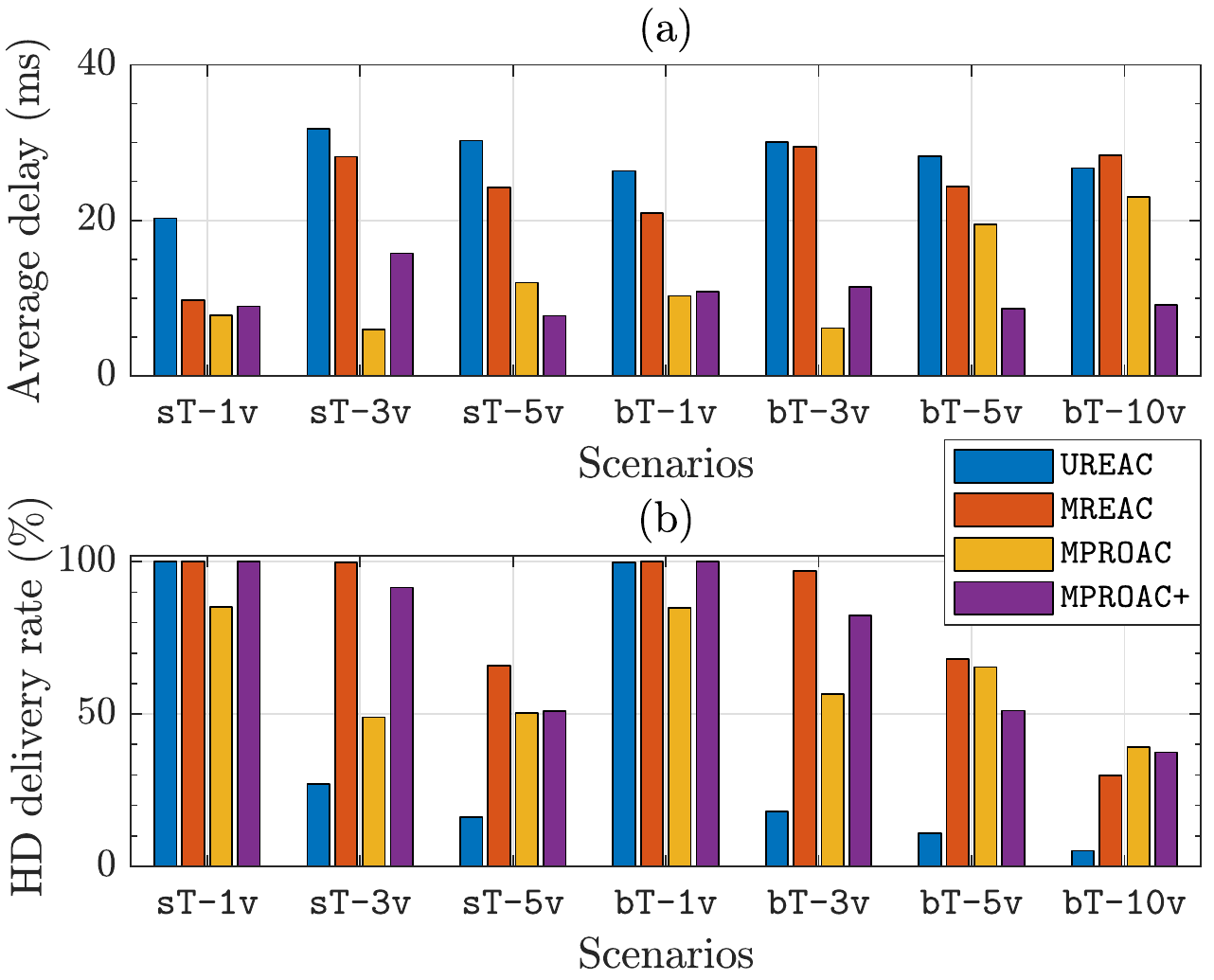}
		\caption{(a) Average delay and (b) HD delivery rate performance for different network scenarios with a load of $1$ Gbps per user ($0.972$ Mb chunk size). }
		\label{plot:all_cases}
	\end{figure}

\subsubsection{Impact of the Network Size}\label{pe_networkSize}
	Finally, the effect of the network size is analyzed. 
	To do so, both the small and the big theater configurations are considered under an increasing amount of users and videos. 
	In Fig.~\ref{plot:all_cases}, it is shown that the proposed scheme achieves close to 100\% \gls{hd} streaming rate in scenarios 1, 2, 4, and 5 while maintaining lower frame delay. 
	Moreover, in the congested scenarios with high number of users and videos, i.e., scenarios 3, 6, and 7, the results show that multicasting provides substantial performance improvement through the gains of $\VRbaselineTwo$ over $\VRbaselineOne$.
	This demonstrates the capability of multicasting to minimize the latency of \gls{vr} streaming to multi-user scenarios. 
	Although the large amount of requested data in these congested scenarios limits the available resources to schedule the predicted frames in advance, the results in Fig.~\ref{plot:all_cases} show that the proposed scheme $\VRPropPlus$ can achieve higher \gls{hd} delivery rate and lower delay compared to the baselines.

\section{Conclusions\label{sec:concl}}
	In this paper, we have formulated the problem of maximizing users' \gls{vr} streaming \gls{qoe} as a network-wide \gls{hd} frame admission maximization problem subject to low latency constraints with very high reliability.
	We have proposed a Lyapunov-framework based approach which transforms the stochastic optimization problem into a series of successive instantaneous static optimization subproblems. Subsequently, for each time instant, a matching theory algorithm is applied to allocate \gls{sbs} to user clusters and leverage a \gls{mmwave} multicast transmission of the \gls{hd} chunks. %
	Using simulations, we have shown that the proposed \gls{drnn} can  predict the \gls{vr} users' future \gls{fov} with high accuracy. 
	The predictions of this model are used to cluster users and proactively schedule schedule the multicast transmission of their future video chunks.
	Furthermore, simulations have provided evidence of considerable gains achieved by the proposed model when compared to both reactive baseline schemes, while notably outperforming the unicast transmission baseline.
	\vspace{-2mm}

\begin{appendices} 
\section{Proof of Lemma 1}
\noindent By leveraging the inequality $(\max[x,o])^2\!\leq\! x^2$ for $x\!\geq\!0$, after squaring the physical and virtual queues in \eqref{eq:queue_update}, \eqref{eq:aux_Q} and \eqref{eq:delay_Q} the upper bounds for the above terms are derived as
	\begin{align}
		\queueUe^2(\slot\!+\!1)-\queueUe^2(\slot)&\leq \admissionTotUe^2(\slot)+\!\!\!\!\!\!\!\sum_{\indexFrame=\{\indexFrameRt,\indexFrameP\}}\negmedspace\sum_{\chunk\in\tileSetFOVExt_\ue^{\indexFrame}}\Bigl(\rateUeChunk^2(\slot)
		-2\admissionTotUe(\slot)\mu_{uc_{f}}(t)\Bigr)\nonumber\\
		&-2\queueUe(\slot)\bigl(\mu_{uc_{f}}(t)-\admissionTotUe(\slot)\bigr),\label{eq:app_ineq1}\\
		\auxVQueueUe^2(\slot\!+\!1)-\auxVQueueUe^2(\slot)&\leq \admissionTotUe^2(\slot)+\auxVbleUe^2(\slot)-2\admissionTotUe(\slot)\auxVbleUe(\slot)\nonumber\\
		&-2\auxVQueueUe(\slot)\bigl(\admissionTotUe(\slot)-\auxVbleUe(\slot)\bigr),\label{eq:app_ineq2}\\
		\latencyVQueueUeChunk^2(\slot\!+\!1)-\latencyVQueueUeChunk^2(\slot)&\leq \queueUe^2(\slot+1)\big(\indicator{\tMTP}\!-\!\epsilon_{d}\big)^2\nonumber\\
		&+2\latencyVQueueUeChunk( \slot)\big(\indicator{\tMTP}\!-\!\epsilon_{d}\big)\queueUe(\slot\!+\!1),\label{eq:app_ineq3}
	\end{align}

\noindent with the one time slot Lyapunov drift given by
	\begin{align}\label{eq:app_drift}
		\lyapunovDrift&\triangleq\lyapunov{\lyapunovQueueCombined(\slot+1)}-\lyapunov{\lyapunovQueueCombined(\slot)}\nonumber\\
		&=\frac{1}{2}\sum_{\ue\in\ueSet}\bigg\{ \Bigl(\queueUe^{2}(\slot+1)-\queueUe^{2}(\slot)\Bigr)+\Bigl({\auxVQueueUe^{2}(\slot+1)}-{\auxVQueueUe^{2}(\slot)}\Bigr)\nonumber\\
		&+\sum_{\indexFrame\in\indexFrameSet}\Bigl({\latencyVQueueUeChunk^{2}(\slot+1)}-{\latencyVQueueUeChunk^{2}(\slot)}\Bigr)\bigg\}.
	\end{align}

Replacing the term $\queueUe(\slot+1)$ in \eqref{eq:app_ineq3} with $\queueUe(\slot+1)\allowbreak = \queueUe(\slot) +  \sum_{\indexFrame=\{\indexFrameRt,\indexFrameP\}}\sum_{\chunk\in\tileSetFOVExt_\ue^{\indexFrame}}\admissionTotUe(\slot) - \rateUeChunk(\slot)$ due to the fact that having \smash{$\indicator{\tMTP}=1$} entails a non-empty queue guarantee, and combining \eqref{eq:app_ineq1}-\eqref{eq:app_ineq3}, an upperbound on the drift function can be expressed as \eqref{eq:app_ineq4}.
	\begin{figure*}[!ht]
		\normalsize  
		\begin{align}\label{eq:app_ineq4}
		\lyapunovDrift&\!\leq\!\frac{1}{2}\!\!\sum_{u\in\ueSet}\Bigg[\!
		\!-\!2\big(\indicator{\tMTP}\!-\negmedspace\epsilon_{d}\big)^2\bigg\{\queueUe(\slot)\Bigl(\!\!\!\!\sum_{\indexFrame=\{\indexFrameRt,\indexFrameP\}}\!\sum_{\chunk\in\tileSetFOVExt_\ue^{\indexFrame}}\!\!\!\rateUeChunk(\slot)-\admissionTotUe(\slot)\Bigr)\bigg\}_{\#a}
		\!+\big(\indicator{\tMTP}\!-\negmedspace\epsilon_{d}\big)^2\bigg\{\queueUe^2(\slot)+\Bigl(\!\!\!\!\sum_{\indexFrame=\{\indexFrameRt,\indexFrameP\}}\!\sum_{\chunk\in\tileSetFOVExt_\ue^{\indexFrame}}\!\!\!\rateUeChunk(\slot)-\admissionTotUe(\slot)\Bigr)^2\bigg\}_{\#b}\nonumber\\
		&\!+\bigg\{\Bigl(\!\!\!\!\sum_{\indexFrame=\{\indexFrameRt,\indexFrameP\}}\!\sum_{\chunk\in\tileSetFOVExt_\ue^{\indexFrame}}\!\rateUeChunk(\slot)-\admissionTotUe(\slot)\Bigr)^2+\Bigl(\admissionTotUe(\slot)-\auxVbleUe(\slot)\Bigr)^2 \bigg\}_{\#e}
		\!\!\!-2\bigg\{\auxVQueueUe(\slot)\Bigl(\admissionTotUe(\slot)-\auxVbleUe(\slot)\Bigr)\bigg\}_{\#f}
		\!\!\!-2\bigg\{\queueUe(\slot)\Bigl(\!\!\!\!\sum_{\indexFrame=\{\indexFrameRt,\indexFrameP\}}\!\sum_{\chunk\in\tileSetFOVExt_\ue^{\indexFrame}}\rateUeChunk(\slot)-\admissionTotUe(\slot)\Bigr)\bigg\}_{\#g}\nonumber\\
		&\!+2\big(\indicator{\tMTP}\!-\!\epsilon_{d}\big)\bigg\{\queueUe(\slot)\!\!\!\!\sum_{\indexFrame=\{\indexFrameRt,\indexFrameP\}}\latencyVQueueUeChunk(\slot)\bigg\}_{\#c}
		\!\!\!-2\big(\indicator{\tMTP}\!-\!\epsilon_{d}\big)\bigg\{\!\!\sum_{\indexFrame=\{\indexFrameRt,\indexFrameP\}}\sum_{\chunk\in\tileSetFOVExt_\ue^{\indexFrame}}\latencyVQueueUeChunk(\slot)\Bigl(\rateUeChunk(\slot)-\admissionTotUe(\slot)\Bigr)\bigg\}_{\#d}\Bigg].
		\end{align}
		\hrulefill 
		\vspace*{-4pt}
	\end{figure*}

	Note that the terms $\#b$, $\#c$, and $\#e$ in \eqref{eq:app_ineq4} are quadratic, therefore upper bounded to comply with the assumption of queue stability. Hence, let 
	
	\begin{align}
		\lyapunovConst(t)\!&\!\ge\frac{1}{2}\!\sum_{u\in\ueSet}\Bigg\{
		2\big(\indicator{\tMTP}\!-\negmedspace\epsilon_{d}\big)\bigg\{\queueUe(\slot)\!\!\!\!\sum_{\indexFrame=\{\indexFrameRt,\indexFrameP\}}\latencyVQueueUeChunk(\slot)\bigg\}\\
		&+\bigl(\indicator{\tMTP}\!-\!\epsilon_{d}\bigr)^2\bigg\{\queueUe^2(\slot)\!+\!\Bigl(\!\!\!\!\sum_{\indexFrame=\{\indexFrameRt,\indexFrameP\}}\!\sum_{\chunk\in\tileSetFOVExt_\ue^{\indexFrame}}\!\!\rateUeChunk(\slot)\!-\!\admissionTotUe(\slot)\Bigr)^2\bigg\}\nonumber\\
		&+\!\bigg\{\Bigl(\!\!\!\!\sum_{\indexFrame=\{\indexFrameRt,\indexFrameP\}}\!\sum_{\chunk\in\tileSetFOVExt_\ue^{\indexFrame}}\!\rateUeChunk(\slot)-\admissionTotUe(\slot)\Bigr)^2\!+\!\Bigl(\admissionTotUe(\slot)-\auxVbleUe(\slot)\Bigr)^2 \bigg\}\Bigg\}\nonumber
	\end{align}
	be the constant parameter at each time instant $t$ collecting the aforementioned terms from the drift above. 
	After subtracting the penalty term $\lyapunovTradeoff\mathbb{E}\bigl[U\bigl(\bigl\{\auxVbleUe(\slot)\bigr\}\bigr)\bigr]$ on both sides of \eqref{eq:app_ineq4}, and further operating on $\#a$, $\#d$ and $\#g$ to denote $\lyapunovWeight$ the term  

	\begin{equation}
	\lyapunovWeight(\slot)=\big[\big(\indicator{\tMTP}-\epsilon_{d}\big)\queueUe(\slot)+
	\sum_{\indexFrame\in\indexFrameSet}\latencyVQueueUeChunk(\slot)\big]\big(\indicator{\tMTP}-\epsilon_{d}\big)+\queueUe(\slot), \nonumber
	\end{equation}
	we have that
	\begin{align*}
		\lyapunovDrift\!-\!\lyapunovTradeoff\mathbb{E}\Bigl[U\Bigl(\Bigl\{\!\auxVbleUe(\slot)\Bigr\}\Bigr)\Bigr]\!&\leq\lyapunovConst(t)-\!\!\sum_{u\in\ueSet}\!\lyapunovTradeoff\mathbb{E}\Bigl[U\Bigl(\bigl\{\auxVbleUe(\slot)\bigr\}\Bigr)\Bigr]\\
		&-\!\!\sum_{u\in\ueSet}\!\alpha_\ue(\slot)\Bigl(\!\!\!\!\sum_{\indexFrame=\{\indexFrameRt, \indexFrameP\}}\!\!\sum_{\chunk\in\tileSetFOVExt_\ue^{\indexFrame}}\!\!\!\!\!\rateUeChunk(\slot)\!-\!\admissionTotUe(\slot)\Bigr)\\
		&-\!\!\sum_{u\in\ueSet}\!\auxVQueueUe(\slot)\Bigl(\admissionTotUe(\slot)\!-\!\auxVbleUe(\slot)\Bigr),
		\label{eq:app_ineq5}
	\end{align*}
	which after some more rearrangements yields equation \eqref{eq:drift_plus_penalty_bound}.
\end{appendices}

\bibliographystyle{IEEEtran}
\bibliography{IEEEabrv,bibfile_VRJournal}

\begin{thebibliography}{10}
\providecommand{\url}[1]{#1}
\csname url@samestyle\endcsname
\providecommand{\newblock}{\relax}
\providecommand{\bibinfo}[2]{#2}
\providecommand{\BIBentrySTDinterwordspacing}{\spaceskip=0pt\relax}
\providecommand{\BIBentryALTinterwordstretchfactor}{4}
\providecommand{\BIBentryALTinterwordspacing}{\spaceskip=\fontdimen2\font plus
\BIBentryALTinterwordstretchfactor\fontdimen3\font minus
  \fontdimen4\font\relax}
\providecommand{\BIBforeignlanguage}[2]{{%
\expandafter\ifx\csname l@#1\endcsname\relax
\typeout{** WARNING: IEEEtran.bst: No hyphenation pattern has been}%
\typeout{** loaded for the language `#1'. Using the pattern for}%
\typeout{** the default language instead.}%
\else
\language=\csname l@#1\endcsname
\fi
#2}}
\providecommand{\BIBdecl}{\relax}
\BIBdecl

\bibitem{ejder_VR_2017}
E.~Ba\c{s}tu\u{g}, M.~Bennis \emph{et~al.}, ``Toward interconnected virtual
  reality: Opportunities, challenges, and enablers,'' \emph{{IEEE} Commun.
  Mag.}, vol.~55, no.~6, pp. 110--117, June 2017.

\bibitem{VisualEffectInPresence}
C.~Hendrix and W.~Barfield, ``Presence within virtual environments as a
  function of visual display parameters,'' \emph{Presence: Teleoperators and
  Virtual Environments}, vol.~5, no.~3, pp. 274--289, 1996.

\bibitem{ciscoVisualIndexGlobalMobile}
Cisco, ``Cisco visual networking index: Global mobile data traffic forecast
  update, 2017–2022,'' institution, Tech. Rep., 2 2017.

\bibitem{Network_VR}
M.~S. Elbamby, C.~Perfecto \emph{et~al.}, ``Toward low-latency and
  ultra-reliable virtual reality,'' \emph{{IEEE} Netw.}, vol.~32, no.~2, pp.
  78--84, March 2018.

\bibitem{jrn:URLLC_Mehdi}
M.~Bennis, M.~Debbah, and H.~V. Poor, ``Ultra-reliable and low-latency wireless
  communication: Tail, risk and scale,'' \emph{Proc. {IEEE}}, vol. 106, no.~10,
  pp. 1834--1853, 10 2018.

\bibitem{PIEEE_VR}
M.~{Zink}, R.~{Sitaraman}, and K.~{Nahrstedt}, ``Scalable 360$^\circ$ video
  stream delivery: Challenges, solutions, and opportunities,''
  \emph{Proceedings of the IEEE}, vol. 107, no.~4, pp. 639--650, April 2019.

\bibitem{doppler_EUCNC_2017}
K.~Doppler, E.~Torkildson, and J.~Bouwen, ``On wireless networks for the era of
  mixed reality,'' in \emph{Proc. Eur. Conf. on Netw. and Commun. (EuCNC)},
  June 2017, pp. 1--6.

\bibitem{jnl:GazeTrack_2018}
P.~Lungaro, R.~Sjoberg \emph{et~al.}, ``Gaze-aware streaming solutions for the
  next generation of mobile vr experiences,'' \emph{{IEEE} Trans. Vis. Comput.
  Graphics}, vol.~24, no.~4, pp. 1535--1544, 2018.

\bibitem{JChakareski_ICC17_viewport}
X.~Corbillon, G.~Simon \emph{et~al.}, ``Viewport-adaptive navigable 360-degree
  video delivery,'' in \emph{Proc. of {IEEE} Int. Conf. on Commun. (ICC)},
  2017, pp. 1--7.

\bibitem{hosseini_divideconquer_2016}
M.~Hosseini and V.~Swaminathan, ``Adaptive 360 {VR} video streaming: Divide and
  conquer,'' in \emph{{IEEE} Int. Symp. on Multimedia (ISM)}, Dec 2016, pp.
  107--110.

\bibitem{qian_optimCell_2016}
F.~Qian, L.~Ji \emph{et~al.}, ``Optimizing 360$^\circ$ video delivery over
  cellular networks,'' in \emph{Proc. Int. Conf. Mobile Comp. and Netw.
  (MOBICOM)}, New York, NY, USA, 2016, pp. 1--6.

\bibitem{Xiao_OptTiling360_2017}
M.~Xiao, C.~Zhou \emph{et~al.}, ``Optile: Toward optimal tiling in 360-degree
  video streaming,'' in \emph{Proc. of ACM Conf. on Multimedia}, ser. MM
  '17.\hskip 1em plus 0.5em minus 0.4em\relax New York, NY, USA: ACM, 2017, pp.
  708--716.

\bibitem{Ghosh_tileRateAdapt_2017}
A.~Ghosh, V.~Aggarwal, and F.~Qian, ``A rate adaptation algorithm for
  tile-based 360-degree video streaming,'' \emph{CoRR}, vol. abs/1704.08215,
  2017.

\bibitem{walid_machine_learning}
M.~Chen, U.~Challita \emph{et~al.}, ``Machine learning for wireless networks
  with artificial intelligence: {A} tutorial on neural networks,'' \emph{CoRR},
  vol. abs/1710.02913, 2017.

\bibitem{Cornia2018PredictingLSTMSaliency}
M.~Cornia, L.~Baraldi \emph{et~al.}, ``Predicting human eye fixations via an
  lstm-based saliency attentive model,'' \emph{IEEE Trans. Image Proc.},
  vol.~27, pp. 5142--5154, 2018.

\bibitem{Bao_MotionPredic360VR_2016}
Y.~Bao, H.~Wu \emph{et~al.}, ``Shooting a moving target:
  Motion-prediction-based transmission for 360-degree videos,'' in \emph{{IEEE}
  Int. Conf. on Big Data}, Dec 2016, pp. 1161--1170.

\bibitem{saliency:HowDoPeopleExploreInVR}
V.~{Sitzmann}, A.~{Serrano} \emph{et~al.}, ``Saliency in vr: How do people
  explore virtual environments?'' \emph{IEEE Trans. Visualization and Comp.
  Graphics}, vol.~24, no.~4, pp. 1633--1642, 4 2018.

\bibitem{nguyen2018yourAttentionIsUnique}
A.~Nguyen, Z.~Yan, and K.~Nahrstedt, ``Your attention is unique: Detecting
  360-degree video saliency in head-mounted display for head movement
  prediction,'' in \emph{Proc. ACM Int. Conf. on Multimedia}, 2018, pp.
  1190--1198.

\bibitem{nguyen2019_360saliencyDataset}
A.~Nguyen and Z.~Yan, ``A saliency dataset for 360-degree videos,'' in
  \emph{Proc. ACM Multimedia Systems Conference}.\hskip 1em plus 0.5em minus
  0.4em\relax ACM, 2019, pp. 279--284.

\bibitem{VR_video_dataset_Appl}
C.-L. Fan, J.~Lee \emph{et~al.}, ``Fixation prediction for 360$^\circ$; video
  streaming in head-mounted virtual reality,'' in \emph{Proc. Wksp on Network
  and Op. Sys. Support for Dig. Audio and Video}.\hskip 1em plus 0.5em minus
  0.4em\relax New York, NY, USA: ACM, 2017, pp. 67--72.

\bibitem{Li2019_LongTermFoVPrediction}
C.~Li, W.~Zhang \emph{et~al.}, ``Very long term field of view prediction for
  360-degree video streaming,'' \emph{CoRR}, vol. abs/1902.01439, 2019.

\bibitem{Xu2018DRLPredictingHM}
M.~Xu, Y.~Song \emph{et~al.}, ``Predicting head movement in panoramic video: A
  deep reinforcement learning approach.'' \emph{IEEE Trans. Pattern Analysis
  and Machine Intelligence}, 2018.

\bibitem{Zhang2019DRLfor360VR}
Y.~{Zhang}, P.~{Zhao} \emph{et~al.}, ``Drl360: 360-degree video streaming with
  deep reinforcement learning,'' in \emph{IEEE Conf. on Comp. Commun. (INFOCOM
  2019)}, April 2019, pp. 1252--1260.

\bibitem{Dey_VRAR_2017}
X.~Hou, Y.~Lu, and S.~Dey, ``Wireless {VR/AR} with edge/cloud computing,'' in
  \emph{Int. Conf. Comp. Commun. and Netw. (ICCCN)}, July 2017, pp. 1--8.

\bibitem{mangiante_VREdge_2017}
S.~Mangiante, K.~Guenter \emph{et~al.}, ``{VR} is on the edge: How to deliver
  360$^\circ$ videos in mobile networks,'' in \emph{Proc. ACM SIGCOMM. Wksh. on
  VR/AR Network}, 2017.

\bibitem{JChakareski_cachingEdgeVR_2017}
J.~Chakareski, ``{VR/AR} immersive communication: Caching, edge computing, and
  transmission trade-offs,'' in \emph{Proc. ACM SIGCOMM. Wksh. on VR/AR
  Network}.\hskip 1em plus 0.5em minus 0.4em\relax New York, NY, USA: ACM,
  2017, pp. 36--41.

\bibitem{WCNC_mmWave_VR}
M.~S. Elbamby, C.~Perfecto \emph{et~al.}, ``Edge computing meets
  millimeter-wave enabled {VR}: Paving the way to cutting the cord,'' in
  \emph{2018 {IEEE} Wireless Commun. and Netw. Conf. (WCNC)}, April 2018.

\bibitem{MTao_ComputingCachingVR_2018}
Y.~Sun, Z.~Chen \emph{et~al.}, ``Communication, computing and caching for
  mobile {VR} delivery: Modeling and trade-off,'' in \emph{{IEEE} Int. Conf. on
  Commun. (ICC)}, May 2018, pp. 1--6.

\bibitem{conf:NokiaWCNC_Prasad_2018}
A.~Prasad, M.~A. Uusitalo \emph{et~al.}, ``Challenges for enabling virtual
  reality broadcast using 5g small cell network,'' in \emph{2018 {IEEE}
  Wireless Commun. and Netw. Conf. Wksh.s {(WCNCW)}}, April 2018.

\bibitem{conf:Nokia5G-WF_Prasad_2018}
A.~Prasad, A.~Maeder, and M.~A. Uusitalo, ``Optimizing over-the-air virtual
  reality broadcast transmissions with low-latency feedback,'' in \emph{{IEEE}
  5G World Forum {(5G-WF)}}, July 2018.

\bibitem{carlsson2019hadUlookedWhereIdid}
N.~Carlsson and D.~Eager, ``Had you looked where i'm looking: Cross-user
  similarities in viewing behavior for 360 video and caching implications,''
  \emph{arXiv preprint arXiv:1906.09779}, 2019.

\bibitem{Walid_Asilomar_datacorrelation}
M.~Chen, W.~Saad \emph{et~al.}, ``Echo state transfer learning for data
  correlation aware resource allocation in wireless virtual reality,'' in
  \emph{Asilomar Conf. Signals, Syst., and Comp.}, Oct. 2017.

\bibitem{MulticastPHY2006}
N.~D. Sidiropoulos, T.~N. Davidson, and Z.-Q. Luo, ``Transmit beamforming for
  physical-layer multicasting,'' \emph{{IEEE} Trans. Signal Process.}, vol.~54,
  no.~6, pp. 2239--2251, June 2006.

\bibitem{ChannelModel2018}
3GPP, ``{ETSI TR 138 901 V14.3.0}: {5G}; {S}tudy on channel model for
  frequencies from 0.5 to 100 {GHz} ({3GPP TR 38.901} version 14.3.0 release
  14),'' European Telecommunications Standards Institute (ETSI), Tech. Rep.,
  2018.

\bibitem{jnl:StatBlockageModelling_Raghavan2019}
V.~Raghavan, L.~Akhoondzadeh-Asl \emph{et~al.}, ``Statistical blockage modeling
  and robustness of beamforming in millimeter-wave systems,'' \emph{IEEE
  Transactions on Microwave Theory and Techniques}, 2019.

\bibitem{wildman_2DsecAntenna_2014}
J.~Wildman, P.~H.~J. Nardelli \emph{et~al.}, ``On the joint impact of beamwidth
  and orientation error on throughput in directional wireless poisson
  networks,'' \emph{{IEEE} Trans. Wireless Commun.}, vol.~13, no.~12, pp.
  7072--7085, Dec. 2014.

\bibitem{neely_lyapunov}
M.~J. Neely, ``Stochastic network optimization with application to
  communication and queueing systems,'' \emph{Synthesis Lect. on Commun.
  Netw.}, vol.~3, no.~1, pp. 1--211, 2010.

\bibitem{book:Lyapunov_Georgiadis2006}
L.~Georgiadis, M.~J. Neely, and L.~Tassiulas, \emph{Resource Allocation and
  Cross-Layer Control in Wireless Networks}, ser. Foundations and Trends in
  Networking.\hskip 1em plus 0.5em minus 0.4em\relax Now Publishers Inc., 2006.

\bibitem{gale_shapley_1992}
A.~Roth and M.~Sotomayor, \emph{Two-sided matching: A study in game-theoretic
  modeling and analysis}.\hskip 1em plus 0.5em minus 0.4em\relax Cambridge
  University Press, 1992.

\bibitem{walid_matching_2015}
Y.~Gu, W.~Saad \emph{et~al.}, ``Matching theory for future wireless networks:
  fundamentals and applications,'' \emph{{IEEE} Commun. Mag.}, vol.~53, no.~5,
  pp. 52--59, May 2015.

\bibitem{gale_shapley_1962}
D.~Gale and L.~S. Shapley, ``College admissions and the stability of
  marriage,'' \emph{Am. Math. Mon.}, vol.~69, no.~1, pp. 9--15, 1962.

\bibitem{ZhouMatchingD2D16}
Z.~{Zhou}, K.~{Ota} \emph{et~al.}, ``Energy-efficient matching for resource
  allocation in d2d enabled cellular networks,'' \emph{{IEEE} Trans. Veh.
  Technol.}, vol.~66, no.~6, pp. 5256--5268, June 2017.

\bibitem{DiMatchingNoma16}
B.~{Di}, L.~{Song}, and Y.~{Li}, ``Sub-channel assignment, power allocation,
  and user scheduling for non-orthogonal multiple access networks,''
  \emph{{IEEE} Trans. Wireless Commun.}, vol.~15, no.~11, pp. 7686--7698, Nov
  2016.

\bibitem{GRUseminal}
K.~Cho, B.~van Merri{\"{e}}nboer \emph{et~al.}, ``Learning phrase
  representations using {RNN} encoder--decoder for statistical machine
  translation,'' in \emph{Proc. Conf. Emp. Methods Natural Lang. Process.
  (EMNLP)}, Oct. 2014, pp. 1724--1734.

\bibitem{LSTMseminal}
S.~Hochreiter and J.~Schmidhuber, ``Long short-term memory,'' \emph{Neural
  Comput.}, vol.~9, no.~8, pp. 1735--1780, Nov 1997.

\bibitem{GRUseminal2}
J.~Chung, C.~Gulcehre \emph{et~al.}, ``\BIBforeignlanguage{English
  (US)}{Empirical evaluation of gated recurrent neural networks on sequence
  modeling},'' in \emph{\BIBforeignlanguage{English (US)}{NIPS 2014 Wksh. on
  Deep Learning}}, 2014.

\bibitem{chollet2015keras}
F.~Chollet \emph{et~al.}, ``Keras,'' \url{https://keras.io}, 2015.

\bibitem{werbosBPTT}
P.~J. Werbos, ``Backpropagation through time: what it does and how to do it,''
  \emph{Proc. {IEEE}}, vol.~78, no.~10, pp. 1550--1560, Oct 1990.

\bibitem{AdamAlgo}
D.~P. Kingma and J.~Ba, ``Adam: {A} method for stochastic optimization,''
  \emph{CoRR}, vol. abs/1412.6980, 2014.

\bibitem{VR_video_dataset}
W.-C. Lo, C.-L. Fan \emph{et~al.}, ``360$^\circ$ video viewing dataset in
  head-mounted virtual reality,'' in \emph{Proc. ACM Conf. on Multimedia
  Syst.}, 2017, pp. 211--216.

\end{thebibliography}

	\vspace*{-2cm}
	\begin{IEEEbiography}[{\includegraphics[width=1in,height=1.25in,clip,keepaspectratio]{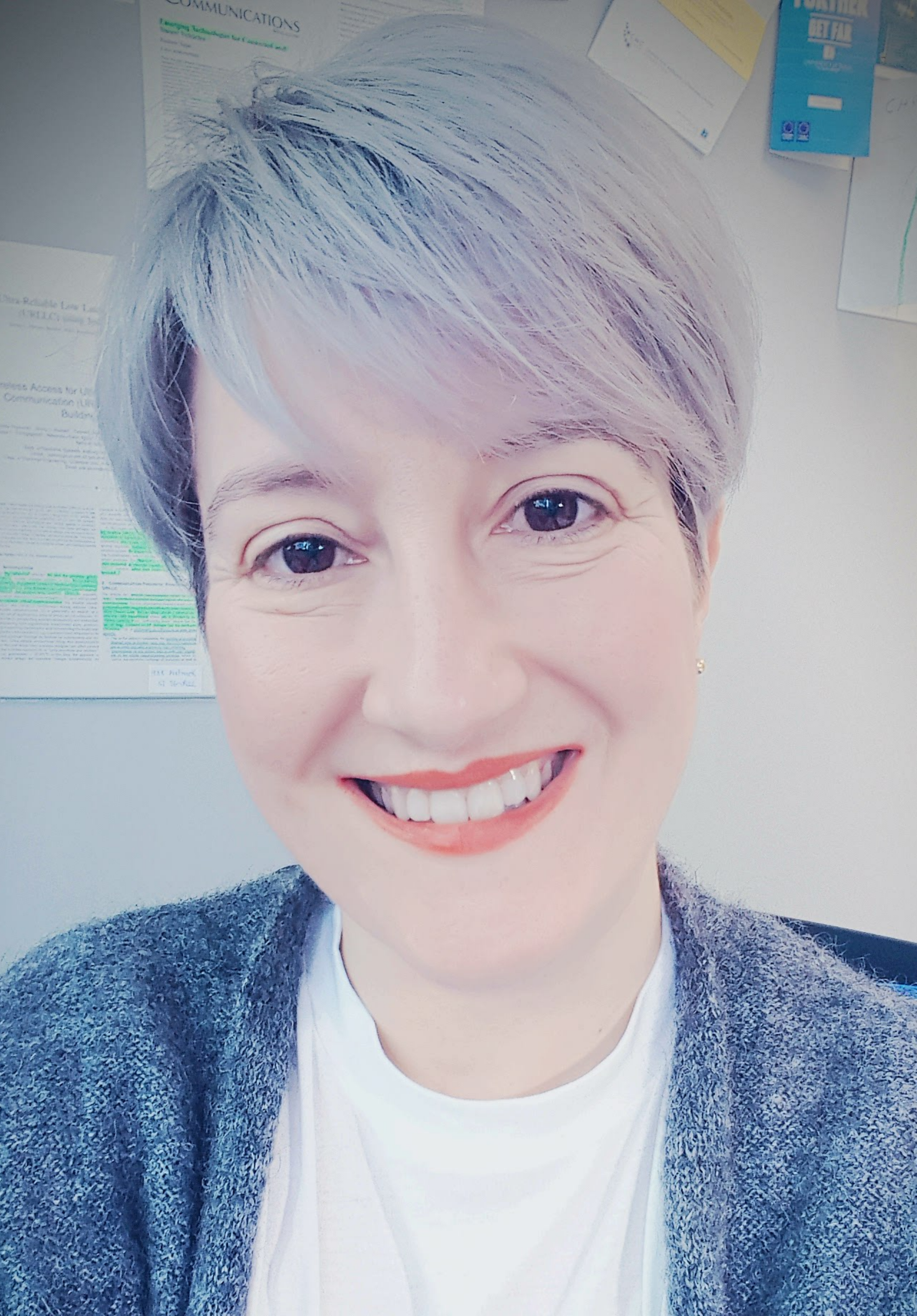}}]{Cristina Perfecto} (S'15-AM'20) received her M.Sc. in Telecommunication Engineering and Ph.D. (with distinction) on Information and Communication Technologies for Mobile Networks from the University of the Basque Country (UPV/EHU) in 2000 and 2019, respectively. She is currently an Associate Professor with the Department of Communications Engineering at this same University. During 2016, 2017 and 2018 she was a Visiting Researcher at the Centre for Wireless Communications (CWC), University of Oulu, Finland where she worked on the application of multidisciplinary computational intelligence techniques in millimeter wave communications radio resource management. Her current research interests lie on the use of machine learning and data analytics, including different fields such as metaheuristics and bio-inspired computation, for resource optimization in 5G networks and beyond.
	\end{IEEEbiography}
	\vspace*{-1cm}
	\begin{IEEEbiography}[{\includegraphics[width=1in,height=1.25in,clip,keepaspectratio]{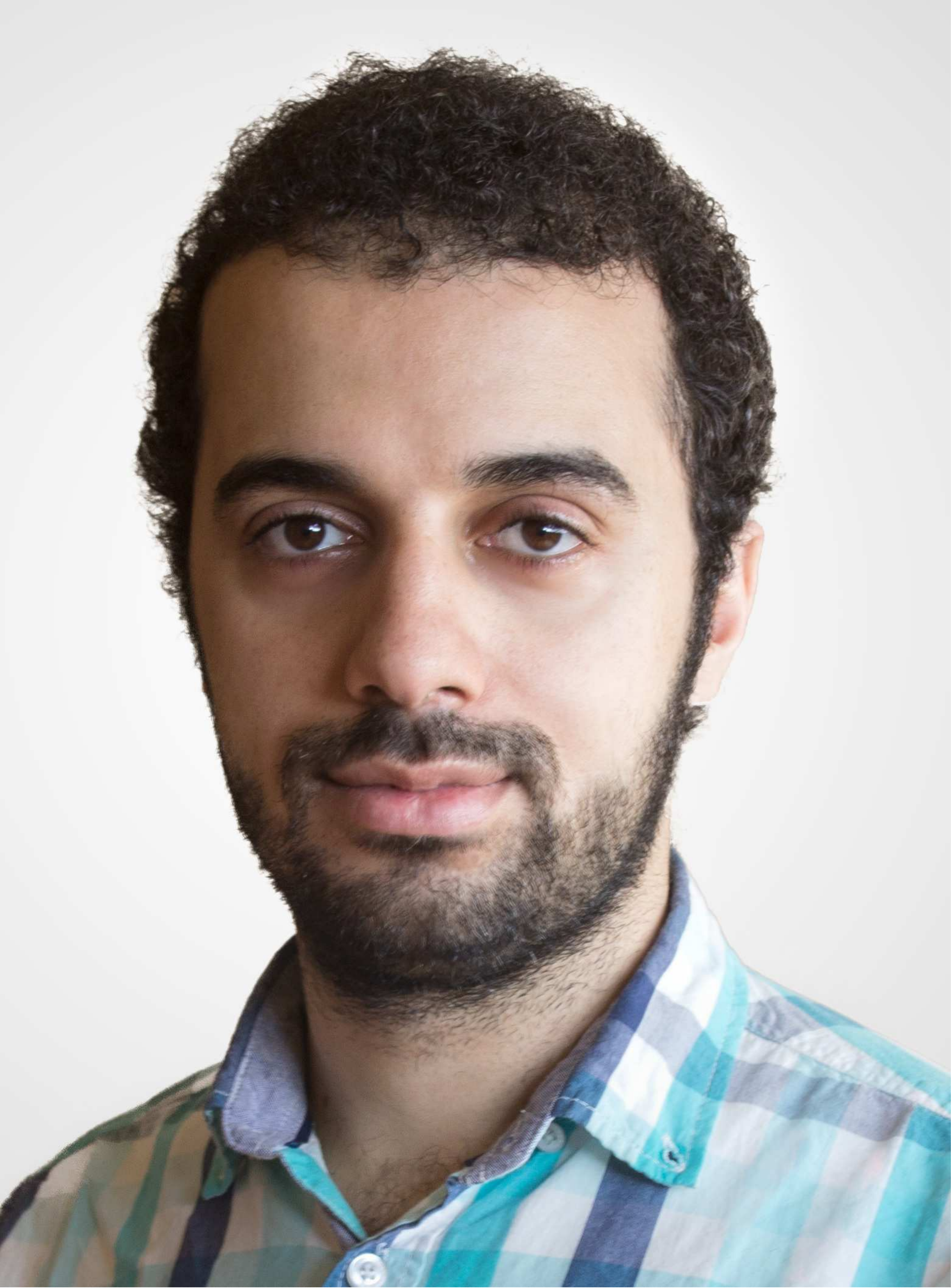}}]{Mohammed S. Elbamby} received the B.Sc. degree (Hons.) in Electronics and Communications Engineering from the Institute of Aviation Engineering and Technology, Egypt, in 2010, the M.Sc. degree in Communications Engineering from Cairo University, Egypt, in 2013, and the Dr.Sc. Degree (with distinction) in Communications Engineering from the University of Oulu, Finland, in 2019. He is currently with Nokia Bell Labs in Espoo, Finland. Previously, he held  research positions at the University of Oulu, Cairo University, and the American University in Cairo. His research interests span resource optimization, network management, and machine learning in wireless cellular networks. He received the Best Student Paper Award from the European Conference on Networks and Communications (EuCNC'2017).
	\end{IEEEbiography}
	\vspace*{-1cm}
	\begin{IEEEbiography}[{\includegraphics[width=1in,clip,keepaspectratio]{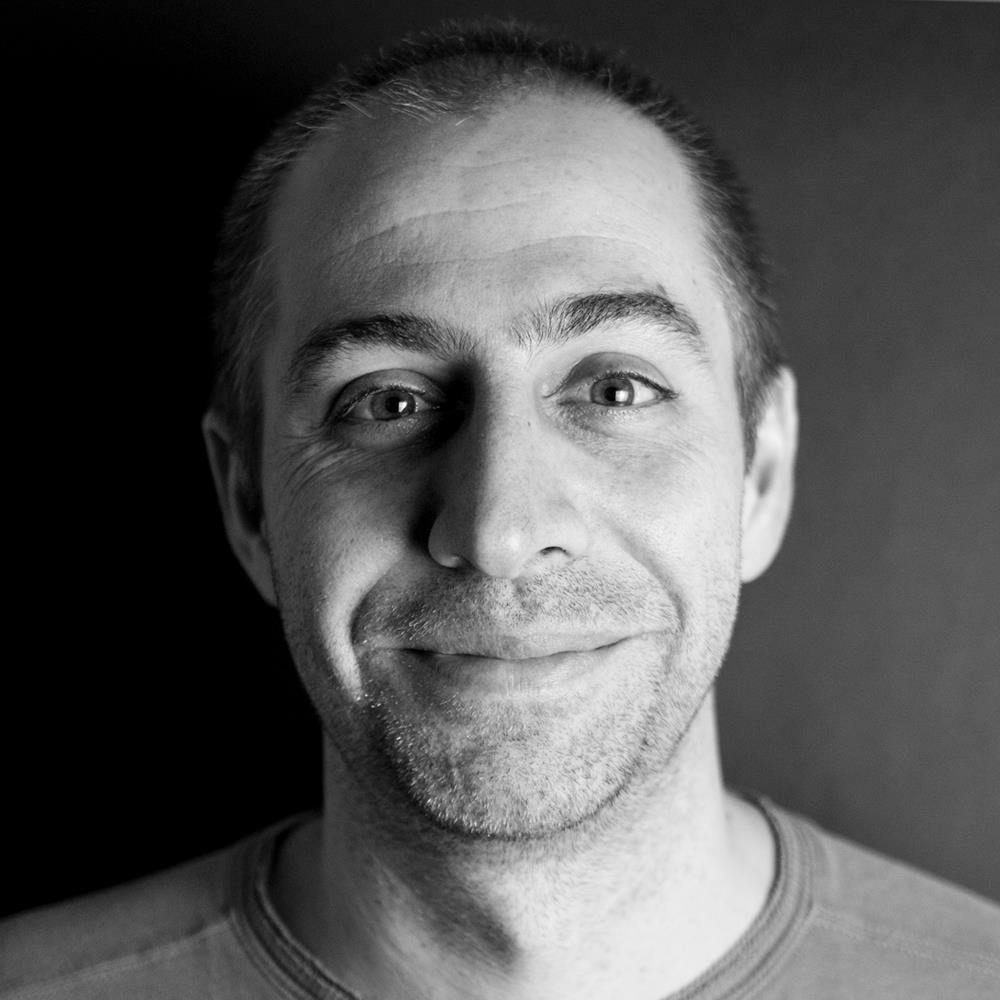}}]{Javier Del Ser} (M'07-SM'12) received his first Ph.D. degree (cum laude) in Electrical Engineering from the University of Navarra (Spain) in 2006, and a second Ph.D. degree (cum laude, extraordinary Ph.D. prize) in Computational Intelligence from the University of Alcala (Spain) in 2013. He is currently a Research Professor in Artificial Intelligence at TECNALIA, Spain. He is also an adjunct professor at the University of the Basque Country (UPV/EHU), an invited research fellow at the Basque Center for Applied Mathematics (BCAM), an a senior AI advisor at the technological startup SHERPA.AI. He is also the coordinator of the Joint Research Lab between TECNALIA, UPV/EHU and BCAM (http://jrlab.science), and the director of the TECNALIA Chair in Artificial Intelligence implemented at the University of Granada (Spain). His research interests gravitate on the design of artificial intelligence methods for data mining and optimization problems emerging from Industry 4.0, intelligent transportation systems, smart mobility, logistics and health, among other fields of application. He has published more than 290 scientific articles, co-supervised 10 Ph.D. theses, edited 7 books, co-authored 9 patents and participated/led more than 40 research projects. He is an Associate Editor of tier-one journals from areas related to data science and artificial intelligence such as Information Fusion, Swarm and Evolutionary Computation and Cognitive Computation.
	\end{IEEEbiography}
	\vspace*{-1cm}
	\begin{IEEEbiography}[{\includegraphics[width=1in,clip,keepaspectratio]{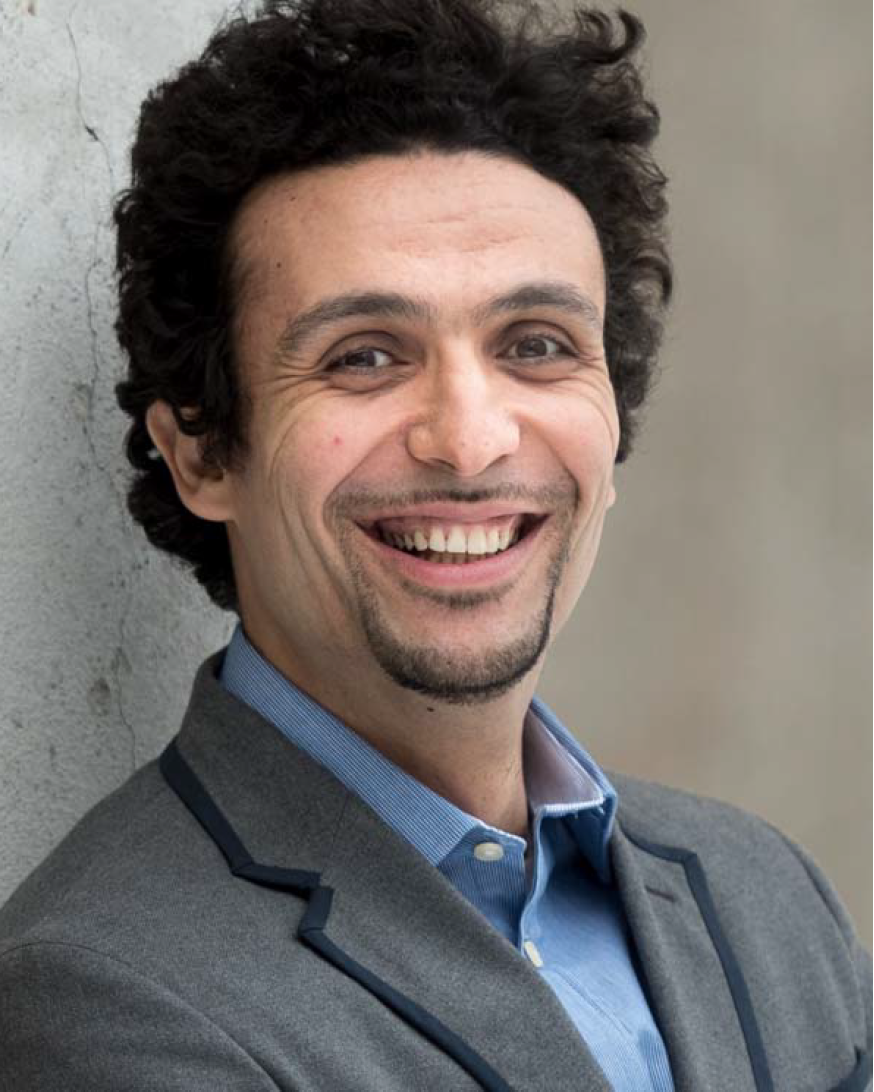}}]{Mehdi Bennis} (S'07-AM'08-SM'15) is an Associate Professor at the Centre for Wireless Communications (CWC), University of Oulu, Finland, an Academy of Finland Research Fellow and head of the Intelligent COnnectivity and Networks/systems group (ICON). His main research interests are in radio resource management, heterogeneous networks, game theory and machine learning in 5G networks and beyond. He has co-authored one book and published more than 200 research papers in international conferences, journals and book chapters. He has been the recipient of several prestigious awards including the 2015 Fred W. Ellersick Prize from the IEEE Communications Society, the 2016 Best Tutorial Prize from the IEEE Communications Society, the 2017 EURASIP Best paper Award for the Journal of Wireless Communications and Networks, the all-University of Oulu award for research and the 2019 IEEE ComSoc Radio Communications Committee Early Achievement Award. Dr Bennis is an editor for the IEEE TRANSACTIONS ON COMMUNICATIONS. 
	\end{IEEEbiography}

	\vfill
\end{document}